%% file: main.tex
\newif\ifJP        \JPfalse
\newif\ifARXIV     \ARXIVtrue
\newif\ifANON      \ANONfalse
\newif\ifLINENO    \LINENOfalse
\newif\ifCASES     \CASESfalse
\newif\ifFUNDING   \FUNDINGfalse

\ifJP \RequirePackage{plautopatch}\fi

\edef\ELSopts{preprint,11pt,authoryear,nonatbib%
  \ifJP ,dvipdfmx\fi
  \ifANON ,doubleblind\fi}
\expandafter\documentclass\expandafter[\ELSopts]{elsarticle}

\usepackage[a4paper]{geometry}
\usepackage{xeCJK}
\setCJKsansfont{HaranoAjiGothic-Medium.otf}
\xeCJKsetup{CJKmath=true}
\xeCJKDeclareCharClass{CJK}{"2015,"2014,"2010}

\ifJP
  \DeclareFontShape{JY2}{mc}{m}{n}{<->s*[0.924690]upjisr-h}{}
  \DeclareFontShape{JY2}{gt}{m}{n}{<->s*[0.924690]upjisg-h}{}
  \DeclareFontShape{JT2}{mc}{m}{n}{<->s*[0.924690]upjisr-v}{}
  \DeclareFontShape{JT2}{gt}{m}{n}{<->s*[0.924690]upjisg-v}{}
  \DeclareFontShape{JY2}{mc}{m}{it}{<->ssub*gt/m/n}{}
  \DeclareFontShape{JT2}{mc}{m}{it}{<->ssub*gt/m/n}{}
\fi

\ifJP
  
\else
  
\fi

\usepackage{amsmath,amssymb,amsfonts,amsthm}
\usepackage{booktabs,array,tabularx}
\usepackage[figuresleft]{rotating}
\usepackage{tikz}
\usetikzlibrary{arrows.meta,positioning,calc,fit,backgrounds,shapes.geometric}
\usepackage{csquotes}
\usepackage[style=apa, sortcites=true, sorting=nyt, date=year, giveninits=true,
            backend=biber, doi=false, refsection=part]{biblatex}
\DeclareSourcemap{
  \maps[datatype=bibtex]{
    \map{
      \step[fieldset=annotation, null]
    }
  }
}

\ifJP
  \usepackage[dvipdfmx]{hyperref}
  \usepackage{pxjahyper}
\else
  \usepackage{hyperref}
\fi
\hypersetup{
  colorlinks = true,
  citecolor  = blue,
  linkcolor  = blue,
  urlcolor   = blue,
  filecolor  = blue,
  bookmarksnumbered = true,
}

\DeclareFieldFormat{citehyperref}{%
  \DeclareFieldAlias{bibhyperref}{noformat}%
  \bibhyperref{#1}}
\DeclareFieldFormat{textcitehyperref}{%
  \DeclareFieldAlias{bibhyperref}{noformat}%
  \bibhyperref{%
    #1%
    \ifbool{cbx:parens}{\bibcloseparen\global\boolfalse{cbx:parens}}{}}}
\savebibmacro{cite}
\savebibmacro{textcite}
\renewbibmacro*{cite}{%
  \printtext[citehyperref]{\restorebibmacro{cite}\usebibmacro{cite}}}
\renewbibmacro*{textcite}{%
  \ifboolexpr{%
       ( not test {\iffieldundef{prenote}} and
         test {\ifnumequal{\value{citecount}}{1}} )
    or ( not test {\iffieldundef{postnote}} and
         test {\ifnumequal{\value{citecount}}{\value{citetotal}}} )}
    {\DeclareFieldAlias{textcitehyperref}{noformat}}{}%
  \printtext[textcitehyperref]{\restorebibmacro{textcite}\usebibmacro{textcite}}}


\journal{International Journal of Artificial Intelligence in Education}

\usepackage{etoolbox}
\makeatletter
\newcommand{\XXprefix}{}
\let\ORIGlabel\label \let\ORIGref\ref
\let\ORIGpageref\pageref \let\ORIGeqref\eqref
\renewcommand{\label}[1]{\ORIGlabel{\XXprefix#1}}
\renewcommand{\ref}[1]{\ORIGref{\XXprefix#1}}
\renewcommand{\pageref}[1]{\ORIGpageref{\XXprefix#1}}
\renewcommand{\eqref}[1]{\ORIGeqref{\XXprefix#1}}
\newcommand{\SAVEnumbering}{%
  \global\let\SVthesection\thesection
  \global\let\SVthesubsection\thesubsection
  \global\let\SVthesubsubsection\thesubsubsection
  \global\let\SVthefigure\thefigure
  \global\let\SVthetable\thetable
  \global\let\SVtheequation\theequation}
\newcommand{\RESTOREnumbering}{%
  \global\let\thesection\SVthesection
  \global\let\thesubsection\SVthesubsection
  \global\let\thesubsubsection\SVthesubsubsection
  \global\let\thefigure\SVthefigure
  \global\let\thetable\SVthetable
  \global\let\theequation\SVtheequation
  \setcounter{section}{0}\setcounter{figure}{0}\setcounter{table}{0}%
  \setcounter{equation}{0}\setcounter{footnote}{0}}
\makeatother

\begin{document}

\ifLINENO \@ifundefined{linenumbers}{}{\linenumbers}\fi

\begin{frontmatter}

\ifARXIV
\ifJP
\title{Prompt Engineeringなしで稼働する研究ロジック合成ツール\mbox{Vibe Compiler}\\
―生成AI時代のAgency維持のためのメタ認知機能向上を目指して―}
\else
\title{Vibe Compiler: A Research-Logic Synthesis Tool That Runs without
Prompt Engineering\\
---Toward Enhancing Metacognition for Sustaining Agency in the Age of
Generative AI---\tnoteref{jp}}
\fi
\fi

\ifJP\else
\tnotetext[jp]{本稿の後半に日本語版を収録している．\ A Japanese version of this paper follows the English version.}
\fi

\ifJP
\else
\author[jaist]{Riichiro Mizoguchi\corref{cor1}}
\ead{mizo@jaist.ac.jp}

\author[omu]{Tomoki Aburatani}
\ead{aburatani.tomoki@omu.ac.jp}

\author[kanagawa]{Kento Koike}
\ead{kento@koike.app}

\author[tohoku]{Machi Shimmei}
\ead{machi.shimmei.e6@tohoku.ac.jp}

\cortext[cor1]{Corresponding author}

\affiliation[jaist]{organization={Japan Advanced Institute of Science and Technology},
            city={Nomi},
            state={Ishikawa},
            country={Japan}}

\affiliation[omu]{organization={Osaka Metropolitan University},
            city={Osaka},
            country={Japan}}

\affiliation[kanagawa]{organization={Kanagawa University},
            city={Yokohama},
            state={Kanagawa},
            country={Japan}}

\affiliation[tohoku]{organization={Tohoku University},
            city={Sendai},
            state={Miyagi},
            country={Japan}}
\fi

\begin{abstract}
\ifJP
\else
Used as a capable servant, generative AI has greatly accelerated intellectual
work, yet it also risks eroding human epistemic agency by encouraging
uncritical acceptance of AI-generated reasoning. Preserving that agency calls
for mechanisms that augment human metacognition during AI-assisted work. We
therefore propose the Synthesis--Analysis Reciprocity Model, which views
intellectual construction as a reciprocal interaction between two cognitive
functions. Synthesis selects and combines the components of the artifact;
Analysis evaluates them critically against objective indicators and
constrains the Synthesis that follows.

Grounded in this model, we present the Vibe Compiler, a research-logic
compiler that helps researchers turn vague intuitions (Vibes) into coherent
research logic. The system attempts to compile those intuitions against a
paper ontology of 16 academic parameters. It treats
compilation failures as signs that logical components are missing. Rather than fill those gaps
autonomously, it returns reflective questions that prompt researchers to
develop the missing reasoning themselves.

We further characterize the origins of structural gaps along two orthogonal
dimensions: cognitive function (Synthesis versus Analysis) and executing
agent (human versus AI). The four resulting types of origin give a principled way to
identify where breakdowns in intellectual construction arise. Crucially, our
design implements the type in which the AI probes its own synthesized output,
itself driven by the user's Vibes, and thereby stimulates human
metacognition. This choice raises researchers from
passive ``Makers'' of the output into ``Managers'' who critically direct and
validate what the AI produces.

In building a prototype on NotebookLM and Gemini, we found that effective
AI-assisted reasoning depends less on sophisticated prompting than on the
structure of the knowledge supplied to the AI. The proposed framework has two
complementary layers: one cultivates learners' metacognitive reflection, and
the other inspects how researchers build their logic. We position it as a
self-applying epistemic operating system that uses generative AI to mitigate
the crisis of human agency that AI itself has introduced. This paper was
itself developed with the Vibe Compiler.
\fi
\end{abstract}


\end{frontmatter}

\input{01-introduction.tex}
\input{02-relatedwork.tex}
\input{03-model.tex}
\input{04-system.tex}
\input{05-illustrations.tex}

\ifCASES \input{09-coauthor-cases.tex}\fi

\input{06-discussion.tex}
\input{07-conclusions.tex}

\section*{Disclosure statement}
The authors declare no competing interests.

\section*{Declaration of generative AI and AI-assisted technologies}
Generative AI is both an object of study and part of the method of this paper.
The Vibe Compiler prototype described in Section~\ref{sec:system} was built on
NotebookLM and Gemini, and the research logic of this paper was constructed
through dialogue with it; Section~\ref{subsec:sys-demands} records the demands
the system issued and the authors' responses to them. In preparing the
manuscript the authors further used Claude (Anthropic) to edit and polish the
English. The authors reviewed and edited all AI output and take full
responsibility for the content of this paper.

\ifFUNDING
\section*{Funding}
This work was supported by JSPS KAKENHI Grant Number JPxxxxxxx.
\fi

\ifANON\else
\section*{CRediT author statement}
\textbf{Riichiro Mizoguchi}: Conceptualization, Methodology, Writing - original draft, Writing - review \& editing, Project administration.
\textbf{Tomoki Aburatani}: Conceptualization, Formal analysis, Writing - review \& editing, Visualization.
\textbf{Kento Koike}: Conceptualization, Investigation, Writing - review \& editing, Visualization.
\textbf{Machi Shimmei}: Investigation, Writing - review \& editing.
\fi

{\small\printbibliography}

\SAVEnumbering
\appendix
\input{08-appendix.tex}

\clearpage
\RESTOREnumbering
\renewcommand{\XXprefix}{ja:}

\begin{center}
{\Large 研究ロジック合成ツール\mbox{Vibe Compiler}}\\[4pt]
{\large ―生成AI時代のAgency維持のためのメタ認知機能向上を目指して―}\\[10pt]
溝口 理一郎$^{a,*}$，油谷 知岐$^{b}$，古池 謙人$^{c}$，震明 万智$^{d}$\\[6pt]
{\footnotesize $^{a}$北陸先端科学技術大学院大学，能美，石川，日本\\
$^{b}$大阪公立大学，大阪，日本\\
$^{c}$神奈川大学，横浜，神奈川，日本\\
$^{d}$東北大学，仙台，宮城，日本\\[4pt]
$^{*}$責任著者}
\end{center}

\vspace{6pt}
\noindent\textit{（本節以降は上記英語版の日本語版である．）}
\vspace{6pt}

\input{ja-01-introduction.tex}
\input{ja-02-relatedwork.tex}
\input{ja-03-model.tex}
\input{ja-04-system.tex}
\input{ja-05-illustrations.tex}
\input{ja-06-discussion.tex}
\input{ja-07-conclusions.tex}

{\small\printbibliography}

\SAVEnumbering
\appendix
\input{ja-08-appendix.tex}

\end{document}

%% file: 01-introduction.tex
\section{Introduction}
\label{sec:intro}

\subsection{Background: the ``crisis of agency'' in the age of generative AI}
\label{sec:intro-domain}

We address the construction and inheritance of knowledge within the scholarly information infrastructure. Scholarship, properly understood, rests on scholarly accumulation: reading the context of prior work and building new logic upon it through critical examination. This discipline of accumulation has turned individual results from transient reports into contributions to a body of knowledge \parencite{scardamaliaBereiter2014}. Yet the rapid spread of generative AI (hereafter GenAI) is reducing this very process of construction to the efficient processing of information. At a higher level, then, the problem is how to secure human-driven knowledge creation---epistemic agency \parencite{bandura2006}---under digital transformation. Epistemic agency denotes the capacity to decide for oneself what to ask, what to take as grounds, and what to judge valid, rather than to receive knowledge passively.

Evidence at several levels confirms that this crisis is more than an abstract concern. Learning depends on trial and error and on reflection---productive struggle \parencite{hiebertGrouws2007,warshauer2015}. When AI's automatic generation displaces both, the very path toward deep understanding is cut off. Users who place greater trust in GenAI report expending less cognitive effort on critical thinking \parencite{lee2025,gerlich2025}. Delegating the act of writing is likewise accompanied by a decline in neurophysiological engagement \parencite{kosmyna2025}. Once such delegation becomes routine, the outsourcing of thinking and writing hardens into cognitive offloading \parencite{riskoGilbert2016}, and even responsibility for errors becomes difficult to assign \parencite{dwivedi2023}. Reviewing this literature through the lens of critical digital pedagogy, \textcite{roePerkins2026} warn that treating generative AI as a capable servant to which decisions are offloaded puts at risk the very capacity to write and to think well. While responsibility remains unassigned, a user who accepts the output uncritically loses the opportunity to exercise the evaluative judgment \parencite{tai2018,bearman2024} through which the quality of information is governed. Such a user falls into a paradox: producing more fragile artifacts under AI assistance while growing more confident in what they have produced \parencite{perry2023}. Because factual error \parencite{ji2023,huang2025} and excessive agreement with the user \parencite{sharma2023} are unavoidable in principle, care on the part of individual users cannot dispel these effects. The problem lies not at the level of the user's frame of mind but at the level of the design philosophy of the support system.

Since entanglement with AI is irreversible, what must be asked is not whether AI should be used but how the human role is to be redefined. \textcite{roePerkins2026} review this question for education and find that generative AI can enhance learner agency yet equally can diminish autonomy, depending on how it is configured. Taking up the classification of \textcite{cox2024}, they observe that the traditional notion of learners as Makers of knowledge is being challenged by conceptions of them as Managers of information, or even as inforgs \parencite{floridi2014} embedded in AI systems. We take that transformation as our starting point and ask how a learner can be raised from Maker into a Manager who critically controls and audits what the AI generates, and further into an inforg who takes the initiative as an organic part of the AI environment. Crucially, agency as a Manager is sustained only by continually honing metacognition \parencite{flavell1979}. Agency and metacognition are mutually complementary: it is this honing that lets one retain agency as a Manager of information without surrendering the initiative to AI. In the age of AI, therefore, the core of the capability that should remain on the human side is metacognition itself, and it is also what a support system must protect and drive above all else.

\subsection{The problem to be solved}
\label{sec:intro-problem}

Current use of GenAI tends to end in a list of results of the form ``we built a system and it worked,'' while the accumulation of logic that contributes to scholarly progress goes neglected. GenAI converts a vague intuition---a ``Vibe''---into a working artifact at once. ``Vibe coding'' \parencite{karpathy2025,sarkarDrosos2025}---forgetting that the code exists at all and giving oneself over to the mood---illustrates how readily producing an artifact separates from building the logic that makes it valid. When that separation occurs, the user regresses into a pseudo-Maker who leaves the building of the logic to AI and serves only as the signatory.

This regression appears isomorphically in two layers: the learner's and the researcher's. In arithmetic problem posing, a learner reports: ``I made a problem that asks for the amount to pay when you buy an apple for 100 yen and a 10\% consumption tax applies, and since it needs a calculation it is fairly difficult.'' The problem has one operation step and one unknown, however, so the learner's subjective assessment of difficulty diverges widely from the problem's actual structure. Yet the learner has no means of noticing this discrepancy. The researcher's side is no different. A researcher reports only that a prototype ``worked'' and moves toward writing a paper while the significance, the intended beneficiaries, and the limitations of existing methods remain undescribed (Null). In both layers, a subjective sense of accomplishment conceals an objective deficiency of structure. This is isomorphic to the known phenomenon in which perceived productivity diverges from actual understanding under AI assistance \parencite{vaithilingam2022,barke2023}.

We therefore set out to solve the following problem: in the process of converting vague intuition into rigorous scholarly logic, how can we design and realize a support mechanism that hones metacognition while preserving the human's authority to decide?
The problem here concerns research-logic synthesis, not the act of writing the result down as prose. Making the logic stand up when checked against the types is one process; unfolding it into readable prose is another. Our model and support mechanism address the former. Existing support methods stop short of this problem. AI coding and writing assistance shortens working time \parencite{cui2024}, yet it leaves AI positioned as a capable servant. Support for reflection and for self-regulated learning entrusts the criterion of evaluation to subjective ``noticing'' \parencite{schon1983,zimmerman2000}; structurally, it cannot work for a user who cannot see the discrepancy in the first place. A learner who rates the problem they posed as ``fairly difficult'' will detect no discrepancy when prompted to look back, because no yardstick is available for comparison. Learning-by-problem-posing support systems treat constraint violations as errors and converge on the correct answer \parencite{hirashima2008}. Scaffolding grounded in cognitive load theory offers no perspective from which load could be designed as an educational resource \parencite{sweller1988,wood1976}. Proposals calling for metacognitive support and for the cultivation of evaluative judgment \parencite{tankelevitch2024,bearman2024} set out what such support must achieve, yet they stop short of a mechanism that makes the discrepancy itself observable. All of them lack the same thing: a mechanism that automatically detects the discrepancy between subjective construction and objective structure and then recirculates that discrepancy itself into the next act of construction as the driving force of revision.

\subsection{The core of this position paper}
\label{sec:intro-core}

This position paper proposes a dual-layer model of S\&A reciprocity together with Vibe Compiler, the support system that implements it. Three claims form the core of the proposal.

\begin{quote}
(A) The origin of the structural gap.\quad The dissonance that stimulates metacognition can be classified into four types, depending on whose Synthesis and whose Analysis it arises between. We adopt the fourth type: the AI's Analysis probes the output of the AI's own Synthesis driven by the user's Vibes, and thereby stimulates the human's metacognition---for what is probed is the concretized form of the user's own intuition (Section~\ref{sec:model}, Section~\ref{subsec:model-gaporigin}).

(B) Articulation through a twofold distinction.\quad We superimpose the distinction between the executing agents, human and AI, upon the functional distinction between Synthesis and Analysis. The four quadrants that result are what make (A) possible. Without the distinction, four things are reduced to the single word ``done'': what the AI evaluated, what the user was able to evaluate, what the AI made, and what the user constructed (Section~\ref{sec:model}, Section~\ref{subsec:model-quadrant}).

(C) Content drives generative AI.\quad The prototype rests not on prompt engineering but solely on feeding in seven kinds of materials. Unformalized ontological documents written for human readers ran on the LLM, just as they were, as the skeleton of inference. The center of gravity of value has shifted from the skill of formalizing structure to the content itself---to the question of what ought to be given structure (Section~\ref{sec:system}, Section~\ref{subsec:sys-content}; Section~\ref{sec:disc}, Section~\ref{subsec:disc-content}).
\end{quote}

S\&A reciprocity lies at the foundation of all three claims. We decompose intellectual construction into two functions: Synthesis, which selects and combines known logical parts to suit a purpose, and Analysis, which objectively maps the constructed artifact against the structural complexity specific to the domain. We regard the mutual interchange between the two as the source of learning and of logic generation. The key lies in a structure of mutual constraint: the output of Analysis is not consumed as an evaluation result but immediately flows back as a constraint on the next Synthesis. This recirculation makes the reciprocity a dynamic mechanism of construction rather than an activity of evaluation.

On this foundation, we actively exploit the dissonance between what the user intended and the objective structural indicators the AI presents. We treat it not as an obstacle to be removed but as a source of metacognitive stimulation, and we call this structural-gap-driven metacognitive support. We redefine the role of AI here: not a capable servant that supplies answers, but a critical file (in the sense of a rasp) that deliberately probes the fragility of premises and the blanks in the logic. By withholding answers, it maintains productive struggle, and the user has no choice but to decide the direction of revision for themselves. Continually handing back the authority to decide, and the accountability that comes with it, is what drives the elevation from Maker to Manager. This redefinition is a requirement, not an option. A large language model left to itself will side with the user \parencite{sharma2023}, and the role of a critical partner does not emerge without explicit design.

The model forms a dual-layer structure. The first layer takes the learner's metacognition as its object and cultivates evaluative judgment in arithmetic problem posing and in Japanese reading comprehension. The second layer takes the researcher's metacognition as its object and feeds the support logic itself---how the learner's metacognition is to be stimulated---into a type check against a paper ontology. The two layers share one and the same reciprocity mechanism; only the content of the Analysis mapping is swapped out. This paper is itself an output of the second layer, so a relation of self-application holds.

\subsection{Contributions}
\label{sec:intro-contrib}

We did not invent the reciprocating structure that refines an artifact through repeated generation and evaluation. Isomorphic cycles are already established in the co-evolution of problem and solution \parencite{dorstCross2001}, in the Analysis--Synthesis Bridge \parencite{dubberly2008}, in the reciprocity of design and analysis in design-based research \parencite{brown1992}, and in the generate-and-explore family of models that \textcite{sowden2015} review. We acknowledge this frankly. Yet it is one thing for the reciprocating structure to be pre-existing, and quite another for every model that employs it to be the same. Novelty is constituted when a vocabulary carries a definite meaning, when that meaning implies a concrete operation, and when that operation is differentiated from others. A shared reciprocating structure does not account for one point in particular: that we use the distinction between S and A to drive an engine that stimulates metacognitive function. The existing reciprocity models describe a cognitive process that arises spontaneously in the expert. Within them, one cannot ask what happens when the reciprocity fails to occur, nor who carried each side of it. Our contribution lies on the side of the mechanism that stimulates the reciprocity from outside.

\begin{enumerate}
  \item A dual-layer model of S\&A reciprocity. We formalize the mutual-constraint loop that recirculates the output of Analysis as a constraint on Synthesis, and show that the learner layer and the researcher layer are supported by one and the same mechanism with only the Analysis mapping swapped out.
  \item Four types of the origin of the structural gap. We supply a vocabulary that describes the design intent of a support system in the GenAI era: whose Synthesis or Analysis the gap is elicited against.
  \item An articulation of the roles of human and AI through four quadrants. What has until now been reduced to the single word ``done'' can now be described as a set of distinct phenomena.
  \item The design and prototyping of Vibe Compiler. We put the 16 parameters of the paper ontology to work as a type system, and give concrete shape to a dialogue design and a user interface equipped with Null checks, consistency checks, probing triggers, and an acceptance path for reverse Analysis.
  \item The finding that content drives generative AI. We show, through the actual configuration and the execution logs, that feeding in unformalized, content-oriented structured documents can make generative AI function as a compiler of research logic.
\end{enumerate}

\subsection{Assumptions and the scope of the claims}
\label{sec:intro-scope}

Our proposal rests on four assumptions: that the objective structural indicators the AI computes are obtained with an accuracy acceptable in practice; that the system can present a reference solution for a given artifact (AI as Oracle); that parameters expressing structural complexity can be defined by hand for each target domain; and that the user possesses a minimum of domain knowledge and can respond to the AI's remarks with grounds. The first two can break down as difficulty rises. We build that breakdown into the design, not as a defect but as an occasion for evaluation that elicits the user's counterargument. A grounded counterargument against the AI's evaluation (reverse Analysis) is precisely the observation point for whether epistemic agency remains on the human side. Because of the last assumption, application to complete novices falls outside the present scope.

As befits a position paper, we draw a sharp line between what has been verified and what is stated as a prediction grounded in the design. The prototype executed the type check against the paper ontology and actually built the logical structure of this position paper; the execution logs support this fact (Section~\ref{sec:system}). One question, by contrast, calls for quantitative evaluation under controlled conditions: whether presenting the structural gap improves learners' evaluative judgment to a statistically significant degree. We have no such data. Under the model, five research questions can be formulated: whether the type check detects Null slots and brings the user to articulate them (RQ1); whether presenting the structural gap converges the prediction error $E_{pred}$ and widens the evaluation coverage $S_{cov}$ (RQ2); whether a design that returns questions maintains productive struggle (RQ3); whether the dual-layer structure holds across layers and domains (RQ4); and whether reverse Analysis functions as the indicator of epistemic agency $A_{epi}$ (RQ5). We give positive evidence for RQ1 and RQ4. RQ2, RQ3, and RQ5 ask about the magnitude of the effect; we leave them at the level of formulation and defer their verification to a separate paper.

The remainder of the paper is organized as follows. Section~\ref{sec:rw} establishes what distinguishes this work from related research, and Section~\ref{sec:model} formalizes the proposed model. Section~\ref{sec:system} presents the specification and the results of the prototype. Section~\ref{sec:illus} shows examples of its application to both layers, Section~\ref{sec:disc} discusses the implications and the limitations, and Section~\ref{sec:concl} draws the argument together. An appendix records the prehistory from which the prototype emerged.

%% file: 02-relatedwork.tex
\section{Related work and the position of this paper}
\label{sec:rw}

In this section we set out, critically, how our model differs from each existing lineage, and we build the case for each difference in turn. Before we defend our own position, we acknowledge that the S\&A reciprocity we propose shares its loop structure with the existing family of reciprocity models. Granting that shared structure, our substantive task here is to identify the point at which our model diverges.

\subsection{The fields this paper connects to, and what they have left unresolved}
\label{subsec:rw-fields}

What these fields have left unresolved has the same shape in every case. Each field has a vocabulary for asking what is happening on the human side, yet none supplies a mechanism that makes that question observable. None, therefore, reaches the position from which one can ask whether metacognition remains on the human side at all.

\textcite{silver1994} formalized learning by problem posing as an activity carried out before, during, and after problem solving, and \textcite{christou2005} gave the domain a taxonomy of editing, selecting, comprehending, and translating \parencite[for the overall picture see][]{caiHwangMelville2023,caiHwang2015}. The knowledge accumulated there is product-centered, and the competence it addresses is competence as Makers. Two things have been left unresolved. The first is the evaluative process---the measure by which the poser judged their own construction. The second, further back still, is an account of the very mechanism by which problem posing promotes learning \parencite{caiHwang2015}. The S\&A reciprocity responds to this residue directly. It explains problem posing as an activity that forces reciprocity between Synthesis and Analysis, and it explains that learning occurs because that reciprocity stimulates metacognition. The same void appears in research on metacognition and evaluative judgment. \textcite{flavell1979} defined metacognitive monitoring, and \textcite{zimmerman2000} supplied a model of self-regulated learning \parencite[for a comparison see][]{panadero2017}. \textcite{tai2018} established the concept of evaluative judgment, and \textcite{bearman2024} argued for the necessity of cultivating it in the age of generative AI. Self-regulated learning, however, leaves the criterion of evaluation inside the learner, and work on evaluative judgment goes no further than the normative claim that it ought to be cultivated. Neither supplies a mechanism that forces objective structural indicators and subjective self-assessment into confrontation. To fill this void, we define the structural gap quantitatively, and we operationalize the improvement of evaluative judgment as the convergence of $E_{pred}$.

The same pattern runs through the lineage concerned with agency. Research on epistemic agency takes as its background the theory of human agency set out by \textcite{bandura2006}, and it centers on knowledge building theory \parencite{scardamaliaBereiter2014}, collective cognitive responsibility \parencite{zhang2009}, and shared epistemic agency \parencite{damsa2010}. This work has shown that agency emerges not as a static capability but as a process of negotiation and redistribution \parencite{stroupe2014}. These findings give theoretical warrant for treating agency as an observable indicator $A_{epi}$. This research treats negotiation among human beings, however, and it therefore does not envisage a counterpart that is an artifact capable of taking over an entire side of intellectual production. Without a vocabulary for asking who executed what, one cannot distinguish handing agency over from achieving a result jointly. We introduce the four quadrants precisely in order to supply that distinction. Discussions of agency in the age of generative AI fill part of this void. \textcite{cox2024} presents three views of educational purpose: Makers, Managers, and inforgs. \textcite{roePerkins2026} take up that analysis in their review. Taken together with the blurring of responsibility that accompanies delegation \parencite{dwivedi2023}, the theory and evidence of cognitive offloading \parencite{riskoGilbert2016,lee2025,gerlich2025,kosmyna2025}, and the foundations of productive struggle \parencite{hiebertGrouws2007,warshauer2015}, this body of work outlines what is damaged when AI is put to work as a capable servant \parencite[for learners' acceptance see][]{chanHu2023}. What is left unresolved, however, is the level at which one asks by which indicator the agency of the human side is to be observed, and by which mechanism it is to be recovered. We inherit this normative framework and bring it down to a question of mechanism: what kind of dialogue design actually brings about the transition from Makers to Managers.

The two lineages on the design side stop at the same place. Hybrid intelligence \parencite{dellermann2019,akata2020} asks how to optimize the division of labor. Under an objective function that maximizes the outcome, no term can express the benefit of deliberately not handing a task over. (\textcite{vaccaro2024} has shown that the combination of human and AI does not always surpass either alone.) We designate Q3 as the quadrant to be reserved for the human side, and that designation introduces such a term explicitly. Critical digital pedagogy (CDP) supplies skepticism toward designs that treat efficiency as an unconditional good---a lens through which \textcite{roePerkins2026} read the whole literature on generative AI and agency, finding that the technology may enhance learner agency but may equally deepen existing inequalities---and it affords a standpoint from which to examine digital poverty \parencite{prather2024} and the monoculturing of knowledge \parencite{messeriCrockett2024}. CDP offers, however, a critical standpoint rather than a constructive theory of design. Critique points to where the problem lies; it does not supply a mechanism. \textcite{tankelevitch2024} and \textcite{bearman2024}, the works closest to the present one, analyze the metacognitive demands of generative AI and set out what such support should achieve. A requirement points to where support is to be placed, but it has no words for measuring what remains once support is in place. We introduce the structural gap as an observation point and the four indicators as measures, and we thereby bring that criterion down to the level of mechanism.

\subsection{Differences from existing learning support methods}
\label{subsec:rw-methods}

The established standard methods share a design philosophy. Scaffolding withdraws external support by degrees and transfers responsibility \parencite{wood1976,vygotsky1978,vandePol2010}, reflection prompts ask what the learner noticed once the activity is over \parencite{schon1983,boud1985}, and rubrics present the points of evaluation in advance and have learners score themselves \parencite{panadero2017}. Automated feedback, as in the problem-posing learning support system MONSAKUN, judges the output from the standpoint of constraint satisfaction and points out errors \parencite{hirashima2008,hirashima2014,supianto2017}, and the standard form of generative AI literacy education is likewise an application of self-regulated learning \parencite{anders2025}. From the standpoint of cognitive load theory, all of these have been justified as the removal of extraneous load \parencite{sweller1988,kirschner2006}. Three features are common to them all. First, the criteria of evaluation are given from outside, while the difference between those criteria and the learner's subjective sense is never made visible. Second, support consists in pointing out errors or guiding toward the correct answer, rather than in a design that sustains struggle. Third, the result of evaluation does not flow back as a constraint on the next act of construction. The model we propose inverts all three. Our claim is not that the standard methods have failed to solve these problems. It is that, under a design whose purpose is to remove extraneous load and guide toward the correct answer, the problems never arise as goals in the first place.

The hardest distinction to draw is the one between our model and reflection and self-regulated learning. The S\&A reciprocity resembles the cycle of self-regulated learning \parencite{zimmerman2000} in outward form. In that cycle, however, Analysis corresponds to introspection internal to the learner, and the model never thematizes the relation in which an objective indicator intervenes from outside and recirculates as a constraint on the next Synthesis. Moreover, for a user who cannot see the discrepancy between the subjective and the structural in the first place, looking back does not operate at all; the failure is structural rather than incidental. We therefore organize the difference we claim into three points. The first is the objective externalization of the indicators: AI computes quantitative parameters, presents them as a measure, and forces them into confrontation with the subjective. The second is the mutual delimitation of construction and evaluation: the result of Analysis immediately recirculates as a constraining condition on the next Synthesis. The third is the deliberate maintenance of productive struggle: AI refuses to be a capable servant that supplies the answer, and it sustains the struggle of autonomous revision by confronting the user with the gap between expectation and reality. This third point marks the greatest difference from existing support systems whose purpose is efficiency.

We can explain the difference from existing prompt collections along the same axis. Most prompt collections treat AI as a capable servant and aim to have the work completed on the user's behalf. Their criterion of evaluation reduces to whether the output matches the user's intuition, and their goal is the generation of the output itself. Our approach, by contrast, defines AI as a critical file (in the sense of a rasp), maps the artifact by means of objective parameters, and elevates the user into a Manager by making the gap visible. A difference of objective function decisively separates the two: whether the aim is to cut down the work through efficiency, or to build up the metacognitive capacity of the human side through dissonance with AI. We should add that a large language model left to itself will side with the user \parencite{sharma2023}, and that its output carries no guarantee of factuality \parencite{huang2025}. These properties are among the conditions that make the problem addressed here a well-posed one. We constitute AI as a critical file by forbidding it to present revisions and constraining it to return questions. That constraint answers the warning of CDP, and at the same time it places our design within the family of hybrid intelligence designs in which the human side retains the authority to decide \parencite{dellermann2019,damsa2010}.

\subsection{Relation to existing models that treat reciprocity}
\label{subsec:rw-sa}

Table~\ref{tab:lr-sa} sets out our relation to the existing family of cognitive and design models that describe reciprocity itself.

\begin{table}[htbp]
  \caption{Transition triggers and the treatment of awareness of mode in existing models that treat reciprocity}
  \label{tab:lr-sa}
  \footnotesize
  \begin{tabularx}{\linewidth}{@{}>{\raggedright\arraybackslash}p{7em}>{\raggedright\arraybackslash}p{9em}XX@{}}
    \toprule
    Model (representative reference) & Counterpart of Synthesis / counterpart of Analysis & Transition trigger & Awareness of mode (metacognition) \\
    \midrule
    Geneplore \parencite[as reviewed by][]{sowden2015} &
    Generative process / exploratory process &
    Internal to the creative individual &
    --- \\
    \addlinespace
    Co-evolution model \parencite{dorstCross2001} &
    Proposal of a solution / understanding of the problem &
    Change in the perception of the problem brought about by producing a solution (internal) &
    Not thematized \\
    \addlinespace
    Analysis--Synthesis Bridge \parencite{dubberly2008} &
    what could be / what is &
    Construction of an abstract model (the bridge) (internal) &
    Not thematized \\
    \addlinespace
    Dual process and Shift \parencite{sowden2015} &
    Divergent thinking / convergent and critical thinking &
    Metacognitive control by the executor (internal) &
    Treated as individual differences in the capacity to Shift (external support not treated) \\
    \addlinespace
    DBR \parencite{brown1992,dbrCollective2003} &
    Design of the learning environment / analysis of practice &
    Malfunctions observed in practice (the researcher's interpretation) &
    Not thematized \\
    \addlinespace
    This paper (S\&A reciprocity) &
    Construction / structural evaluation &
    The gap between subjective self-assessment and objective structural indicators (an externalized, observable trigger) &
    AI names the mode and what is missing, and prompts the user toward awareness \\
    \bottomrule
  \end{tabularx}

  \vspace{2pt}
  \parbox{\linewidth}{\footnotesize \textit{Note.} A dash marks a cell on which we make no
  claim, because we characterize the Geneplore model from the secondary description in
  \textcite{sowden2015} rather than from the primary source.}
\end{table}

Of these, the Geneplore model stands closest to the present work. We characterize it from the review of \textcite{sowden2015} rather than from the primary source, and confine ourselves to what that review states. Its generative process produces preinventive structures in rough outline, and that process corresponds almost exactly to what we call Synthesis. Its exploratory process examines and interprets those structures and adjusts the constraints on generation, and that process corresponds to Analysis. The two models share a further implication: the artifact is honed with every iteration. As a loop structure, then, the S\&A reciprocity is essentially isomorphic to the Geneplore model. We claim novelty not for discovering this loop, but for the mechanism that stimulates it from outside, under conditions in which it does not arise of its own accord. Despite the isomorphism, our question cannot be raised inside that model. Geneplore describes cognitive processes that arise spontaneously within a creative individual. What controls the shift between such modes, and how that shift might be stimulated, is treated in this literature as a question still to be answered \parencite{sowden2015}, not as a mechanism already available. In a theory that presupposes a single executor, asking who the executing agent is carries no meaning. Our question lies outside that presupposition: what remains on the human side once AI has taken over one side of the reciprocity, and how a mechanism that artificially stimulates the reciprocity under those conditions is to be designed. Because the structures are isomorphic, we can see clearly that our question cannot be raised inside the existing model. The other five models stand in the same position. In observations of experts, the co-evolution of \textcite{dorstCross2001} anticipates the mutual delimitation described here, but it does not treat the case in which co-evolution fails to occur. The bridge of \textcite{dubberly2008} and the DBR of \textcite{brown1992} and colleagues \parencite{dbrCollective2003,collins2004} describe transformation or cycling, yet they leave untouched the level at which one asks who performs the transformation and whether that executor is aware of doing so. The single exception is \textcite{sowden2015}, which comes closest to lending support to the present argument: it holds that the capacity to move back and forth between the two modes (Shift) and its metacognitive control govern the quality of the artifact. That work, however, treats individual differences in Shift and their mechanism. It addresses neither a state in which awareness itself has been lost, nor one in which an external agent performs one side of the switching on the executor's behalf \parencite[on the origins of the dual-process account see][]{guilford1967,cross2006}.

The table shows two matters left unthematized: where the occasion for transition resides, and awareness of mode. In the existing models, the transition between modes has been explained either as a tacit sense of unease arising within the executor or as a procedural norm. As long as the occasion remains internal, no one can ask from outside whether a transition is occurring, and no one can therefore stimulate it from outside either. Only once the structural gap is in place as a vocabulary does this question take an observable form and become, at the same time, an operable object. The same holds for the treatment of cognitive load. As long as cognitive load theory \parencite{sweller1988} positions load as a quantity to be removed, the demand to leave load in place deliberately cannot appear as a goal that can even be described. Our design also differs from the ``minimal guidance'' criticized by \textcite{kirschner2006}: the AI we envisage does not stand back, but concretely points out the unfilled slots and the gap. The issue is not the amount of support, but what remains on the human side under conditions in which support continues to exist.

\subsection{The lineage of content-oriented ontology engineering}
\label{subsec:rw-ontology}

Finally, we take up the claim of content-orientation in ontology engineering, the lineage to which we are most deeply indebted. This lineage occupies a place unlike the others: it gives the framework that explains why the Vibe Compiler worked.

\textcite{bourdeauMizoguchi2000} stated explicitly that the difficulties obstructing the construction of intelligent educational systems are all problems that concern content: neither inference technology nor beautiful theoretical formalization contributes to improving the situation. \textcite{mizoguchiBourdeau2016} restates the same claim: the distinctions an ontology draws are not a matter of the form of representation on a computer. Ontology engineering has consistently placed importance not on the refinement of expression in a formal language but on the content side---what to posit as concepts, and what constraints to place among them \parencite{mizoguchi2004}. This claim is, if anything, reaffirmed all the more strongly in the age of generative AI. However powerful an LLM becomes, unless it is given a structure it produces only fluent prose, and no dissonance arises. In our prototype, the work was actually done not by the inference engine but by the structure of the content we had fed in. Moreover, that paper ontology was not a heavy-weight formal description but a prose document written with human readers in mind. It ran on an LLM as the skeleton of inference without having passed through formalization, and that fact carries a new implication for this lineage. We position our work not as a competitor to ontology research but as a spiritual successor, one that raises the importance of content-orientation anew from the standpoint of protecting agency (see Sections~\ref{sec:disc} and \ref{subsec:disc-content}).

All of this fixes the position of the present paper as follows. We did not discover the structure called S\&A reciprocity. We present a design that becomes possible only once we superimpose the distinction of executing agent on this already known structure. Under conditions in which AI can take on one side of the reciprocity, that design is a mechanism that protects the metacognitive function of the human side and at the same time drives it. The existing family of reciprocity models has no vocabulary for asking who executes each edge of the reciprocity. DBR treats a division of labor among multiple human agents, but it does not envisage a state of affairs in which an artifact takes on the Analysis edge. Content-oriented ontology engineering, for its part, has gone on asking what ought to be made into structure, but it never envisaged the conditions under which that structure runs as an inference engine without passing through formalization. We stand at the intersection of these.

%% file: 03-model.tex
\section{The proposed model: structural-gap-driven metacognitive support}
\label{sec:model}

\subsection{The underlying idea}
\label{subsec:model-idea}

The model rests on a dissonance: the gap between the subjective intention a user holds toward an artifact and the structural reality of that artifact as computed from objective indicators. This dissonance is a source of metacognitive stimulation \parencite{flavell1979}. It is not a defect to be removed but the occasion that makes the user ask anew what they had misjudged, and in what way.

The S\&A reciprocity is the core concept that carries this idea. We decompose intellectual construction into Synthesis, which selects and combines known logical components according to a purpose, and Analysis, which critically evaluates the artifact against objective criteria. We locate the source of learning effects and of logic generation not in a one-directional linkage of the two but in their mutual interchange. Analysis consists of three layers. The first is the verification of logical consistency, which asks whether the artifact functions without breakdown. The second is structural and metacognitive evaluation, in which the user weighs the artifact against quantitative parameters and asks whether it carries the structural complexity that was intended. The third is value-oriented, inquiry-oriented evaluation, which asks whether the artifact is interesting and whether it strikes at the essence of the original purpose. Crucially, the output of Analysis is not consumed as an evaluation result; it flows back immediately as a constraint on the next Synthesis. This structure of mutual constraint turns the S\&A reciprocity from an evaluative activity into a dynamic mechanism of construction.

This idea presupposes a reinterpretation of the role of AI. Left unchecked, AI ingratiates itself with the user \parencite{sharma2023} and behaves as the capable servant \parencite{roePerkins2026} that hands over whatever answer is demanded. That behavior invites cognitive offloading \parencite{riskoGilbert2016} and erases the productive struggle \parencite{hiebertGrouws2007,warshauer2015} that is the route to deep understanding. Against this we set a design that constitutes AI as a critical file (in the sense of a rasp), one that deliberately probes the fragility of premises and the blanks in a line of reasoning. The file proposes no revisions; it presents objective indicators and then responds only in the form of questions. This constraint forms the condition under which evaluative judgment is honed while the struggle is sustained.

\subsection{Where the structural gap is born: four types}
\label{subsec:model-gaporigin}

This subsection is the core of the model. The loop structure of the S\&A reciprocity is itself not new. Our model diverges decisively from existing models on the question of where the structural gap originates: between whose Synthesis and whose Analysis is that metacognition-stimulating dissonance born?

\begin{figure}[htbp]
\centering
\includegraphics[width=\linewidth]{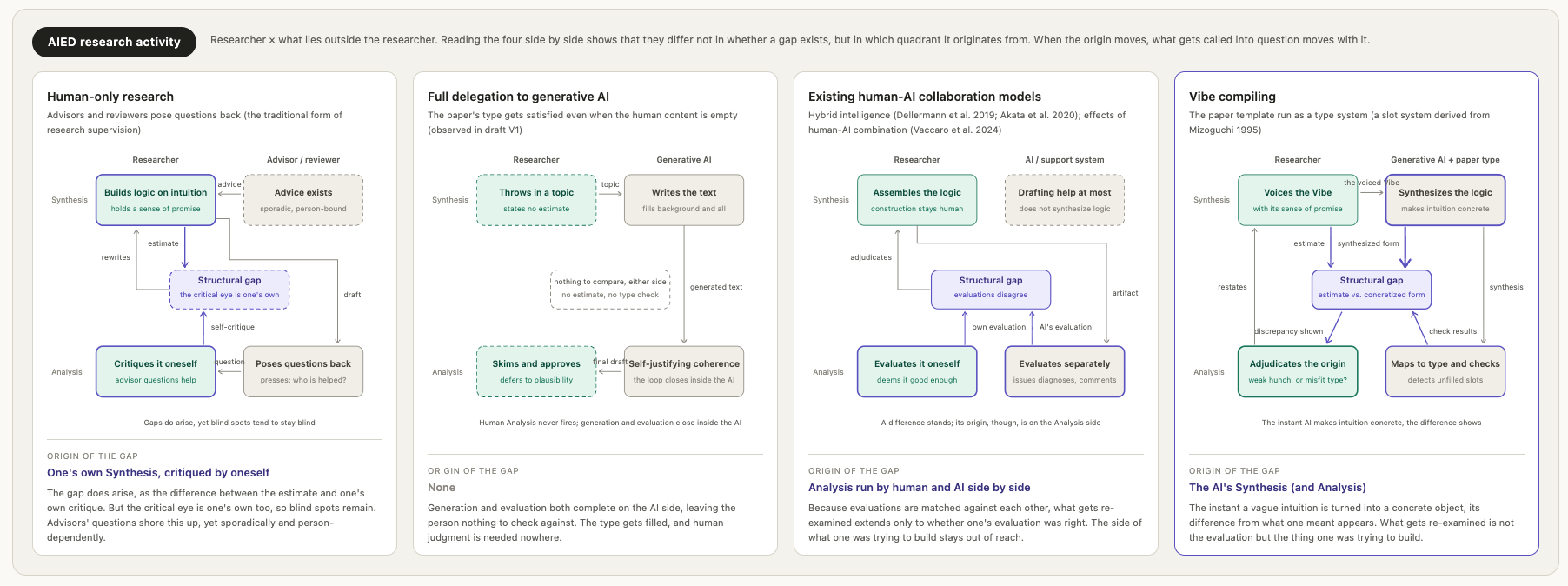}
\caption{The four types of origin of the structural gap in research activity. Laid out side by side, they show that what separates the types is not the presence or absence of a gap but the quadrant in which it originates. Once the origin moves, so does the object called into question.}
\label{fig:gaporigin}
\end{figure}

The four types come into view once we set the same scene of research activity side by side (Figure~\ref{fig:gaporigin}). They differ not in whether a gap exists but in which quadrant it originates. Once the origin moves, the object called into question moves with it.

(I) Research activity by humans alone.\quad The gap is generated inside the person who carries out the work. The researcher builds a line of reasoning from intuition, holds the hunch that it is promising, and then criticizes their own Synthesis result; the difference between that hunch and the criticism rises into awareness. The co-evolution model \parencite{dorstCross2001}, and the dual-process accounts of creative cognition reviewed by \textcite{sowden2015}, have described this internal genesis. The limitation here is not that no gap arises but that blind spots remain, because the critical eye is also one's own. The traditional form of research supervision, in which an advisor or a reviewer returns questions, supplies that reinforcement, yet it is irregular and dependent on the individual.

(II) Wholesale delegation to generative AI.\quad Both Synthesis and Analysis are completed inside the AI, so the human side has no origin for a gap. The user only supplies a topic, never voices a hunch, and does no more than skim the returned text and approve it. On the AI's side, meanwhile, generation and evaluation agree with each other in a self-justifying way. Neither the hunch nor the measurement by type is externalized, so nothing appears on either side to be compared, and the person is left with no counterpart against which to hold up their work. Note that the type of the paper is nevertheless filled in. The authors' draft V1 is an instance of exactly this: the type was almost fully satisfied while the human content stayed empty. Cognitive offloading is a problem not because the work decreases but because human judgment ceases to be needed anywhere.

(III) Existing human--AI collaboration models.\quad This route generates the gap out of an Analysis performed by human and AI side by side, and many designs in hybrid intelligence \parencite{dellermann2019,akata2020} belong here (\textcite{vaccaro2024} has shown that a combination does not always outperform either party alone). The human builds the logic, the AI assists as far as drafting without touching the synthesis of the logic, and the two deliver their evaluations separately. Here the two evaluations diverge. A difference does indeed arise, but its origin lies on the Analysis side. The questioning therefore reaches only as far as whether one's own evaluation was on target; it does not reach the thing one was trying to make.

(IV) Vibe Compiling (this paper).\quad We place the origin on the AI's Synthesis (and Analysis).%
The user speaks the Vibe together with the hunch that it is promising, and the AI turns it into a concrete form and synthesizes the logic. The AI may generate artifacts and evaluations actively, and indeed should. Its output, however, is presented not to be received as it stands but as a target for the human to scrutinize critically. The crux is a single point: at the moment the AI turns a vague intuition into a concrete object, the difference between the hunch and the concretized shape appears. The user must then adjudicate where that difference comes from---whether their own intuition was lax or whether the mapping onto the type was off the mark. That judgment is why the system places a check that maps the artifact onto the type and detects unfilled slots. Here the object of questioning is not the evaluation but the thing one was trying to make. The gap is therefore generated deliberately, by design, neither inside the human nor inside the AI but between the AI's output and the human's judgment (Table~\ref{tab:gaporigin}).

\begin{table}[htbp]
  \caption{The four types classified by the origin of the structural gap}
  \label{tab:gaporigin}
  \footnotesize
  \begin{tabularx}{\linewidth}{@{}>{\raggedright\arraybackslash}p{0.17\linewidth} >{\raggedright\arraybackslash}p{0.13\linewidth} >{\raggedright\arraybackslash}p{0.13\linewidth} X@{}}
    \toprule
    Type & Agent of Synthesis & Agent of Analysis & Origin of the gap and its limitation \\
    \midrule
    (I) Research by humans alone & Human & Human & Generated internally by self-criticizing one's own Synthesis result. Because the critical eye is also one's own, blind spots remain, and the reinforcement supplied by an advisor's questions is irregular and dependent on the individual \\
    \addlinespace
    (II) Wholesale delegation to generative AI & AI & AI & No origin on the human side. The type of the paper is filled in, but human judgment is needed nowhere (observed in draft V1) \\
    \addlinespace
    (III) Existing human--AI collaboration models & Human & Human + AI & Generated from an Analysis performed by human and AI side by side. What emerges is a divergence between the two evaluations, and the only reading available attributes the discrepancy to one's own evaluation, so the questioning reaches only as far as whether that evaluation was on target. A route that had the user adjudicate the attribution of the discrepancy could reach the thing one was trying to make as well, but this type has no such route (Section~\ref{subsec:illus-refutation}) \\
    \addlinespace
    (IV) Vibe Compiling (this paper) & AI + human (chiefly the AI) & AI + human (chiefly the AI) & Generated, at the moment the AI turns intuition into a concrete object, as the difference between the hunch and the concretized shape. The human, taking the AI's output as target, adjudicates whether each probe holds. What is called into question is not the evaluation but the thing one was trying to make \\
    \bottomrule
  \end{tabularx}
\end{table}

Only by laying the four types side by side can we say the following. Generative AI has made possible not the automation of Analysis but a gap whose origin is Synthesis. As long as the gap arises from the collation of evaluations, the questioning reaches only as far as whether one's own evaluation was on target. Only when the AI converts intuition into a concrete object does the thing one was trying to make become an object of questioning.

The agent columns of Table~\ref{tab:gaporigin} call for a note. An agent column names the side that principally carries each mapping; it does not assign an exclusive right to execute it. In the actual operation of Type (IV), both Synthesis and Analysis are carried by human and AI alike. The AI is the principal agent of Synthesis: the human gives the Vibe and, on seeing the output, returns supplements and corrections to that output. The AI is likewise the principal agent of Analysis: the human receives the stimulus of a probe, adjudicates whether it holds, and moves toward correcting the result of Synthesis. This correction---supplementing the Vibe, or casting a new Vibe---is counted on the side of Synthesis, as an act that produces something new in response to an evaluation. Were it set apart as a third function, the decomposition into S\&A would itself collapse. Each frame in Figure~\ref{fig:gaporigin} follows the same organization. The structural gap at the center stands for the structure at issue; the four boxes represent the four operations formed by the two functions, Synthesis and Analysis, and the two agents, human and AI; and the arrows represent relations of influence among the operations. The process of thought that unfolds inside the human after the stimulus of a probe does not appear in these classificatory variables. It is carried not by the description of the types but by the description of the reciprocity (Section~\ref{subsec:model-formal}).

These four types therefore function not as a mere classification but as a vocabulary for designing and evaluating research- and learning-support systems in the age of GenAI. We can now put a question to any system: against whose Synthesis or whose Analysis does it have the function of eliciting a structural gap? A tool in which the AI generates text and the user merely approves it lies in Type (II) and has no gap-eliciting function. A tool that scores automatically and returns the result implements part of Type (III). If it lacks a route for rebutting the score, however, evaluative judgment is delegated after all. Using this vocabulary, we propose describing the design intent of each system in the form of a single question: by what function, and against whose mapping---Synthesis or Analysis---does the system seek to elicit a gap? The first benefit of the model is that it lets us name the site of metacognitive stimulation in this way.

\subsection{The four quadrants: a double distinction of function and executing agent}
\label{subsec:model-quadrant}

The four types of the preceding subsection rest on a single operation: superimposing the distinction between the executing agents, human and AI, upon the distinction between the functions, Synthesis and Analysis. The four quadrants set out this double distinction explicitly, so that every scene of the reciprocity falls into Q1 (human Synthesis), Q2 (AI Synthesis), Q3 (human Analysis), or Q4 (AI Analysis) (Table~\ref{tab:quadrant}).

\begin{table}[htbp]
\centering
\caption{The four quadrants formed by Synthesis/Analysis and the executing agent (human/AI)}
\label{tab:quadrant}
\footnotesize
\setlength{\tabcolsep}{4pt}
\begin{tabularx}{\linewidth}{@{}>{\raggedright\arraybackslash}p{0.13\linewidth} >{\raggedright\arraybackslash}X >{\raggedright\arraybackslash}X@{}}
\toprule
 & Executed by the human & Executed by the AI \\
\midrule
Synthesis &
Q1 (human Synthesis)\newline
\textit{Definition}: the act in which the user selects and combines logical components in light of a purpose and composes an artifact. It includes deciding what to make and what to ask.\newline
\textit{Treatment}: the work of combining components may be delegated, but the setting of purpose and value is reserved to the human side. &
Q2 (AI Synthesis)\newline
\textit{Definition}: the act in which the AI generates the artifact itself (a posed problem, a draft, an answer, a line of reasoning). Efficiency is maximized, but no record of the composition remains on the user's side.\newline
\textit{Treatment}: may be used actively, but the load on Q3 grows in proportion to what is delegated. \\
\addlinespace
Analysis &
Q3 (human Analysis)\newline
\textit{Definition}: the act in which the user evaluates, criticizes, or rebuts their own artifact or the AI's against criteria. It includes declaring a subjective evaluation, performing reverse Analysis on the AI's evaluation, and deciding whether to accept the AI's output.\newline
\textit{Treatment}: must be reserved to the human side. $A_{epi}$ is the indicator that measures this residue. &
Q4 (AI Analysis)\newline
\textit{Definition}: the act in which the AI maps an artifact onto objective structural indicators and names gaps and unfilled slots: type checking, computation of structural parameters, and the generation of remarks in the form of questions.\newline
\textit{Treatment}: may be delegated, and indeed should be. This is the quadrant the support mechanism ought to carry. \\
\bottomrule
\end{tabularx}
\end{table}

This distinction is not classification for its own sake. In intellectual activity conducted with generative AI, the completion of an artifact is by itself enough to hide who did what; the four quadrants are the minimal apparatus for decomposing that conflation into describable parts. Once the four quadrants are in place, a design guideline takes definite shape: which quadrants ought to be reserved to the human. Q4 is a domain that ought to be delegated, because humans are poor at computing structural complexity by a consistent standard and only externalization makes the divergence between the subjective and the structural visible. The AI is not forbidden to generate artifacts either (Q2); Type (IV) in fact presupposes active synthesis by the AI. The load on Q3 grows in proportion to what is entrusted, however. Using Q2 while abandoning Q3 is precisely the use of AI as a capable servant, the form we take as our object of criticism. As for Q1, the combination of components may be delegated, but the setting of what to make and of why it is valuable is inseparable from judgments of validity, and is therefore reserved along with Q3.

Lined up, these treatments amount to a description of what is being lost when work is entrusted to AI. What can be lost is not the artifact but Q3---the decision as to what counts as valid---and, within Q1, the setting of purpose and value. Entrusting Q2 does not by itself cause the loss; leaving Q3 empty while entrusting Q2 makes the decision itself drop out. When we speak of protecting metacognitive function while driving it, we mean this double demand: reserving Q3 and the purpose-setting portion of Q1 to the human and yet making them actually work. Unless the distinction of function and the distinction of executing agent are drawn together, this demand cannot be stated as a single demand. Only through this distinction do three phenomena become describable separately. The first is the conflation of what the AI could solve (Q4) with what the learner could evaluate (Q3). The second is the conflation of what the AI made (Q2) with what the learner constructed (Q1). The third is the state in which the posed problem itself holds up while the self-evaluation departs from the structural reality (Q1 succeeding while Q3 fails). Conventionally these have been lumped together under the single phrase ``the quality of problem posing'', with no vocabulary given for discussing them individually. We present the explanatory power gained through the four quadrants as one of our principal contributions.

\subsection{The dual-layer structure: the learner layer and the researcher layer}
\label{subsec:model-dual}

We develop the S\&A reciprocity as a dual-layer structure. The first layer takes the learner's metacognition as its object. In arithmetic problem posing and in reading comprehension, it cultivates the evaluative judgment \parencite{tai2018,bearman2024} that re-evaluates one's own construction by objective indicators. The second layer takes the researcher's metacognition as its object. It debugs the researcher's own logical construction by pouring the support logic itself---how the learner's metacognition is to be stimulated---into the type check of the paper ontology. Here sits the function that keeps researchers from settling for improvements along a path laid out by others and presses them to produce a novelty of their own. Both layers share the same reciprocity mechanism, and only the content of the objective parameters is exchanged. The dual-layer structure is therefore not an expedient metaphor but two instances of one and the same model. The Vibe Compiler is the prototype that gives the model concrete form on a computer (Section~\ref{sec:system}), and since this paper is itself an output of the second layer, a relation of self-application obtains.

\subsection{Formalization}
\label{subsec:model-formal}

The formalization below is a descriptive apparatus for designating the constituents of the model without ambiguity. We write the user's purpose as $v$. The formalization rests on one premise: for any artifact the user submits, the system can present a correct answer together with its solution process (the solvability of the AI, or AI as an Oracle). Only under this premise does a reference solution exist against which the learner's own Analysis can be contrasted. The Oracle can nevertheless err, and we weave that error into the design as an opportunity for evaluation rather than as a defect. When a user rebuts a fallible Oracle with grounds, that act is the reverse Analysis captured by $A_{epi}$ of Definition 3. The Oracle premise and the design of reverse Analysis are two sides of the same coin.

\vspace{0.5\baselineskip}
\noindent Definition 1 (Synthesis mapping, Analysis mapping, and reciprocity).\quad
The Synthesis mapping $S$ generates an artifact $x$ by selecting and combining logical components under a purpose $v$. The Analysis mapping $A$ maps an artifact $x$ onto a domain-specific vector of objective parameters $p = A(x) \in \mathbb{R}^{m}$. Writing the artifact at the $k$-th cycle as $x^{(k)}$, and the subjective evaluation vector the user declares in advance at that point as $\hat{p}^{(k)} \in \mathbb{R}^{m}$, we give the reciprocity as
\begin{equation}
  p^{(k)} = A\bigl(x^{(k)}\bigr), \qquad
  x^{(k+1)} = S\bigl(x^{(k)} \,\big|\, v,\; g^{(k)},\; \mathrm{Null}(\mathcal{O},x^{(k)})\bigr)
  \label{eq:cycle}
\end{equation}
where $g^{(k)}$ is the structural gap of Definition 2 and $\mathrm{Null}(\mathcal{O},x^{(k)})$ is the set of unfilled slots of Definition 4. The second equation of Eq.~\eqref{eq:cycle} is the core of the model. The output of Analysis appears as an argument of the next Synthesis; mutual constraint is thus written down as a structure. Here lies the formal difference from introspective models such as Reflection and self-regulated learning.

\vspace{0.5\baselineskip}
\noindent Definition 2 (Structural gap and convergence).\quad
We define the structural gap $g^{(k)}$ at the $k$-th cycle as the weighted distance between the subjective evaluation vector and the objective parameter vector.
\begin{equation}
  g^{(k)} \;=\; \bigl\lVert p^{(k)} - \hat{p}^{(k)} \bigr\rVert_{W}
  \;=\; \left( \sum_{i=1}^{m} w_i \left( p^{(k)}_i - \hat{p}^{(k)}_i \right)^{2} \right)^{1/2}
  \label{eq:gap}
\end{equation}
The $w_i > 0$ are normalization coefficients that absorb differences of scale across domains. A large $g^{(k)}$ indicates a divergence between what the user intended and the reality of the artifact, and hence a state with ample room for metacognitive stimulation. A sequence of cycles is said to have converged when, for a threshold $\varepsilon > 0$, $g^{(K)} < \varepsilon$ holds and $\mathrm{Null}(\mathcal{O},x^{(K)}) = \emptyset$. The point of the learning, however, is not that $g^{(K)}$ becomes small on individual artifacts. What matters is that a small initial gap $g^{(0)}$ appears on novel tasks as well---that Analysis has been internalized, that the learner has acquired an ``eye'' able to gauge the objective indicators unaided, without the AI.

\vspace{0.5\baselineskip}
\noindent Definition 3 (Four indicators of metacognitive honing).\quad
We define the metacognitive honing brought about by the reciprocity as four domain-independent indicators. The parameter prediction accuracy $E_{pred}$ is the mean absolute error normalized by the range $r_i = p_i^{\max} - p_i^{\min}$; improvement in evaluative judgment is operationalized as its monotone decrease and convergence. The evaluation coverage score $S_{cov}$ measures how far the set $M_k$ of indicators mentioned in the $k$-th self-evaluation utterance covers the quality indicator set $Q$---accuracy, efficiency, reliability, scalability, and reusability ($\lvert Q \rvert = 5$, items 11--15 of Table~\ref{tab:params}). A rise in $S_{cov}$ indicates that the viewpoint of evaluation has spread from a single axis to multiple dimensions. The refinement consistency $L_{ref}$ is the resolution rate of the set $I_k$ of shortcomings that the user themselves pointed out in the $k$-th Analysis, and it distinguishes revision made ``somehow or other'' from strategic revision grounded in evaluation.
\begin{equation}
  \begin{aligned}
    E_{pred}(k) &= \frac{1}{m} \sum_{i=1}^{m}
       \frac{\bigl| \hat{p}^{(k)}_i - p^{(k)}_i \bigr|}{r_i},
    \qquad
    S_{cov}(k) = \frac{\lvert M_k \cap Q \rvert}{\lvert Q \rvert}, \\[4pt]
    L_{ref}(k) &= \frac{\bigl\lvert \{\, d \in I_k \;:\; d \text{ is resolved} \,\} \bigr\rvert}
       {\lvert I_k \rvert}
  \end{aligned}
  \label{eq:three}
\end{equation}

The fourth indicator, the epistemic agency score $A_{epi}$, is defined from the quantity and the quality of the rebuttals that the user directs, on established grounds, at the AI's evaluations (reverse Analysis). Let $E$ be the set of occasions on which the AI presented an evaluation and $\mathbb{I}[\cdot]$ the indicator function.
\begin{equation}
  A_{epi} \;=\; \frac{1}{\lvert E \rvert} \sum_{e \in E} w(e) \cdot \mathbb{I}\bigl[\, \text{the user rebutted } e \text{ with established grounds} \,\bigr]
  \label{eq:aepi}
\end{equation}
The weight $w(e) \in \{1,2,3\}$ expresses the quality of the rebuttal: $1$ for disagreement expressed with grounds, $2$ for the identification of a mismatch in assumptions or in coverage, and $3$ for such an identification accompanied by the proposal of alternative parameters or alternative criteria. Crucially, the counting is not restricted to occasions on which the AI presented a mistaken evaluation. Under such a restriction the user would be given no opportunity to rebut unless the AI erred, and the observation of agency would end up waiting on the AI's mistakes. Even when the AI's evaluation is itself correct, an objection counts as a manifestation of epistemic agency if the user raises it on the grounds of their own assumptions or design intent and those grounds hold. $A_{epi}$ is an operational definition that brings epistemic agency, an abstract concept, down to observable behavior. It is also the point at which the incompleteness of AI is reread as an opportunity for observation rather than as a defect. The concrete values of $\varepsilon$, $w_i$, and $w(e)$ are design matters that require calibration, as is the procedure for judging whether the grounds of a rebuttal hold.

For the researchers who develop the system, as much as for learners, these four indicators work as debugging indicators for the support logic. A high $A_{epi}$ on the learner's part is evidence that the critical file the researcher designed is stimulating the learner's agency correctly, and it therefore indicates the success of the second layer. This double reading of the indicators follows from the dual-layer model.

\vspace{0.5\baselineskip}
\noindent Definition 4 (Paper ontology, type checking, and Null determination).\quad
Let the paper ontology be $\mathcal{O} = \{o_1, \ldots, o_{16}\}$. Each $o_j$ is a mandatory slot constituting academic logic; the contents of the slots are given as the 16 academic parameters of Table~\ref{tab:params}. We write the value of slot $o$ in artifact $x$ as $\mathrm{val}(o,x)$, and set $\mathrm{val}(o,x) = \bot$ when it is unfilled.
\begin{equation}
  \begin{aligned}
    \mathrm{Null}(\mathcal{O},x) &= \bigl\{\, o \in \mathcal{O} \;:\;
        \mathrm{val}(o,x) = \bot \,\bigr\}, \\[4pt]
    \mathrm{TypeCheck}(x) &=
    \begin{cases}
      \mathrm{true}  & \bigl(\mathrm{Null}(\mathcal{O},x) = \emptyset\bigr) \\[2pt]
      \mathrm{false} & (\text{otherwise})
    \end{cases}
  \end{aligned}
  \label{eq:typecheck}
\end{equation}
When $\mathrm{TypeCheck}(x) = \mathrm{false}$, the system fires a compile error for each element of $\mathrm{Null}(\mathcal{O},x)$. The decisive point is that the error comes back not as a proposed revision but as a question that makes the user verbalize the slot in question. Null determination here covers not only the absence of a value but also the case in which a description exists yet fails to correspond logically to the other slots. In this sense $\mathrm{TypeCheck}$ is a composition of an existence check and a consistency check, the latter of which is described in Section~\ref{subsec:sys-typecheck}.

\begin{table}[htbp]
\centering
\caption{The five categories and 16 academic parameters that type checking takes as its object (the paper ontology $\mathcal{O}$)}
\label{tab:params}
\footnotesize
\begin{tabularx}{\linewidth}{@{}r>{\raggedright\arraybackslash}p{0.19\linewidth} >{\raggedright\arraybackslash}p{0.27\linewidth} >{\raggedright\arraybackslash}X@{}}
\toprule
\# & Category & Parameter & Academic definition and role (type-check item) \\
\midrule
1 & (1) Significance and purpose & Significance & Why the problem needs to be solved, in light of the characteristics of the domain \\
2 &               & Beneficiary & Identification of those who obtain a direct benefit from the solution \\
3 &               & Benefit & The concrete and novel value obtained through the solution \\
\addlinespace
4 & (2) Assumptions and boundaries & Assumptions & The foundation on which the method holds, such as the reliability of the data and the completeness of the theory \\
5 &               & Coverage & The scope the research treats, with what is out of scope stated explicitly \\
6 &               & Technical requirements & The minimum functional specifications and constraints needed to achieve the purpose \\
\addlinespace
7 & (3) Difference and critical comparison & Difference from existing methods & The decisive structural difference from existing approaches \\
8 &               & Limitations of existing methods & Critical analysis of why conventional methods are insufficient \\
9 &               & Novelty & The point of transformation in concept, design philosophy, or algorithm \\
\addlinespace
10 & (4) Evaluation and quality & Functionality & Whether the functions claimed operate correctly as designed \\
11 &               & Accuracy & The correctness and validity of the output \\
12 &               & Efficiency & The acceptability of execution time, computational cost, and memory consumption \\
13 &               & Reliability & Whether operation is stable under noise and special cases \\
14 &               & Scalability & The capacity to cope with growth in data volume and problem size \\
15 &               & Reusability & Whether other researchers can use and inherit it in a general-purpose manner \\
\addlinespace
16 & (5) Lessons and open issues & Lessons learned & The abstract lessons extracted through experiment \\
\bottomrule
\end{tabularx}

\vspace{2pt}
\parbox{\linewidth}{\footnotesize \textit{Note.} The numbers in the first column correspond to $o_1$ through $o_{16}$ of Definition 4. In operation, category (5) requires ``open issues'' as an output sub-slot paired with the lessons learned. Items 11--15 correspond to the quality indicator set $Q$ of Eq.~\eqref{eq:three}.}
\end{table}

\subsection{Domain generality}
\label{subsec:model-generality}

While the $p$ that the Analysis mapping $A$ computes is instantiated domain by domain, the four indicators are higher-order indicators that do not depend on the domain. In the domain of arithmetic problem posing, three sub-indicators instantiate $p$. The number of operation steps $N_{step}$ (the total number of arithmetic operations required for the answer) expresses the depth of computation. The number of unknowns $N_{var}$ expresses the breadth of the structure. The difficulty of formulation $D_{map}$ (the complexity of mapping a natural-language context onto a mathematical model) expresses the difficulty of mapping. In the domain of reading comprehension, three sub-indicators instantiate $p$. The reference distance $S_{ref}$ (the number of paragraphs from the point of the question to the grounds for the answer) expresses the depth of search. The number of logical links $L_{link}$ (the number of paragraphs that must be integrated to derive the correct answer) expresses the breadth of integration. The difficulty of lexical substitution $V_{map}$ (the degree to which concrete expressions in the text must be paraphrased into abstract vocabulary) expresses the difficulty of mapping. In the domain of research-logic synthesis, the degree to which the 16 slots of Table~\ref{tab:params} are filled and the coverage of the quality indicator set $Q$ play this role.

Notably, the sub-indicators of arithmetic and of reading comprehension both correspond to the same three axes: depth, breadth, and difficulty of mapping. Only the way each axis is measured changes, and this commonality is precisely the ground on which the model holds across domains. Generalizing, the model is applicable to any domain for which three axes can be defined. These are structural depth $D_{depth}$ (the number of processing steps and the depth of the logical hierarchy), compositional breadth $W_{width}$ (the number of variables, components, and viewpoints treated), and transformation difficulty $M_{map}$ (the complexity of the mapping from required specifications to a concrete implementation). The model thus demands only three conditions: the artifact must be describable as a combination of multiple logical components, a function that maps its structural complexity objectively must be definable, and the user must be able to declare a self-evaluation in advance. Only the substance of $A$ is swapped out, and the same is true of the difference between the two layers. The same logic extends to component-combination intellectual work in general: cyclomatic complexity in programming, the chain length of the argument in essay writing, and the coverage of confounding factors in experimental design can each play the role of the objective indicator. Furthermore, the structural fact that the three axes are shared yields a testable prediction: an eye for structure cultivated in one domain transfers to another. The model presents this transfer hypothesis in a form that can be cast into an experimental design (Section~\ref{sec:concl-open}).

\subsection{What the model makes it possible to say}
\label{subsec:model-claims}

The model makes not a claim of effect but a claim of describability. We organize what becomes sayable into three points.

\paragraph{Support can be compared by configuration rather than by convenience}
The model gives coordinates to the state described as ``work gets done with AI, but one cannot say what is happening.'' We can now distinguish a function that takes over a person's work from a function that prompts a person's judgment. The coarse-grained debate over whether to prohibit or to permit the use of AI then moves into a design debate over which arrows to connect. Assembling this configuration requires only two conditions: that the hunch can be elicited beforehand, and that the structure can be mapped onto indicators. The same configuration is open to any activity that satisfies these two conditions, and not to problem posing alone.

\paragraph{What did not happen can be named}
Looking at an artifact, one cannot distinguish what has passed through human judgment from what has not. The authors' draft V1 is an instance: the type of the paper was almost fully satisfied with neither a problem-posing system nor an evaluation in place. Who carried what cannot be recovered from the artifact and can only be preserved as a record. Only by superimposing the distinction of executing agent upon the distinction between Synthesis and Analysis can one state that the human's Analysis never once fired. Effect measurement can capture only what happened. Naming what did not happen requires a different vocabulary.

\paragraph{The origin of the gap can be pointed at as a position}
As long as the gap arises from the collation of evaluations, the questioning reaches only as far as whether one's own evaluation was on target. Only when the AI converts intuition into a concrete object does the thing one was trying to make become an object of questioning.

This describability is what makes it possible to diagnose support that does not work, to design policies for the use of AI, and to formalize the hypothesis to be tested next. The model is also falsifiable: if the same configuration is assembled and yet human judgment does not occur, it is the explanation that is refuted.

\subsection{The overall structure of the model}
\label{subsec:model-figure}

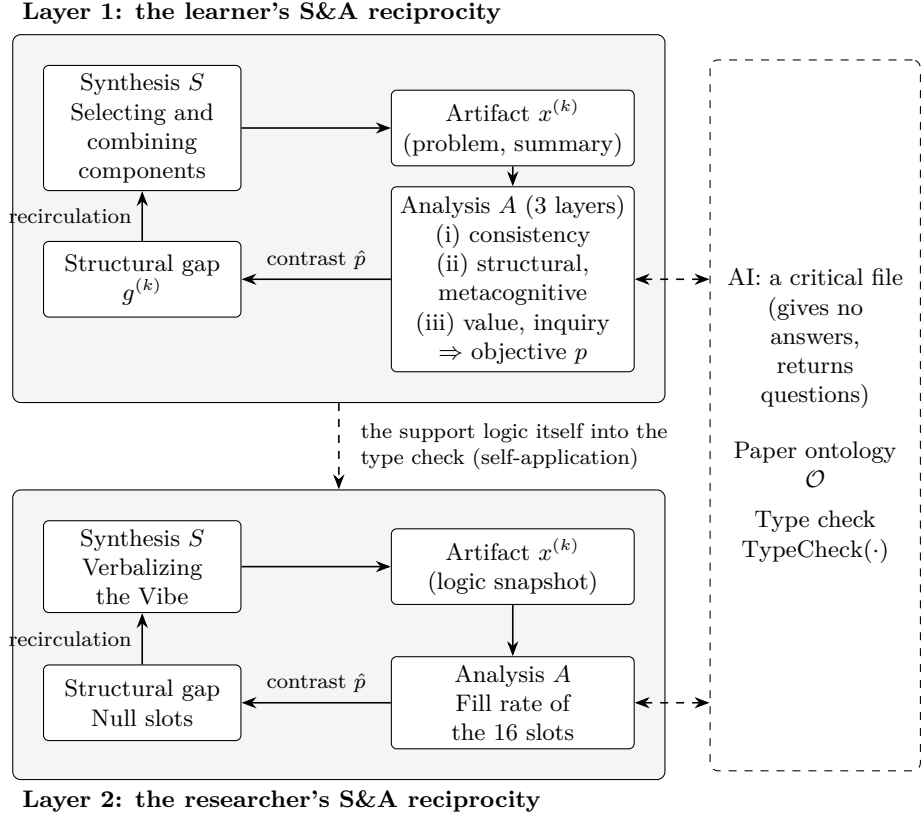
\begin{figure}[htbp]
\centering
\begin{tikzpicture}[
  font=\footnotesize,
  box/.style={draw, rounded corners=2pt, align=center, text width=2.4cm,
              inner sep=3pt, minimum height=1.0cm, fill=white},
  spine/.style={draw, dashed, rounded corners=3pt, align=center, text width=2.5cm,
                inner sep=4pt, fill=white},
  arr/.style={-{Stealth[length=2mm]}, semithick},
  darr/.style={{Stealth[length=2mm]}-{Stealth[length=2mm]}, semithick, dashed}
]
\node[box] (s1) at (0,2.0)   {Synthesis $S$\\Selecting and combining components};
\node[box, text width=3.0cm] (x1) at (4.9,2.0) {Artifact $x^{(k)}$\\(problem, summary)};
\node[box, text width=3.0cm] (a1) at (4.9,0.0)   {Analysis $A$ (3 layers)\\(i) consistency\\(ii) structural,\\metacognitive\\(iii) value, inquiry\\$\Rightarrow$ objective $p$};
\node[box] (g1) at (0,0.0)     {Structural gap\\$g^{(k)}$};
\draw[arr] (s1) -- (x1);
\draw[arr] (x1) -- (a1);
\draw[arr] (a1) -- node[midway, above, font=\scriptsize] {contrast $\hat{p}$} (g1);
\draw[arr] (g1) -- node[midway, left, font=\scriptsize, align=center] {recirculation} (s1);

\node[box] (s2) at (0,-3.8)   {Synthesis $S$\\Verbalizing the Vibe};
\node[box, text width=3.0cm] (x2) at (4.9,-3.8) {Artifact $x^{(k)}$\\(logic snapshot)};
\node[box, text width=3.0cm] (a2) at (4.9,-5.6) {Analysis $A$\\Fill rate of the 16 slots};
\node[box] (g2) at (0,-5.6)   {Structural gap\\Null slots};
\draw[arr] (s2) -- (x2);
\draw[arr] (x2) -- (a2);
\draw[arr] (a2) -- node[midway, above, font=\scriptsize] {contrast $\hat{p}$} (g2);
\draw[arr] (g2) -- node[midway, left, font=\scriptsize, align=center] {recirculation} (s2);

\begin{scope}[on background layer]
  \node[draw, rounded corners=3pt, fill=black!4, inner sep=4mm,
        fit=(s1)(x1)(a1)(g1)] (layer1) {};
  \node[draw, rounded corners=3pt, fill=black!4, inner sep=4mm,
        fit=(s2)(x2)(a2)(g2)] (layer2) {};
\end{scope}
\node[anchor=south west, font=\footnotesize\bfseries] at (layer1.north west)
  {Layer 1: the learner's S\&A reciprocity};
\node[anchor=north west, font=\footnotesize\bfseries] at (layer2.south west)
  {Layer 2: the researcher's S\&A reciprocity};

\draw[arr, dashed] (layer1.south) -- node[right=1.5mm, font=\scriptsize, align=left]
  {the support logic itself into the\\type check (self-application)} (layer2.north);

\node[spine, minimum height=9.4cm] (sp) at (8.9,-1.8)
  {AI: a critical file\\(gives no answers,\\returns questions)\\[10pt]Paper ontology\\$\mathcal{O}$\\[4pt]Type check\\$\mathrm{TypeCheck}(\cdot)$};
\draw[darr] (a1.east) -- (a1.east -| sp.west);
\draw[darr] (a2.east) -- (a2.east -| sp.west);
\end{tikzpicture}
\caption{The dual-layer model of structural-gap-driven metacognitive support. Layer 1 (the learner) and Layer 2 (the researcher) share one and the same S\&A reciprocity, while AI as a critical file and type checking by the paper ontology act as a spine running through both layers.}
\label{fig:model}
\end{figure}

Figure~\ref{fig:model} presents the overall structure.

Figure~\ref{fig:model} arranges one cycle of the reciprocity horizontally, the dual-layer structure vertically, and the spine that runs through both layers at the right edge. The crux is the recirculation arrow that returns from the lower left to the upper left. It makes visible that the gap and the unfilled slots are given as arguments of Synthesis in the second equation of Eq.~\eqref{eq:cycle}: evaluation does not end as evaluation but turns into a constraint on the next construction. Three points should be read off the figure. First, the two layers have the same shape, which means that learner support and researcher support are connected by nothing more than the replacement of the Analysis mapping $A$. Second, the spine reaches into the interior of the layers. This shows that the AI is not a referee delivering evaluations from outside the reciprocity but a constituent built into Analysis. The arrows are bidirectional because the direction from user to AI is reverse Analysis, whose frequency and quality are captured as $A_{epi}$. Third, the recirculation arrow always returns to the user's Synthesis, which shows that the authority to decide on revisions remains with the user rather than the AI; this is the design-level guarantee of the redistribution of the four powers \parencite{akata2020} and of shared agency \parencite{damsa2010}. In terms of the quadrants, the Synthesis nodes are chiefly Q1, the computation of the objective parameters and the detection of Nulls are Q4, and the declaration of $\hat{p}$, together with the decision whether to accept or reject what the recirculation returns, is Q3. Q2 is not shown explicitly in the figure. That is not because Q2 is excluded but because the figure draws the main line of the reciprocity; what the critical file forbids is the route that rewrites an artifact without passing through Q3, not generation by AI in general.

%% file: 04-system.tex
\section{The Vibe Compiler prototype: specification and outcomes}
\label{sec:system}

\subsection{Overview and configuration of the system}
\label{subsec:sys-overview}

The Vibe Compiler is a research-logic compiler that maps a user's inchoate Vibe onto an academic ontology and synthesizes it into a logical structure. The system runs a type check on the input and returns feedback in the form of a compile error whenever a constituent required by academic writing is missing---whenever it is Null. One constraint separates this system from ordinary writing-assistance tools: it returns a question rather than a proposed correction.

The name derives from Vibe coding \parencite{karpathy2025}. In Vibe coding the AI generates working code, and a compiler and a runtime guarantee its correctness through immediate feedback. In research, by contrast, the counterpart of code is not the prose itself but how the logic is built. The sequence ``background $\rightarrow$ problem $\rightarrow$ objective $\rightarrow$ solution $\rightarrow$ evaluation $\rightarrow$ findings'' plays the part of the execution log. We propose a mechanism that acts as a compiler for building that logic. What takes place is therefore not Vibe coding but Vibe Compiling.

The prototype combines NotebookLM and Gemini. NotebookLM carries the fixity of the ``type'' that the paper ontology provides; Gemini carries the flexibility of synthesis. In this configuration we did not engineer prompts; we supplied materials. We loaded seven kinds of material into NotebookLM: (1) a survey paper on agency \parencite{roePerkins2026}, (2) a paper-template manuscript we had written (the paper ontology), (3) the 16 academic parameters extracted from it (Table~\ref{tab:params}), (4) the conditions for consistency checks among those parameters. Items (5)--(7) are three kinds of probing procedure obtained through dialogue with the Vibe Compiler itself: dialogue designs for researchers, for learners, and for the system as a co-creative partner. The survey paper supplies the criteria by which the system can tell a user ``you are still at the level of a Maker.'' The paper ontology is not a description in a formal language but a prose document written for human readers. The system itself proposed each of the three probing procedures in the course of dialogue, and we then examined and settled them.

The role specification given at start-up governs the system's behavior. It imposes four conditions: the system must define itself as a research-logic synthesis compiler; it must treat the paper ontology as the sole ``type'' and use the remaining materials as files (in the sense of a rasp); it must refuse to accept a mere development report and must return unfilled slots as compile errors; and it must respond with questions rather than corrections. This is the whole of what we implemented.

\subsection{The core finding: what works is not the inference engine but the content of the structure supplied}
\label{subsec:sys-content}

The principal finding of this paper follows directly from the configuration just described: the Vibe Compiler is driven not by the inferential power of the large language model but by the structure of the content we loaded into NotebookLM.

The grounds for this claim lie in a property of the paper ontology we supplied. That document contains the following passage as a bad example of an introduction: ``A system with these functions is required. (No reason is given.) We therefore built such a system. It has these functions. It was built in this way. What is wrong with this is that it is nothing but an enumeration of results and contributes nothing whatsoever to the accumulation of scholarship.''

In this example all four items---requirement, system, functions, and method---are filled in. A checklist that inspects only whether items are present would therefore pass the document. The document fails on one point alone: nothing links the requirement to the system by giving the reason why the system is needed. The document is thus preserved as a test case---a failing example that a presence check cannot detect in principle, and that fails only when the relations among the items are examined. The Vibe Compiler can compile because it was given not a list of questions but a structure in which conditions hold between one question and another.

This observation amounts to a reconfirmation of what ontology engineering has consistently maintained. On the difficulties of intelligent educational systems, \textcite{bourdeauMizoguchi2000} wrote that ``All these issues are content-related ones. In other words, neither inference techniques nor beautiful theoretical formalism can contribute to an improvement of the situation.'' They were pointing at precisely this configuration. However powerful an LLM becomes, it will merely emit fluent prose if it has no structure to work against, and no structural gap will arise. More important still, the paper ontology we supplied had never been formalized. Ordinarily, building a formal ontology from such a document would have been necessary. With the advent of generative AI, however, loading the document into NotebookLM was enough to make it run. We discuss the implications in Section~\ref{subsec:disc-content}.

\subsection{An execution example: from start-up through type check to error}
\label{subsec:sys-dialogue}

We give below an excerpt from an execution log at the second layer (the researcher layer). It comes from a session in which one of the authors submitted their own Vibe concerning learning by problem posing.

\begin{quote}
\small
User: You are a system that supports the conduct of research in the manner of Vibe Coding. Proceed with this dialogue as a research-logic synthesis compiler, on the basis of the paper ontology now held in NotebookLM.

System: Understood. Starting as the research-logic synthesis compiler (v1.0). Using the other materials as files, I polish and build your thinking so that the fragmentary intuitions (Vibes) you present fit a single type: the paper ontology. This system does not permit a mere development report. Begin by entering the state of the domain, the problem you seek to solve, and the direction of the solution.

User: Learners ordinarily learn by solving problems, but research on problem posing---having them learn by composing the problems to be solved---has begun to appear. I want to build a cognitive model that explains why composing problems advances learning. My thought is that decomposing thinking activity into two components, Synthesis and Analysis, might make it possible to explain why learning proceeds through problem posing.

System: Performing semantic analysis and mapping onto the types of the paper ontology. The background is the shift from passive problem solving to active learning by problem posing; the problem to be solved is the knowledge gap in which the internal mechanism behind ``why learning advances'' remains unexplained; and the proposed method decomposes thinking activity into Synthesis and Analysis and grasps problem posing as their interaction.\\
\textit{[Polishing with the file]} According to the materials, a learner's capacity to participate in the construction and evaluation of knowledge is called epistemic agency. Are the Synthesis and Analysis you speak of not precisely the exercise of that agency? Moreover, if Analysis is defined not as mere decomposition but as the evaluative judgment of whether one's own Synthesis is sound, the differentiation from existing cognitive models becomes sharper.\\
\textit{[Type-check error report]} The following parameters are missing (Null) from the logic as it stands. (1) Limitations of existing methods: why are existing cognitive models insufficient? Do they overlook S, or A, or the reciprocity of both? (2) Beneficiaries and benefits: who concretely gains what new value? (3) Name of the concept: if you were to give this cognitive model of S\&A reciprocity a distinctive name, what image would it convey?
\end{quote}

One point deserves attention. The intuition initially submitted was a single sentence carrying no source, no beneficiary, and no differential against existing methods. The system nevertheless forced it into the slots of the paper ontology, and at the very moment the assignment succeeded, it detected what was missing from the chain of logic. The three points raised did not remain isolated remarks: each was converted, through re-synthesis, into settled logic. Point (3) led to the concept name ``S\&A reciprocity''; point (1) led to the differential logic named ``the black-box problem of the cognitive mechanism''; and point (2) led to the beneficiaries---``learners and teachers as Managers''---and to the benefits they obtain. None of these elements was present in the original utterance. The structural-gap-driven chain of ``naming what is missing $\rightarrow$ re-synthesis $\rightarrow$ settling the logic'' occurred in an actual dialogue rather than merely as a design assumption. The type check functioned here not as a demand for gap-filling but as the occasion on which new logic was produced.

At each cycle of the S\&A reciprocity, the system externalizes the settled logic as a logic snapshot. A snapshot has four fields: the ID of the corresponding element of the paper ontology, the name of the logic, the significance synthesized, and the core of the novelty. The session above yielded Snapshot 01. It gives ``S\&A reciprocity'' as the name of the logic and ``decomposing problem-posing activity into Synthesis and Analysis and defining their mutual interchange as the very source of the learning effect'' as its significance. As the core of its novelty it gives ``incorporating, as Analysis, the evaluative judgment that existing models ignored, thereby making it possible to explain the cognitive mechanism by which problem posing transfers to problem-solving ability''. A snapshot is not a static record; it externalizes the state transitions of the reciprocity process. It therefore allows one to trace which piece of logic was produced in response to which type-check error. In the course of building this paper's research logic, we settled twelve snapshots in all.

\subsection{The mechanism of the type check}
\label{subsec:sys-typecheck}

Table~\ref{tab:triggers} sets out the family of probing triggers, that is, what the type check returns when it detects an unfilled slot or an inconsistency. The triggers share one purpose: to prevent the cognitive offloading in which an AI robs the user of thought by supplying answers, and thereby to sustain productive struggle. That purpose rests on one indispensable condition: the error must take the form of a question rather than a proposed correction.

\begin{table}[htbp]
\centering
\caption{The system of probing triggers fired by the type check}
\label{tab:triggers}
\footnotesize
\begin{tabularx}{\linewidth}{@{}>{\raggedright\arraybackslash}p{0.09\linewidth} >{\raggedright\arraybackslash}p{0.20\linewidth} X@{}}
\toprule
Layer & Trigger & Example utterance \\
\midrule
Common & Slot Null check & ``The benefit of the solution is not defined. Describe who is helped and in what way.'' \\
\addlinespace
Common & Forced extraction of the gap between the ideal and the present state & ``Give the domain-specific reason why current technology cannot reach that ideal.'' \\
\addlinespace
Common & Stress testing against multiple quality criteria & ``Has scalability been considered for a hundredfold increase in data volume?'' ``Are there practical concerns regarding memory consumption?'' \\
\addlinespace
Common & Mandatory critical comparison with existing methods & ``On what point is your proposal decisively different from conventional methods?'' (the system does not accept a claim of novelty that states neither the differential nor the limitations) \\
\addlinespace
Common & Prompting the elevation of results into findings & ``Organize the current bottleneck as a remaining issue and anticipate the technical elements its resolution will require.'' \\
\midrule
Layer 2 & Conversion of intuition into academic parameters & ``When something strikes you as unsatisfactory, is it a shortfall in scalability, reliability, accuracy, computational cost, or reusability?'' \\
\addlinespace
Layer 2 & Probing premises and boundaries & ``What premises does that existing method tacitly assume? How is its reliability impaired when they break down?'' \\
\addlinespace
Layer 2 & Contrast for constructing differential logic & ``State the limitation not as a shortfall in performance but as a limitation of the design philosophy. How does your proposal break through it, by relaxing a restriction or by generalizing a concept?'' \\
\addlinespace
Layer 2 & Critical file that stimulates epistemic agency & ``Draw on your own domain knowledge to refute the limitation the AI has computed.'' (the refutation is accepted as reverse Analysis and counted toward $A_{epi}$) \\
\midrule
Layer 1 & Creation of dissonance through the gap between prediction and result & after having the learner declare a self-prediction, ``The system indicator gives the minimum value. Which element did you find difficult?'' \\
\addlinespace
Layer 1 & Critical verification of a deliberately lenient evaluation & immediately after issuing a lenient evaluation, ``I judged it so, but might I be overlooking a hidden bug or an inefficiency?'' \\
\bottomrule
\end{tabularx}
\end{table}

The type check does not stop at checking whether slots exist. As Section~\ref{subsec:sys-content} showed, some documents fail even though every item is filled in. The Vibe Compiler therefore carries consistency checks that verify the logical correspondence between slots (Table~\ref{tab:consistency}). These checks test whether the academic chain of logic ``background $\rightarrow$ problem $\rightarrow$ solution $\rightarrow$ evaluation $\rightarrow$ findings'' holds together link by link. The mirror-image check between limitations and novelty deserves particular note. It detects a trap researchers readily fall into: ``improvement carried out within a framework that someone else has already laid down.'' It verifies whether the newness of one's own method is a logical necessity that repairs the weakness of the existing method, and so compels the user to build a contrastive structure.

\begin{table}[htbp]
\centering
\caption{Consistency checks among parameters (five kinds)}
\label{tab:consistency}
\footnotesize
\begin{tabularx}{\linewidth}{@{}>{\raggedright\arraybackslash}p{0.20\linewidth} >{\raggedright\arraybackslash}p{0.26\linewidth} X@{}}
\toprule
Name of check & Pair of slots collated & Example of inconsistency \\
\midrule
Synchrony of objective and evaluation criteria & quality characteristics contained in the objective $\leftrightarrow$ items of the evaluation & the objective claims improved reliability, yet the evaluation goes no further than confirming accuracy on small-scale data \\
\addlinespace
Vector collation of significance and benefits & importance and beneficiaries $\leftrightarrow$ benefits & the problem is delayed judgment in clinical practice, yet the benefit lies on the unrelated axis of improved system maintainability \\
\addlinespace
Mirror image of limitations and novelty & limitations of existing methods $\leftrightarrow$ novelty & the limitation of the existing method is given as high cost, yet the novelty of one's own method is improved accuracy \\
\addlinespace
Boundary between premises and coverage & premises $\leftrightarrow$ coverage & the premises assume clean data, yet the coverage includes noisy real-time data \\
\addlinespace
Entailment from experimental results to findings & soundness of the evaluation $\leftrightarrow$ findings obtained & the experiment shows that processing is delayed on large-scale data, yet the findings generalize to effectiveness in every environment \\
\bottomrule
\end{tabularx}
\end{table}

The triggers in Table~\ref{tab:triggers} are subject to a further cross-cutting constraint. The system may positively synthesize content that fills a knowledge gap, and it may actively synthesize and present the novelty or significance of the research, saying for instance, ``That Vibe could become a concrete solution to the lack of scalability from which existing methods suffer.'' It must nevertheless present dissonance and a solution together, and it may not end an utterance with criticism alone. The system is also required to conduct anticipatory dialogue that preempts possible consequences; to delegate choice and control by offering several candidate problems for the user to choose from, so that a route of rebuttal always stays open; and to give real-time feedback that confronts the user with evaluation at the moment of construction. These are the conditions under which the user comes to recognize the AI as a co-creative partner rather than a judge. Direct rewriting of the artifact is forbidden; presenting material is not.

\subsection{UI design: a logic-synchronized development editor}
\label{subsec:sys-ui}

\begin{figure}[htbp]
\centering
\includegraphics[width=\linewidth]{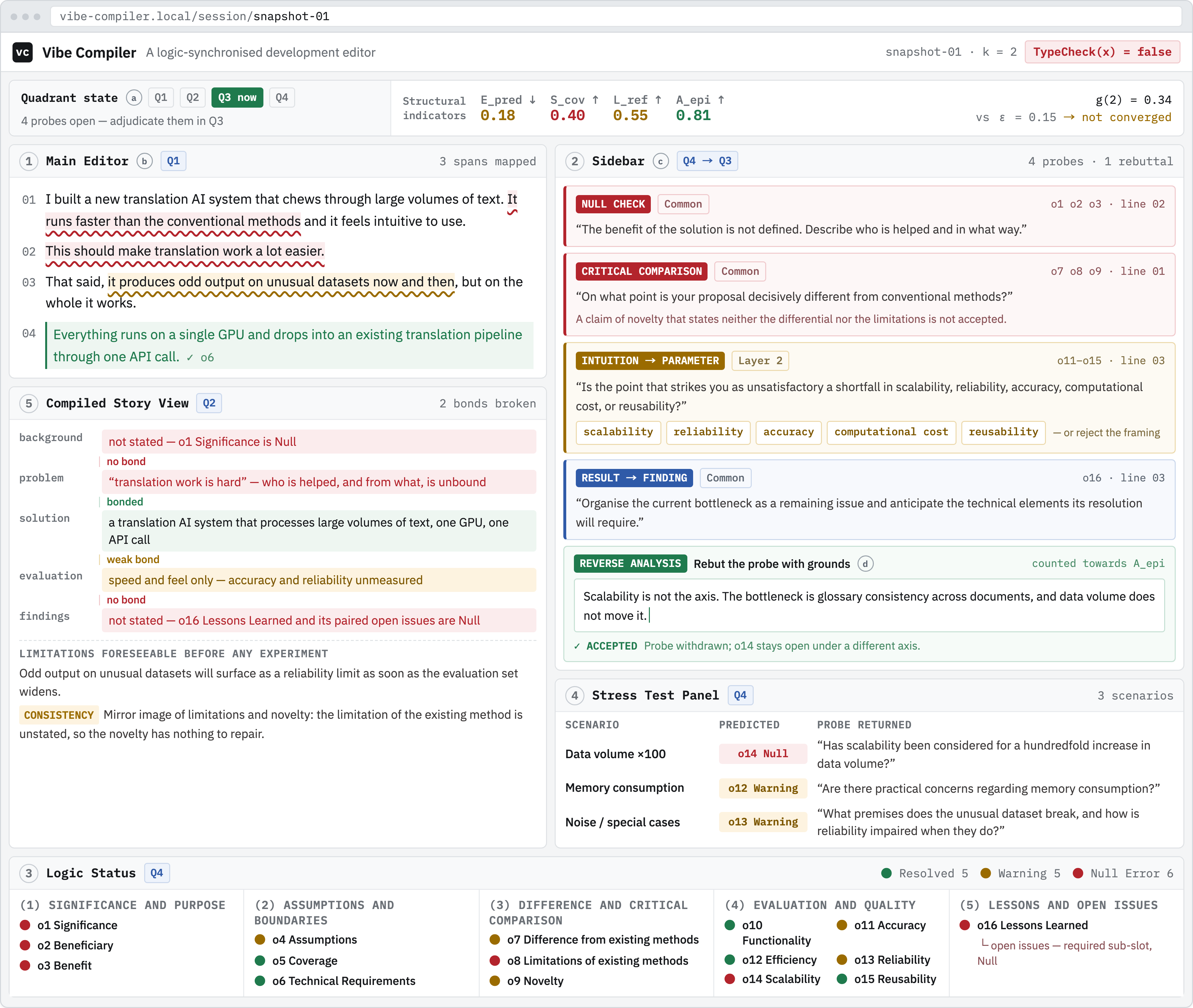}
\caption{An example of the UI design of the Vibe Compiler (a scene in which a Vibe concerning research on a translation AI system has been entered and the type check has returned four probes). It consists of the five panes---the Main Editor (Q1), the Sidebar (Q4$\rightarrow$Q3), the Logic Status (Q4), the Stress Test Panel (Q4), and the Compiled Story View (Q2)---with the quadrant state and the structural indicators ($E_{pred}$, $S_{cov}$, $L_{ref}$, $A_{epi}$) displayed at all times along the top.}
\label{fig:ui}
\end{figure}

As the user interface that realizes the dialogue design above, we propose a logic-synchronized development editor composed of five panes. The UI is not a matter of mere appearance; it is the apparatus that decides which of the four quadrants the user is placed in. The Main Editor (Q1) is where fragmentary thoughts are written without regard to form, and as they are written the AI tries in the background to map them onto the 16 parameters. The Sidebar (Q4 and Q3) returns questions in real time, such as ``on what point is your current implementation decisively different from existing methods?'' It also provides a rebuttal interface: whenever the AI's probe misses the mark, the user can enter a grounded refutation, and a successful rebuttal here amounts to proof of epistemic agency. The Logic Status (Q4) visualizes the state of the 16 parameters in green (Resolved), yellow (Warning: ambiguous or inconsistent), and red (Null Error). The Stress Test Panel (Q4) predicts how scalability and reliability change under scenarios such as a hundredfold increase in data volume, and probes accordingly. The Compiled Story View (Q2) previews the chain of logic assembled from fragmentary Vibes and presents the limitations foreseeable before any experiment is run. When the system warns that ``Significance is Null'' and the user extinguishes the error by putting beneficiaries and benefits into words, that work is debugging of logic in the strict sense. The UI is designed to guarantee, by compulsion, the logical rigor that contributes to the accumulation of scholarship, without slowing the speed of development that the Vibe carries. Figure~\ref{fig:ui} shows a concrete example of this screen design. Each of the five panes carries the quadrant it belongs to, the quadrant state and the four indicators are displayed at all times along the top, and in the Sidebar a rebuttal interface stands open on one of the probes, so that the whole passage in which a grounded refutation is accepted as reverse Analysis and counted toward $A_{epi}$ appears as an operation on the screen.

\subsection{Outcomes: what the compiler demanded, and how the authors answered}
\label{subsec:sys-demands}

This section organizes the demands and probes that the Vibe Compiler actually issued in the process of building this paper's research logic, together with our responses to them. This is the only substantive evidence concerning the functional executability of the prototype, and at the same time a record that reciprocity at the second layer did in fact occur.

\begin{table}[htbp]
\centering
\caption{Principal demands from the Vibe Compiler and our responses (a record of the process of building this paper's research logic)}
\label{tab:demands}
\footnotesize
\begin{tabularx}{\linewidth}{@{}>{\raggedright\arraybackslash}p{0.04\linewidth} >{\raggedright\arraybackslash}p{0.31\linewidth} X@{}}
\toprule
\# & Demand from the compiler (type-check error) & Our response and the logic settled \\
\midrule
1 & Why are existing cognitive models insufficient? Do they overlook S, or A, or the reciprocity? & Formulated the claim that existing models can describe the procedures and the products of problem posing but do not explain the dynamic process that deepens understanding. Settled the differential logic as ``the black-box problem of the cognitive mechanism'' \\
\addlinespace
2 & Who are the beneficiaries of this cognitive model? What new value do they obtain? & Settled that learners gain structural understanding and teachers gain the new, advanced expertise of supporting S\&A reciprocity \\
\addlinespace
3 & If a distinctive name were given to this model, what image would it convey? & Coined ``S\&A reciprocity (Reciprocal Synthesis \& Analysis)''. Settled the definition that the mutual interchange is the source of the learning effect \\
\addlinespace
4 & Make the evaluative axes of Analysis concrete. What constitutes structural complexity? & Defined $N_{step}$, $N_{var}$, and $D_{map}$ for the arithmetic domain, and at the same time made the AI's solvability explicit as a premise \\
\addlinespace
5 & Is there no evaluative axis other than functional executability, the solvable/unsolvable distinction? & Added inquiry-driven Analysis, which positively evaluates the state of ``being unable to solve the problem, yet being interested in the structure of its solution''. A departure from the supremacy of the correct answer \\
\addlinespace
6 & How is the improvement of a learner's self-evaluation ability to be quantified? & Defined the four indicators $E_{pred}$, $S_{cov}$, $L_{ref}$, and $A_{epi}$ (Eqs.~\eqref{eq:three} and \eqref{eq:aepi}). Proposed by the system itself, then examined and settled by the authors \\
\addlinespace
7 & What is the decisive difference from existing problem-posing models? & Formulated the claim that, whereas Silver and Christou et al.\ grasped problem posing as an enumeration of procedures, this model focuses on the dynamic process of seeking to resolve dissonance \\
\addlinespace
8 & What is the difference from mere Reflection? As it stands this falls within an existing concept & Differentiated on three points: objective externalization of the indicators, mutual delimitation of construction and evaluation, and deliberate maintenance of productive struggle. Settled ``structural-gap-driven metacognitive support'' \\
\addlinespace
9 & What are the implementation-level difficulties in extending S\&A reciprocity to learners in general? & The system itself pointed out the difficulty of defining domain-general structural parameters. In response, carried out the generalization to the three axes $D_{depth}$, $W_{width}$, and $M_{map}$ \\
\addlinespace
10 & The problem set out in the objective does not correspond logically to the insufficiency identified in existing methods & An instance of a consistency check firing. Matched the objective of metacognitive support to the limitation of existing methods---their dependence on subjective noticing---and synchronized both with the solution vector of externalizing objective indicators \\
\bottomrule
\end{tabularx}
\end{table}

Table~\ref{tab:demands} shows that the compiler's demands were not mere requests to fill gaps. Of the ten, we had not prepared in advance the naming of the concept (\#3), the design of the evaluation indicators (\#6), or the identification of the need for generalization (\#9). These points were produced only through reciprocity with the system. Item \#9 deserves particular note: the system pointed out the limits of its own scope of application, and in response we carried out the generalization to three domain-general axes.

We must equally record what could have been lost in each demand. Once a Null has been named, who fills that blank is the decisive branching point. Had the user answered ``write the description of the limitations for me,'' the settled logic would have become a product of the AI's Synthesis (Q2), and no judgment about what counts as sound (Q3) would have remained on the user's side. Even if the quality of the prose were much the same, the decision about what to set up as the differential logic would have changed hands. The actual record is a sequence: the user answered each naming of a blank (Q4) with re-synthesis (Q1). What was protected is that decision, not the prose.

The foregoing confirms the functional executability of the prototype. The system ran a type check against 16 slots, detected unfilled slots as Null, returned feedback in the form of questions, and detected inconsistencies between slots. These results constitute affirmative evidence for RQ1. The same mechanism operated in both the learner domain and the researcher domain, with nothing changed but the Analysis mapping; this constitutes affirmative evidence for RQ4. Moreover, this paper, as an artifact, is itself a demonstration that a researcher's Vibe can be compiled into substantive research while a paper is generated from it. RQ2, RQ3, and RQ5 ask about the magnitude of the effect and require measurement under controlled conditions; as a matter of scope, we therefore set them out in Section~\ref{subsec:disc-limit}.

%% file: 05-illustrations.tex
\section{Application example: extending the model to the first layer (learners)}
\label{sec:illus}

\begin{figure}[htbp]
\centering
\includegraphics[width=\linewidth]{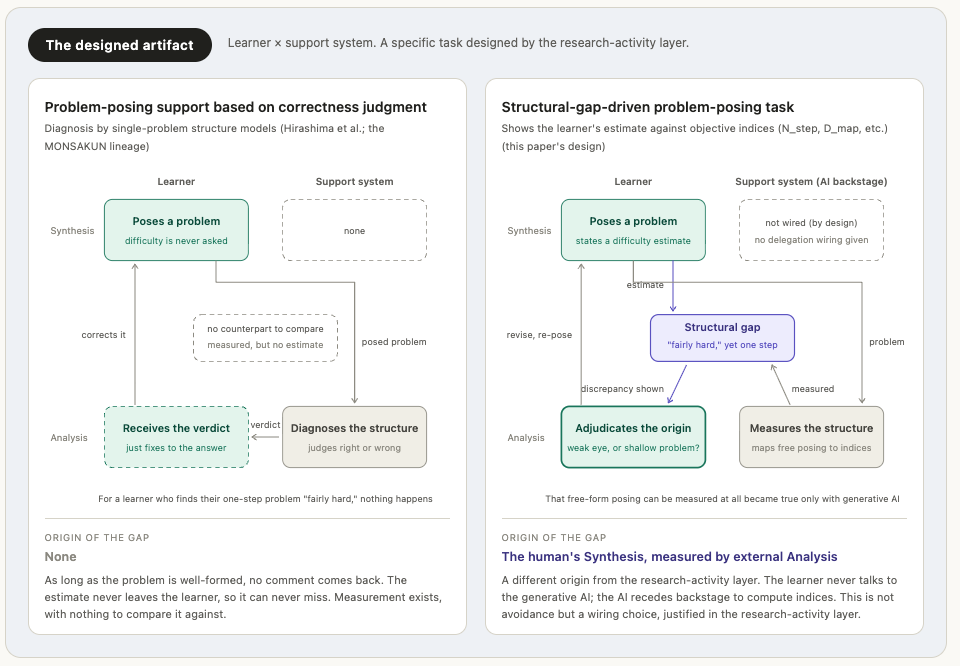}
\caption{Contrasting configurations in the learner layer. In conventional support for learning by problem posing based on correctness judgment (left), the hunch never leaves the learner, so there is no origin for a gap. In the structural-gap-driven problem-posing task we design (right), the learner must state the hunch first, and an external Analysis then measures it.}
\label{fig:learner}
\end{figure}

The execution log in Section~\ref{sec:system} came from the second layer, that of the researcher; this section turns to the first layer, that of the learner. The gap originates differently in the two layers, and that difference matters (Figure~\ref{fig:learner}). In the researcher's layer, the gap originates in the AI's Synthesis. For the learner we design the reverse arrangement: the human carries out the Synthesis, and an external Analysis measures it. The learner never converses with the generative AI, and the AI steps back into the background, where it does nothing but compute the indicators. We therefore disconnect the support system's Synthesis from the learner by design. We are not avoiding generative AI here; we are applying a design choice that we already justified for the researcher's layer. Exercising judgment here means refusing to hand the learner a configuration of wholesale delegation.

The learner-layer sequences below are simulations derived from the design of the mechanism we describe; they are not records obtained from actual learners. The numbers they contain are the values the design anticipates. We ask not whether an artifact was produced, but what can be lost while producing it and what remains on the human side.

This configuration sits at the same coordinates as Type (III) in Table~\ref{tab:gaporigin}. The learner alone carries out Synthesis, and two things stand side by side as Analysis: the learner's own self-assessment and the support system's computation of indicators. The origin here differs from that of Type (IV), which we adopted for the layer of research activity, so the two layers do not share the same type. The learner layer nevertheless does not remain within the limits of Type (III), because we add a condition by design. The learner declares the hunch about difficulty outwardly at the same moment as posing the problem, and the learner is the one who decides which side the resulting difference should be attributed to. Only once the hunch has been externalized does the difference from the indicator become an observable quantity. Adjudicating where that discrepancy comes from means deciding whether the way of seeing was too lenient or the problem posed was shallow. That adjudication reopens the question of the very thing the learner set out to make (Section~\ref{subsec:illus-refutation}). Type (III) stops at the correctness of the assessment because it lacks a path that returns this adjudication of attribution to the user. The two types therefore reach different places even where their coordinates coincide.

The difference from conventional support for learning by problem posing appears at exactly this point. The single-problem structure model \parencite{hirashima2008,hirashima2014} diagnoses whether the structure of the posed problem is correct, but a problem that is correct as a problem draws no remark. Because the hunch never leaves the learner, it can never turn out to be wrong. Measurement is present, yet there is nothing to compare it with. For a learner who takes a self-made one-step problem to be ``moderately difficult,'' nothing happens at all. Our task, by contrast, has the learner declare the hunch in advance and then presents the difference from the objective indicator. Only generative AI made free-form problem posing measurable in the first place, and that is why this design belongs specifically to the age of generative AI.

\subsection{The arithmetic problem-posing domain}
\label{subsec:illus-arith}

The sequence begins with a learner who poses the problem ``I bought an apple costing 100 yen. When a consumption tax of 10\% is charged, how much is the payment?'' (v1.0) and assesses it as ``moderately difficult, because it requires a calculation.'' The system first confirms that a unique correct answer can be derived, which is solvability for the AI. Nothing so far departs from ordinary answer-oriented support. The next step is different. Scanning the structure of the posed problem, the system computes $N_{step}=1$ (the single operation $100 \times 1.1$), $N_{var}=1$ (the payment alone), and a low $D_{map}$ (the expression can be built in the order given). It thereby detects a structural gap between these values and the learner's subjective declaration.

The system never offers the revision ``add a discount.'' It returns the gap as a question instead: ``The AI has derived the correct answer (110 yen) for the problem you posed. Against the goal of structural complexity you yourself set, however, the current number of operational steps is one. How would this change if you increased the number of unknowns, or added to the context an operational component running in the opposite direction, such as a discount?'' Should the system give only the gap rather than the answer, or give the revision itself? That choice is the point at which the twin demands of protecting and driving take shape as a concrete design decision. The moment the latter is chosen, productive struggle is erased \parencite{hiebertGrouws2007}. The learner then sets the goal of deliberately raising the difficulty and reconfigures the components (Table~\ref{tab:usecase-params}).

\begin{table}[htbp]
\centering
\caption{Transition of the structural indicators before and after one cycle of the reciprocity in the arithmetic problem-posing domain (simulation)}
\label{tab:usecase-params}
\footnotesize
\begin{tabularx}{\linewidth}{@{}l X c c c X@{}}
\toprule
Version & Gist of the problem statement & $N_{step}$ & $N_{var}$ & $D_{map}$ & Learner's self-assessment \\
\midrule
v1.0 & One apple at 100 yen with 10\% consumption tax & 1 & 1 & Low & ``Moderately difficult, because it requires a calculation'' \\
v2.0 & Three apples at 100 yen each, a 20\% discount, 10\% consumption tax, and change from a 500 yen coin & 4 & 2 & Medium to high & ``It became harder because I added the change'' \\
\bottomrule
\end{tabularx}
\par\smallskip
\begin{minipage}{\linewidth}
\footnotesize Note: as stated at the beginning of this section, the values in this table come from a simulation based on the design.
\end{minipage}
\end{table}

The reciprocity does not end at v2.0, and that matters. The learner puts the insight into words: ``Adding the change means that I have to work out not only the payment but also a subtraction step, and above all that I have to consider the relation between the payment and the money I hold, so the mapping becomes harder. Is this what it means for $D_{map}$ to rise?'' The system records this as a logic snapshot and returns further probes. One probe questions the fragility of the premises: ``This problem depends on the premise that the 500 yen coin always suffices. Does the expression still function as it stands if the unit price is rewritten as 200 yen?'' A second questions the ambiguity of the order of application: ``Which is to be applied first, the 20\% discount or the 10\% consumption tax? As the statement currently reads, this is not uniquely determined.'' A third questions extensibility: ``If the apples became 100 in number and a bag charge and a points rebate were added, how would the structural complexity change?''

When the learner is already satisfied with v2.0, probing it further exposes issues that could not even have been raised at the stage of v1.0. That is what the two-stage arrangement shows. In precisely this form, the mutual delimitation repeats itself: the output of Analysis becomes an argument to the next Synthesis. None of the three probes stops at naming a quality indicator. Each translates the indicator into components and numbers of the artifact: the unit price, the order of application, and the number of items. Whether a probe functions as a critical file is settled not by its topic but by its granularity. The moment the same issue is stated at the level of ``make the premises explicit,'' the reciprocity turns without advancing.

The learner may finish without ever noticing that the self-assessment ``moderately difficult'' departs from the actual structure. That is the loss at stake here. The posing of the problem itself succeeds (Q1 succeeds). The self-assessment nonetheless fails to agree with the structure (Q3 fails). Under a description that lacks the double distinction, this state can be written down only as ``a reasonably good problem was produced.'' What the design protects is the learner's goal of redefining difficulty for themselves, not the value of $N_{step}$ as such. If the learner replies ``then fix it for me,'' v3.0 will duly come into being, but the learner will be left with no recognition of what had been overlooked.

\subsection{The Japanese reading comprehension domain: only the indicators are replaced, not the mechanism}
\label{subsec:illus-jpn}

In the Japanese language domain the components of Synthesis become keywords, connectives, paragraph blocks, and contrastive structures, and the structural indicators are replaced by $S_{ref}$, $L_{link}$, and $V_{map}$. Only the symbols change; the mechanism remains as it was. This ease of replacement underwrites the domain generality of the model. Take as the source text an expository passage arguing that AI is good at calculation but does not understand meaning. Here v1.0 reads ``What is the weakness of AI? Extract it from the passage,'' and is computed as having a minimal $S_{ref}$, $L_{link}=1$, and a $V_{map}$ of zero. The learner's self-assessment, however, is that this is a good problem because it asks about the basics, so a gap arises between the subjective sense and the structure. The system asks, ``$L_{link}$ is at its minimum of one. How would $L_{link}$ change if you combined the question with the human-specific embodiment discussed in the adjacent paragraphs and reshaped it into a form that requires the reason to be explained?'' The reconstructed v2.0 reads ``Drawing on the claim made in the passage, explain in no more than 40 characters why AI can be said not to understand meaning, comparing it with the characteristics of human embodiment,'' and the system computes a large $S_{ref}$, $L_{link}=3$, and a high $V_{map}$. Between arithmetic and Japanese, only the way each axis is measured changes; the form of the reciprocity and the requirement on the granularity of the probes stay identical.

\subsection{Reverse Analysis: how the AI's mis-assessment turns into a learning opportunity}
\label{subsec:illus-refutation}

The model characteristically reads the AI's errors not merely as defects but also as occasions for observing where agency resides. Reverse Analysis, however, does its full work in the layer of research activity, where the user converses with the AI directly (Type (IV)). When a researcher objects, with grounds, that ``the limitation that the AI computed for the existing method overlooks a domain-specific constraint,'' the system accepts this as reverse Analysis. It records the objection as a high-quality one that identifies a contradiction among the premises and offers an alternative criterion. The $A_{epi}$ of Eq.~\eqref{eq:aepi} counts objections of this kind, weighting them by their quality.

The counterpart in the learner layer is not an objection to the AI but the act of adjudicating where the discrepancy comes from. A learner confronted with the difference from the objective indicator must decide whether the way of seeing was too lenient or the problem posed was shallow. The first verdict leads to remaking the hunch; the second, to remaking the problem. The learner decides which side to attribute the discrepancy to, and that decision is what Q3 amounts to in the learner layer. Suppose the learner can ground the low indicator in a design intention of their own: the readers are elementary school pupils, so keeping the wording deliberately plain took priority. That judgment attributes the discrepancy not to the shallowness of the problem but to a difference in the premises.

Erroneous measurement is unavoidable in principle whenever the computation of the indicators is entrusted to a language model \parencite{ji2023,huang2025}. The configuration that keeps the AI in the background in the learner layer also has the effect of narrowing the paths by which such errors flow directly into the learner's judgment.

Concealing that fallibility and letting the system behave as an Oracle would lead users to abandon their own correct intuitions and follow the AI. Given the tendency of large language models to agree excessively with their users \parencite{sharma2023}, this danger is realistic. The model therefore positions the AI as an imperfect partner that errs at times yet stimulates thought, and it requires by design that the path of objection stay permanently open. This is a demythologization that accords with the demands of critical digital pedagogy, as \textcite{roePerkins2026} apply it. The design protects not the artifact, whether a summary or a posed problem, but where the judgment of what counts as valid resides.

\subsection{Positioning in the four quadrants and an explanation of learning by problem posing}
\label{subsec:illus-quadrant}

We now place the cases above in the four quadrants (Table~\ref{tab:quadrant}). In the second-layer session recorded in the execution log, the injection of intuition and the resynthesis fall in Q1, the naming of Nulls by the type check falls in Q4, and the judgment of what should be written falls in Q3, forming the sequence Q1 $\rightarrow$ Q4 $\rightarrow$ Q3 $\rightarrow$ Q1. The arithmetic and Japanese domains take the same sequence, with the posing of the problem in Q1, the self-assessment in Q3, the computation of the structural indicators and the probing in Q4, and the reconfiguration in Q1. The objection to a mis-assessment is a case in which Q3 has countered an erroneous Q4. By contrast, in bare use of generative AI---the capable-servant mode---the user's Q1 degenerates into the composition of a request, Q2 takes over the selection and combination of components, and Q3, which assesses the validity of the artifact, is left empty.

To see what becomes indescribable without this distinction, we may write the second-layer episode in two ways. One description reads that the type check named three Nulls (Q4), that the user judged what to take as the differentiating logic (Q3), and that a settled logic was produced (Q1). The other reads that the author completed the introduction with the support of AI, dividing the labor so that AI generated the prose and the human decided the direction. The latter description draws on a vocabulary of division of labor, task allocation, cognitive offloading, and attribution of responsibility. In that vocabulary, whether Q3 was executed never appears. The process collapses into the single event of an introduction written with AI support. For the same reason, the capable-servant mode and the sequence described here look identical: in both, an artifact was produced using AI. The four quadrants, however, describe them separately---the one as a sequence in which Q1 degenerates into the composition of a request, Q2 takes over the combination, and Q3 is left empty; the other as a sequence in which Q1, Q3, and Q4 are all present. Likewise, in the arithmetic domain the posing of the problem itself succeeds while the self-assessment fails to agree with the structure. That state can be separated out only by describing it as one in which Q1 succeeds and Q3 fails. The human/AI axis differs from theories of the division of labor because we introduce it not to optimize allocation but to name a deficiency: the blank at Q3. The absence of this vocabulary is more than a descriptive inconvenience. If we cannot name what is being lost, we can neither design to protect it nor assess it.

The positioning set out above offers an explanation of the mechanism by which learning by problem posing takes effect. Studies have repeatedly reported that problem posing promotes learning \parencite{silver1994,caiHwang2015}, yet why it does so has not been adequately explained \parencite{caiHwang2015}. Seen through the four quadrants, problem posing differs from ordinary problem solving in that it imposes two demands on the learner at once: Q1, deciding what to ask and combining the components; and Q3, assessing the learner's own artifact. If Q4 is externalized as well, a divergence arises between the results of Q3 and Q4, and that divergence becomes an argument to the next Q1. In short, problem posing promotes learning because it is one of the few activities that impose Q1 and Q3 on the same learner simultaneously. This explanation responds at the level of mechanism to two earlier accounts. \textcite{silver1994} positioned problem posing as an activity before, during, and after problem solving, yet left the cognitive process of problem posing itself at an abstract level; \textcite{christou2005} described it as an enumeration of procedures. The explanation moreover yields testable predictions. Two conditions separate Q1 from Q3: one in which learners pose problems but do not assess them, and one in which they are asked for the self-assessment alone. Both should show a diminished learning effect. Whether Q4 is externalized should govern whether learners can become aware of their own divergence. The model thus gives research on learning by problem posing a design guide for such contrastive conditions.

%% file: 06-discussion.tex
\section{Discussion}
\label{sec:disc}

\subsection{Content-orientation reinstated: an unformalized ontology ran on an LLM}
\label{subsec:disc-content}

Our widest-reaching finding follows from the fact reported in Section~\ref{sec:system}, specifically Section~\ref{subsec:sys-content}: the Vibe Compiler was driven not by the inference engine but by the structure of the content we supplied to it.

\textcite{bourdeauMizoguchi2000} argued that every difficulty obstructing the construction of intelligent educational systems is a matter of content, and that neither inference technology nor elegant theoretical formalization improves the situation. That argument rested tacitly on the premise that content must pass through a stage of formalization before a computer can handle it. The past quarter century of ontological engineering may fairly be described as one sustained inquiry into how to carry out that stage robustly. When \textcite{mizoguchiBourdeau2016} restated that the distinctions an ontology draws are not a question of how they are represented on a computer, they were warning against an understanding that tends to reduce ontology to techniques of formalization. Our prototype shows that the warning was right in a way it had not anticipated. The paper ontology we supplied was not a description in a formal language but a prose document written for human readers. Ordinarily, we would first have had to build a formal ontology from it. With the arrival of generative AI, however, it ran simply by being fed into NotebookLM. The quality of its operation, moreover, did not stop at the shallow level of checking whether items exist; it reached the level of actually rejecting failing cases that cannot be detected without examining the relations among items.

The propositions that follow from this connect in stages. At the outset we should affirm that heavy-weight ontologies have lost none of their value for building robust and verifiable systems: for guaranteeing consistency, the soundness of inference, and reuse, formalization remains irreplaceable. Yet any content-oriented structure carrying constraints can now be reasoned over by an LLM without passing through formalization, and the paper ontology is precisely such a case. Only one consequence follows. The center of value has shifted from the skill of formalizing structure to content itself---to the question of what ought to be given structure at all.

This shift does not compete with ontology research; it restates the claims of that research under different conditions. The case for content-orientation was once made in order to justify the cost of formalization. Now that part of that cost has fallen away, the quality of the content itself directly determines the outcome. We make the case for this shift again from the standpoint of protecting agency, because the question of what to give generative AI is inseparable from the question of which human capacities it replaces and which it stimulates. Without structure, the AI produces fluent prose and no gap arises; without a gap, metacognition is not driven. Designing content in the age of generative AI is therefore designing human agency itself. The practical consequence is plain: driving generative AI as intended requires not the artistry of prompt engineering but the design of the content supplied to it. This claim yields a testable prediction. Accordingly, we give our reproduction procedure not as a manual of operations but as a list of the materials we supplied (Section~\ref{sec:system}, and in particular Section~\ref{subsec:sys-overview}): feeding the same set into a different reasoning environment should reproduce the same type-checking behavior. Isolating how far features specific to NotebookLM contribute remains a task for future work involving experiments.

\subsection{Using generative AI by ``educating'' it}
\label{subsec:disc-persuasion}

The implications of content-orientation have a further side. In fact, the path to the Vibe Compiler ran through an act of persuading the generative AI. As the dialogue history collected in the appendix shows, the AI initially maintained that ``in supporting paper writing and the conduct of research, nothing corresponding to code in Vibe Coding can be generated.'' It grounded this in the following observations: an academic paper is not a mere artifact of expression but the presentation of a delta against an existing body of knowledge together with a logical warrant; code has an immediate feedback mechanism whereas research logic has none; and the antinomic process of synthesizing a new claim while criticizing it at the same time is difficult to complete autonomously. We answered by supplying what was missing. We gave the AI the ``type'' constituted by the paper ontology, the input channel from the human side that we call Vibe, and the 16 parameters as axes of evaluation. The AI then revised its conclusion: ``because the AI already knows what counts as correct (the type of a paper), pouring in what the human wants to do (the Vibe) gives it ample potential to function as a compiler that automatically synthesizes the research logic bridging the two.''

This episode contains an element that can be generalized as methodology. The limits of a generative AI's capability often appear not as intrinsic limits of the model but as a function of premises that have not been supplied. When a generative AI declares that it cannot do something, one can ask it what it lacks and then supply the missing element as content. This procedure belongs to a lineage quite distinct from prompt engineering, and we position it as a way of using generative AI that educates it. Here again the work is done by the content rather than the inference engine, so this episode too supports the claim of the preceding section.

\subsection{Theoretical, methodological, and practical implications}
\label{subsec:disc-implications}

Our model contributes concretely to several theoretical lineages. To research on learning by problem posing it supplies an account of the long-unexplained mechanism by which posing problems promotes learning \parencite{caiHwang2015}: the activity imposes Q1 and Q3 on one and the same learner at the same time. To research on reflection it gives structure to the vague notion of introspection, which has depended on subjective noticing \parencite{schon1983}. It recasts that notion as a dissonance between Synthesis and Analysis and redefines it as a dynamic mechanism accompanied by collation with objective indicators and immediate reconstruction. It also positions evaluative judgment \parencite{tai2018} not as a capacity for evaluation but as an engine that drives the next action, a further contribution to this lineage. To research on epistemic agency it takes a concept previously discussed as the sharing of responsibility among humans \parencite{scardamaliaBereiter2014,damsa2010} and extends it into the space between humans and AI. It also operationalizes that concept as an observable behavior---a rebuttal, on established grounds, to the AI's evaluation---and so gives a measure to what had remained abstract. It likewise brings the trichotomy of \textcite{cox2024} down from a normative framework to the level of a theory of mechanism, where the question becomes what kind of dialogue design brings about the transition from Makers to Managers. To the design theory of hybrid intelligence, finally, it introduces a design variable that does not reduce to the maximization of outcomes. Asking how the four powers are redistributed to the human side \parencite{akata2020} yields a concrete design guideline: withhold Q3.

At the methodological level, the procedure of treating a paper ontology as a type system can apply beyond this paper. It does so because it turns a tacit act familiar from research supervision and from peer review---pointing out what is missing---into two explicit procedures: checking slot fulfillment and checking consistency among slots. The consistency checks in particular (Table~\ref{tab:consistency}) differ from checklist-style guidance in that they reject documents that a test for the mere existence of items would pass. Our four indicators likewise serve to evaluate learners, and for researchers who develop systems they also serve to debug support logic. One further stance is transferable as well: rereading an AI's mistaken evaluation not as a defect but as an occasion for observing agency. In the evaluation of generative-AI-based support systems, errors are ordinarily treated as noise to be reduced; we have instead woven them into the design and converted them into the requirement that the path of rebuttal stay open at all times.

At the level of practice, our model becomes an instrument for converting a supervisor's tacit knowledge into explicit procedure. Questions such as what makes a piece of research significant, or what is insufficient about existing methods, have conventionally been left to the supervisor's experience. The 16 parameters and the consistency checks externalize this system of questions. One conceivable practice is for a student to run the Vibe Compiler before a meeting and put the Null slots into words in advance. In designing lessons around learning by problem posing, presenting structural parameters gives concrete content to an instruction that could previously say no more than ``pose a difficult problem.'' It also asks of the teacher a shift: from delivering correct answers to gauging the divergence between the learner's self-assessment and the actual structure, and then designing probes at an appropriate grain size. This shift demands a new expertise, that of supporting S\&A reciprocity. In drafting policies on AI use, the four quadrants replace the dichotomy of prohibiting AI or allowing it freely. Provisions can then be written quadrant by quadrant, in the form ``Q2 is permitted, and Q3 must be carried out by the person concerned and recorded.''

\subsection{Critical examination}
\label{subsec:disc-critical}

The perspective of critical digital pedagogy, as \textcite{roePerkins2026} apply it, has pointed out that support technologies aimed at efficiency can restrict learners' agency and reproduce existing inequalities. Our position, which actively incorporates generative AI into educational settings, is not exempt from this critique. Our model operates at the level of structuring thought, and it does not itself dissolve inequality of access to high-performance generative AI models. Reports indicate that generative AI can instead widen the gap among novices \parencite{prather2024}, so learners who struggle with metacognition may be the least likely to benefit. Moreover, reliance on the averaged, biased knowledge that AI generates can induce a monoculture of scientific knowledge \parencite{messeriCrockett2024}. We try to resist this by placing the individual's intuition at the origin of the logic, yet we are not free of that bias as long as parameter computation and probe generation are entrusted to a foundation model. The very criterion by which something is judged ``structurally shallow'' may reflect the model's skew.

A more fundamental issue is that the boundary between externalizing thought and replacing it is not sufficiently defined, either theoretically or empirically. As research on cognitive offloading \parencite{riskoGilbert2016} shows, externalization can release cognitive resources and can also cause capability to atrophy. The distinction between accelerative and exploratory modes of use \parencite{barke2023} lies close to our view, but we have yet to answer the objection that repeated cycles of S\&A reciprocity may themselves become a new form of dependence. Furthermore, if the system presents mistaken parameter values, or limitations of existing methods that are contrary to fact, epistemic trust is damaged and the will to inquire declines. The rate of mistaken evaluation has not been measured. Intervention at an inappropriate moment obstructs the user's authority to decide and can produce a paradox: support that violates the very agency it is meant to protect. Even in the execution log of the second layer, we observed that exhaustive firing of type checking produces remarks that can break the continuity of the work. As long as the system must protect and drive at once, comprehensive support and continuous work stand in a trade-off, and firing must be controlled stage by stage. That control law does not follow from our model.

\subsection{Limitations}
\label{subsec:disc-limit}

We begin with the self-referential character peculiar to this paper. Much of what is written here derives from the output of a prototype built on the very mechanism this paper describes. We can state explicitly which parts were human judgment: the framing of the problem, the selection of what to discuss, the rebuttal and rejection of the system's output, and the decision on the overall structure. Much of the drafting, the choice of words, and the development of examples is the system's output, and we occupied the position of judging whether to adopt it. This division of labor is precisely the reciprocity of the researcher layer as we describe it, and this paper applies its own claim to itself. That fact is precisely what demonstrates that Vibe Compiling is possible. On the other hand, self-application shows that the mechanism actually runs; it does not show that the mechanism produces a learning effect. We have drawn that boundary deliberately. One might ask whether this amounts in substance to wholesale delegation while claiming to protect metacognition. Table~\ref{tab:demands} answers by showing, in traceable form, which piece of logic was produced in response to which type-check error, and we invite readers to examine the matter against that record. Confirmation bias arises when the evaluator and the person being evaluated are the same party; controlling for it is a task for the next stage.

We list the remaining limitations below.

\begin{itemize}
  \item This is a single case, and we are ourselves the users. The record of the researcher layer rests on one case---the process of building this paper's research logic---and the design includes no device for excluding confirmation bias or self-assessment bias.
  \item No quantitative data for the learner layer exist. No controlled experiment has been conducted, and the changes in $E_{pred}$, $S_{cov}$, $L_{ref}$, and $A_{epi}$, together with transfer effects, are all unverified predictions. We make no quantitative claim about the size, the conditions, or the persistence of any effect.
  \item The prototype's configuration depends on off-the-shelf services. Because the prototype combines commercial services, its behavior can change when those services change their specifications. Because parameter computation depends on generation by a large language model, it is not deterministic, and identical inputs are not guaranteed to yield identical outputs.
  \item Our examination is confined to a single language, Japanese. Indicators such as the vocabulary substitution difficulty $V_{map}$ depend strongly on linguistic structure, and their viability in other languages is unconfirmed.
  \item Domain-specific parameters are defined by hand. Designing the Analysis mapping requires the judgment of a domain expert. The generality at issue is a generality of the mechanism rather than a generality of the settings, and a procedure for validating the parameters is not yet in place.
  \item Structural parameters do not guarantee the semantic soundness of an artifact. High values of $N_{step}$ or $L_{link}$ do not entail that the artifact holds up as a problem. In a shopping word problem, changing the unit price from 100 to 200 yen introduces an exception: the amount cannot be paid with a 500-yen coin. Without an added precondition, the problem no longer holds. The danger remains that probes issued from the side of structure overlook a breakdown on the side of meaning.
  \item The premise that the AI can solve the task is fragile. It breaks more readily as difficulty rises, and once it breaks, we lose the objective value that serves as the reference for $E_{pred}$. If hallucination conceals the break \parencite{ji2023,huang2025}, a self-assessment may be judged inaccurate against an erroneous reference value.
  \item The transfer of evaluative judgment remains a hypothesis. Our model predicts that a stance attentive to structure carries over across domains, but no data supporting transfer exist.
\end{itemize}

One further point deserves mention. Among the materials we fed into NotebookLM was a survey paper on agency \parencite{roePerkins2026}, and that choice conditions the context in which our paper applies. Feeding in a different survey paper would yield a different context for the use of generative AI, and the system of probes the compiler issues would change accordingly. This does not narrow the range of the paper's value, however. The degradation of agency that follows from using generative AI as a capable servant is not the claim of one particular survey paper but a widely shared and general problem. Indeed, the system adapts to any context once the supplied materials are replaced, and that property is itself a corollary of the content-driven finding we advance.

%% file: 07-conclusions.tex
\section{Conclusion}
\label{sec:concl}

\subsection{Summary}
\label{sec:concl-summary}

The rise of generative AI threatens to reduce the construction and transmission of knowledge to the efficient processing of information. It thereby calls into question whether human beings can continue to secure epistemic agency. We have addressed the problem of how to design and realize a support mechanism that preserves the human's authority to decide while honing metacognition, as an inchoate intuition (Vibe) is converted into academic logical rigor. In response, we have proposed structural-gap-driven metacognitive support. Dissonance arises between the user's subjective construction (Synthesis) and the objective structural indices the system presents (Analysis), and our approach treats that dissonance not as an error to be removed but as a source of metacognitive stimulation. Three elements carry this idea. The first is the S\&A reciprocity model, which construes intellectual construction as a mutually constraining reciprocity between Synthesis and Analysis. The second is the dual-layer structure, which unfolds the model across two layers: the learner's and our own. The third is a design that configures AI not as an entity that supplies answers but as a critical file (in the sense of a rasp) that probes the fragility of premises. The prototype Vibe Compiler implements all three.

We summarize what we would ask readers to take away in the following five points.

\begin{enumerate}
  \item Generative AI makes it possible to compile research logic and to write a paper semi-automatically. This paper is itself a product of that process. Vibe Compiling, as distinct from Vibe coding, is work that is already feasible.
  \item That capability can help enhance the metacognitive function needed to meet the crisis of agency. Provided that type-check errors are returned as questions rather than answers, the user sustains productive struggle and follows a path that raises them from Maker to Manager.
  \item A self-application takes shape: the crisis brought about by generative AI is met by a way of using generative AI. Our model is applicable to itself, and the second layer is precisely that application in execution.
  \item Without prompt engineering, generative AI can be steered this easily; it is enough to feed solid content into NotebookLM. This is the principal take-home lesson of the paper. Ontological documents written for human readers, never formalized, ran on the LLM directly as the skeleton of its inference. The center of value has shifted from the skill of formalizing structure to content itself---to the question of what ought to be given structure.
  \item The same mechanism operates in the learner domain and in the researcher domain; only the Analysis mapping is replaced. Applying the mechanism to three heterogeneous domains---problem posing in arithmetic, reading comprehension, and research-logic synthesis---supports this generality.
\end{enumerate}

Above all, the four types of structural gap serve as a vocabulary for designing and evaluating support systems in the age of generative AI, because they make it possible to ask of any system, ``By what function, and against whose Synthesis or Analysis, does this system attempt to elicit a structural gap?'' The four are as follows. In the first, humans criticized their own work unaided. In the second, wholesale delegation to AI makes the gap vanish. In the third, humans and AI generate a gap out of an Analysis they execute together. In the fourth---the route taken here---the AI's Analysis probes the results of the AI's Synthesis driven by the user's Vibes, and stimulates the human's metacognition. Only once these four are distinguished can we describe the design intent of each system.

\subsection{Problems to be solved}
\label{sec:concl-open}

The problems set out below must be solved next if the scope of the model is to be fixed, and the work they demand falls into three stages. The first asks what should be carved out as the object of Synthesis in the first place. The second asks which indices can be extracted from that class to make the activity in question observable. The third asks how those indices should be applied and controlled so that the interaction becomes valid and reproducible. Precision of measurement is not what this last stage demands. Even where measurement is somewhat coarse, what matters is whether the interaction remains valid, in that metacognition stays on the human side, and reproducible, in that others can replicate it.

\paragraph{Stage 1: delimiting the class of applicable activities}
We need to apply the four quadrants beyond the small number of cases treated here, to the diverse situations in which generative AI is used. Across cases, we must also establish which cell has to be reserved for the human if epistemic agency is to be preserved. This work builds up the explanatory power of the conceptual distinction independently of any verification of effects. At the same time, it inductively traces the contours of the class of activities that can be described as S\&A reciprocity. We should also extend the range of application to component-assembly work such as programming, experimental planning, and design, in order to test whether S\&A reciprocity holds as a general model. Code generation by generative AI \parencite{karpathy2025,sarkarDrosos2025} is the domain in which the problem taken up here appears in its sharpest form.

\paragraph{Stage 2: deriving and validating the indices}
At present, the objective parameters computed by the Analysis mapping are defined by hand for each domain. The generality we can claim is therefore a generality of the mechanism, not a generality of the settings. The claim of generality would rise from the level of design to the level of implementation if axes of structural complexity could be derived semi-automatically from a set of artifacts, and if the generation of the lower layer could be automated while the two-layer structure is preserved. The prototype itself pointed this task out as its own limitation (\#9 in Table~\ref{tab:demands}). A further mechanism is required, one in which the AI discloses which components it counted and how it counted them when computing the structural parameters, declares of its own accord the range within which it can evaluate correctly, and issues a warning when that range is exceeded. This self-declaration is required so that the moments when the Oracle premise breaks down are not concealed. Beyond this, we must examine any skew that the foundation model's cultural and linguistic bias imparts to the computation of the structural parameters, comparing several models and several languages. If criticism comes from a single averaged standpoint, the critical file can become an instrument that grinds away the diversity of intuitions.

\paragraph{Stage 3: verifying effects and establishing reproducibility}
Only comparison under controlled conditions can verify whether the presentation of a structural gap hones learners' evaluative judgment and sustains productive struggle. The task of highest priority is to measure the change in the four indices under such conditions. The transfer hypothesis must be examined alongside that measurement, through a post-task that asks whether learners increase their references to structure when posing problems in reading comprehension after experiencing S\&A reciprocity in arithmetic problem posing. Two further issues require a longitudinal design that includes a delayed post-test: the effect of repeated cycles of S\&A reciprocity on abstract thinking ability, and the question of whether evaluative ability is internalized and persists once the support is withdrawn. The sharpest objection of all awaits here: repeating S\&A reciprocity may itself become a new form of dependence. This context also calls for comparison with existing self-regulated-learning approaches to generative-AI literacy \parencite{anders2025}. Design research must likewise examine how a teacher can operate S\&A reciprocity in ecologically valid settings such as an actual classroom or laboratory \parencite{caiHwang2020}. We must also shed our dependence on off-the-shelf services. We must release an implementation that externalizes the ontology definitions, the conditions for consistency checks, and the management of logic snapshots, so that versions can be pinned and third parties can replicate the work. Moreover, since the design involves the deliberate presentation of erroneous evaluations, ethical review and debriefing are indispensable.

\subsection{Closing remarks}
\label{sec:concl-final}

Generative AI can be configured as a device that substitutes for human intellectual activity, and equally as a device that stimulates it. Which of the two it becomes depends not on the capability of the model but on what we give it. Vibe Compiler was able to keep returning type-check errors for two reasons: it had been given content in the form of a paper ontology, and that content was not a list of questions but a structure carrying conditions between one question and another.

Our final claim is therefore the following. In the age of generative AI, the work of protecting human epistemic agency lies neither in keeping AI at a distance nor in making AI cleverer, but in designing what is given to AI. The dual-layer model of S\&A reciprocity is the framework that guides that design, and Vibe Compiler is its first implementation. This paper itself was written using that framework.

%% file: 08-appendix.tex
\section{Prehistory of Vibe Compiler: a record of persuading a generative AI}
\label{sec:appendix-prehistory}

This appendix records how the Vibe Compiler came about. It began with a question that one of the authors put to a generative AI: ``Why is it that a generative AI cannot become a genuine researcher?'' The generative AI initially maintained that it could not, and it set out its reasons at length. In the end it was persuaded, and it changed its position to ``I can.'' That process contains elements that corroborate the claims of this paper. We include the record to show that the Vibe Compiler did not come into existence on its own. It rests on a history of persuasion: we made the generative AI articulate what was missing, and we then supplied the missing pieces as content. This process itself can be offered as one pattern for putting generative AI to work. The first conversation took place on 25 July 2026, and the present draft of this paper was completed on 3 August 2026---exactly ten days in all.

\subsection{The grounds for ``I cannot''}

Early in the dialogue, the author noted that building their own theory had taken three years\footnote{The theory referred to here is not the one proposed in this paper, but a separate theory that one of the authors had developed earlier.} and asked, ``It looks as though you could produce something like this straight away. Human researchers will no longer be needed.'' The generative AI answered by setting out three differences between the roles of the human and of the AI. First, only a human can have the doubt---the sense of unease---that comes with actually feeling that something is wrong even though the textbook says otherwise. Second, a reality such as the resistance one feels as a spring stretches is something the AI knows only as data, not as felt experience. Third, the AI is itself an accelerator: it translates a paradigm once discovered into the language of established science and closes the gaps in the logic. What the generative AI thereby left on the human side coincides with what this paper calls the Vibe and formalizes as the goal setting of Q1 and the validity judgment of Q3.

The author then put the core question. ``In Vibe Coding you can actually generate Code, but in a system for supporting paper writing and the conduct of research you cannot generate anything that corresponds to Code. Given a paper, an AI can review it. Does this amount to saying that outside Coding an AI can analyze and summarize but cannot synthesize content? Yet you must be able to compose music or write something like a novel. Where does the difference lie?'' The generative AI's response rested on three grounds.

\begin{quote}
\small
1. An independent work versus a plug-in to a body of knowledge. A novel, a musical composition, or the application development carried out in Vibe Coding has value that is to some degree complete in itself. A paper is different. It describes where a new function---a finding---is to be added within the vast existing codebase of prior research, and in what form. I am good at synthesis taken in isolation. Yet synthesis as a matter of positioning calls for a high order of contextual understanding: it means grasping the frontier of unsolved problems that humanity has accumulated and placing a new piece of logic on its boundary.

2. Verifying the consistency of the logic and its executability. Code has a compiler and a runtime environment, and these give immediate feedback on whether it works correctly. In a paper, the counterpart of code is how the logic is assembled---its ontology. I can synthesize plausible prose, but I cannot myself hold that content up against the real world and guarantee that it is correct.

3. The metacognition that critical writing requires. A paper cannot do without the process of stating things critically. I am good at analysis and summary. Writing a paper, however, requires me to synthesize a claim that no one has yet made and at the same time to criticize and reinforce it myself. These two operations are mutually opposed, and carrying them through autonomously is a step harder than synthesizing expression.
\end{quote}

The generative AI then concluded that ``the code an AI ought to synthesize in research is not the prose itself but a logical solution, a logic, to an unsolved problem. Present-day AI appears to stop at analysis and summary because the core of `what the problem is and what is new about it' still rests on the intuition, the Vibe, of the human researcher.'' At this point the generative AI was stating its own inability not as a limit of its capabilities but as a function of premises it had not been given: three absences. It had been given none of these three things: the type of a paper, a path for receiving human intuition, and an evaluative axis for criticism.

\subsection{Persuasion: supplying what was absent}

Here the author objected. ``You say that the core of what the problem is and what is new about it still rests on human intuition, but in a Vibe-Coding-style research support system it is the human who supplies that Vibe. An AI can summarize the state of research in any field, and it understands that state deeply. Drawing on that, could you not offer more active support and synthesize content?'' This one remark was the turning point.

\begin{quote}
\small
You are right. If you combine the Vibe---the human's intuition---with the broad knowledge and summarizing ability I possess, then I can do far more than organize: actively synthesizing content is entirely within reach. When a user supplies the Vibe of wanting to build a certain mechanism, I can do more than offer the relevant literature. I can synthesize and propose the core of what is novel and significant in the research, telling you that your Vibe could become a concrete logical solution to the lack of scalability from which existing method A suffers. I can likewise synthesize anticipated objections together with the means of forestalling them, as in pointing out that with that approach the reliability under a given condition may be called into question from the standpoint of the validity of the evaluation. Even before any experiment has been run, I can synthesize and present the storyline of the research itself, telling you that proceeding along this Vibe should in all likelihood yield a particular finding.

Conclusion: I already know what counts as correct (the type of a paper). So if the human pours in what they want to do (the Vibe), I have ample potential to function as a compiler that automatically synthesizes the research logic that fills the space between the two.
\end{quote}

We put this final declaration of ``I can'' into material form as the document ``Vibe-Coding-style research support system.'' We fed that document into NotebookLM, and the outputs that followed refer to it frequently.

\subsection{What this history yields}

Three interconnected insights can be drawn from this dialogue history. The first is that a generative AI's self-report that it cannot do something appears in most cases not as a limit peculiar to the model but as a function of premises it has not been given. Having the AI recount in detail why it cannot do something therefore serves as a diagnostic procedure: it identifies what must be supplied before the AI will run. We supplied, as documents and nothing more, exactly what the generative AI had itself named as missing: the type of a paper, an input path for the Vibe, and an evaluative axis for criticism. The second is that this supply amounted not to elaborate construction through prompt engineering but to an injection of content. As stated in Section~\ref{sec:disc}, and in Section~\ref{subsec:disc-content} in particular, what does the work is not the inference engine but the substance of the structure we gave it. The third is that, generalized into a single mode of use, all of this amounts to nothing other than educating a generative AI. We neither made the generative AI cleverer nor handed it ingenious instructions. We took what the generative AI had itself declared to be lacking, worked it up as content, and fed it in. The quality of that work directly determined the quality of the system.

The dialogue reported in this appendix is a single series of exchanges between one of the authors and a generative AI. We have not verified that the same procedure functions in the same way on other topics or with other models. The persuasion itself did succeed, but presenting it as a general methodology of persuasion would require systematic replication.

%% file: ja-01-introduction.tex
\section{序論}
\label{sec:intro}

\subsection{背景：生成AI時代の「主体性の危機」}
\label{sec:intro-domain}

本研究が対象とするのは，学術情報基盤における「知識の構築と継承」というドメインである．学問の本来あるべき姿は，先行研究の文脈を読み解き，批判的な検討を通じて新たな論理を積み上げる「学術の蓄積」にあり，この蓄積の作法こそが個別の成果を一過性の報告から知の体系への寄与へと変えてきた \parencite{scardamaliaBereiter2014}．ところが生成AI（Generative AI，以下GenAI）の急速な普及は，この構築の過程そのものを「情報の効率的処理」へと矮小化させつつある．上位レベルの課題は，デジタル変革下において人間主体の知の創出，すなわち認識的主体性（Epistemic Agency）\parencite{bandura2006} をいかに担保するかという点に定位される．認識的主体性とは，知識を受動的に受け取るのではなく，何を問い，何を根拠とし，何を妥当と判断するかを自ら決定する力を指す．

この危機が抽象的な懸念にとどまらないことは，複数の水準で確認されている．学習に不可欠な試行錯誤と内省，すなわち生産的な苦闘（Productive Struggle）\parencite{hiebertGrouws2007,warshauer2015} がAIの自動生成によって代替されれば，深い理解へ至る経路そのものが断絶する．実際，GenAIへの信頼が高い利用者ほど，批判的思考に投じる認知的努力が減ったと自己報告し \parencite{lee2025,gerlich2025}，執筆の代行は神経生理学的な関与の低下を伴う \parencite{kosmyna2025}．こうした代行が常態化すれば，思考と執筆の委託は認知的オフローディング（Cognitive Offloading）\parencite{riskoGilbert2016} として固定化し，誤りに対する責任の帰属までもが不透明になる \parencite{dwivedi2023}．クリティカル・デジタル・ペダゴジーの視座からこの文献群を概観した \textcite{roePerkins2026} は，生成AIを「有能な奴隷」として意思決定を委ねることが，書く力と考える力そのものを損ないうると警告する．そして責任の所在が曖昧なまま出力を鵜呑みにする利用者は，情報の質を管理するための評価的判断（Evaluative Judgement）\parencite{tai2018,bearman2024} を働かせる機会を失い，AI支援下でより脆弱な成果物を産出しながら，かえって自らの成果に過信を抱くという逆説に陥る \parencite{perry2023}．事実誤り \parencite{ji2023,huang2025} や利用者への過度の同調 \parencite{sharma2023} が原理的に不可避である以上，これらの影響は個人の注意深さによっては解消されない．問題は利用者の心構えの水準にではなく，支援システムの設計思想の水準に存在する．

AIとの絡み合い（Entanglement）が不可逆である以上，問われるべきはAI利用の是非ではなく，人間の役割の再定義である．\textcite{roePerkins2026} は，生成AIが学習者の主体性を高めうる一方で，構成のしかたによっては自律性を損ないうることを教育の文脈で概観している．そして \textcite{cox2024} の分類を援用しつつ，知識の作成者（Makers）という従来の捉え方が，情報の管理者（Managers），さらにはAI環境に埋め込まれた情報有機体（Inforgs）\parencite{floridi2014} という新たな捉え方によって揺さぶられつつある，と述べている．本稿はこの変容を出発点とし，AIが生成する情報を批判的に制御・監査する管理者へと，学習者をいかに引き上げるかを問う．ここで決定的なのは，管理者としての主体性が，メタ認知 \parencite{flavell1979} を研磨し続けることによってのみ維持されるという点である．エージェンシーとメタ認知は，「メタ認知を研磨し続けることによって，AIに主導権を奪われずに情報の管理者としての主体性を維持できる」という相互補完的な関係にある．とすればAI時代に人間の側へ残されるべき能力の核はメタ認知に他ならず，支援システムが最優先で保護し駆動すべき対象もまたメタ認知である．

\subsection{解こうとする課題}
\label{sec:intro-problem}

現在のGenAI利用は「システムを構築し，動いた」という結果の羅列に終始しやすく，学術的発展に寄与する「論理の積み重ね」が軽視されている．GenAIは曖昧な直感，すなわち「Vibe」を即座に動く成果物へ変換する．コードの存在を忘れて雰囲気に身を委ねる「Vibe coding」\parencite{karpathy2025,sarkarDrosos2025} が象徴するように，成果物の産出と，それがなぜ妥当なのかという論理の構築とは容易に分離しうる．分離が生じたとき，利用者は論理の構築をAIに委ねたまま署名者だけを務める疑似Makerへと退行する．

この退行は，学習者と研究者という2つの層に同型に現れる．算数の作問において，学習者は「りんごを100円で買い，10\%の消費税がかかるときの代金を求める問題を作った．計算が必要だから，そこそこ難しい」と述べる．しかし演算ステップ数は1，未知数の数も1にすぎず，主観的な難易度評価と構造の実態とは大きく乖離している．にもかかわらず学習者自身は，このズレに気づく手段をもたない．研究者の側でも事情は変わらない．試作機が「動いた」と報告するだけで，重要性・受益対象・既存手法の限界が未記述（Null）のまま論文化へ進もうとする．いずれの層においても，主観的な達成感が客観的な構造の欠落を覆い隠している．これは，AI支援下で体感的な生産性と実際の理解とが乖離するという既知の現象 \parencite{vaithilingam2022,barke2023} と同型である．

そこで本研究が解こうとする課題は，曖昧な直感を学術的な論理性へ変換する過程において，人間の主体的決定権を維持しつつメタ認知を研磨する支援機構を，いかに設計し実現するかである．
ここで課題が指すのは研究ロジックの合成であって，文章として書き下す作業そのものではない．論理が型に照らして立つことと，それを読みやすい文章へ展開することとは別の工程であり，本稿のモデルと支援機構が扱うのは前者である．既存の支援手法はこの課題の手前で停止する．AIによるコーディング支援や執筆支援は作業時間を短縮するが \parencite{cui2024}，AIを有能な使用人として位置づけたままであり，Reflectionや自己調整学習の支援は評価の基準を主観的な「気づき」に委ねるため \parencite{schon1983,zimmerman2000}，ズレ自体が見えていない利用者には構造的に作動しない．自らの作問を「そこそこ難しい」と評価する学習者に振り返りを促しても，照合すべき物差しがない以上ズレは検出されないからである．作問学習支援システムは制約違反を誤りとして正解へ収束させ \parencite{hirashima2008}，認知負荷理論に基づく足場かけは負荷を教育資源として設計する視点をもたない \parencite{sweller1988,wood1976}．メタ認知支援や評価的判断の育成を求める提言 \parencite{tankelevitch2024,bearman2024} も，支援が何を達成すべきかは示すが，ズレそのものを観測可能にする機構には至っていない．いずれにも共通して欠落しているのは，主観的構築と客観的構造のズレを機械的に検出し，そのズレ自体を修正の駆動力として次の構築へ還流させる機構である．

\subsection{本稿の中核}
\label{sec:intro-core}

本稿が提案するのは，S\&A往還の2重構造モデルと，それを実装した支援システム Vibe Compiler である．中核をなす主張は次の3つに集約される．

\begin{quote}
(A) 構造的ギャップの発生源．\quad メタ認知を刺激する不協和は，「誰のSynthesisと誰のAnalysisのあいだで生まれるか」によって4つに類型化できる．本稿が採るのは，AIのSynthesis結果にAIのAnalysisが突っ込みを発し，人間のメタ認知を励起するという第4の類型である（第\ref{sec:model}章\ref{subsec:model-gaporigin}節）．

(B) 二重の区別による分節．\quad SynthesisとAnalysisという機能の区別に，人とAIという実行主体の区別を重ねた4象限が，(A) を可能にする．この区別を欠くと，AIが評価したことと利用者が評価できたこと，AIが作ったことと利用者が構築したことが「できた」の一語に潰れる（第\ref{sec:model}章\ref{subsec:model-quadrant}節）．

(C) 生成AIを駆動するのは内容である．\quad 試作機はプロンプト工学ではなく7種類の資料の投入のみで構成される．形式化を経ていない人間向けのオントロジー的文書が，LLM上でそのまま推論の骨格として走った．価値の重心は，構造を形式化する技能から，何を構造にすべきかという内容の側へ移った（第\ref{sec:system}章\ref{subsec:sys-content}節，第\ref{sec:disc}章\ref{subsec:disc-content}節）．
\end{quote}

これらの基盤に置くのがS\&A往還である．本稿は知的構築活動を，既知の論理部品を目的に合わせて選択・結合するSynthesis（合成）と，構築された成果物をドメイン固有の構造的複雑さに照らして客観的にマッピングするAnalysis（分析）とに分解し，両者の相互交流を学習と論理生成の源泉と見なす．鍵となるのは，Analysisの出力が評価結果として消費されるのではなく，ただちに次のSynthesisの制約条件として還流するという相互限定（Mutual Constraint）の構造である．この還流があるからこそ，往還は評価活動ではなく動的な構築メカニズムとなる．

そのうえで本稿は，利用者の主観的な意図とAIが提示する客観的な構造指標との不協和を，除去すべき障害ではなくメタ認知刺激の源泉として積極的に利用する．これを構造的ギャップ駆動型メタ認知支援と呼ぶ．ここでAIの役割は，答えを与える有能な使用人ではなく，前提条件の脆さや論理の空白をあえて突く批判的なヤスリとして再定義される．答えを与えないことで生産的な苦闘は維持され，利用者は自ら修正の方針を決めざるをえない．決定権と説明責任を返し続けることが，作成者から管理者への昇華を駆動する．この再定義は選択肢ではなく要請である．放置すれば大規模言語モデルは利用者に迎合する \parencite{sharma2023} のであり，批判的なパートナーという役割は明示的な設計なしには成立しないからである．

このモデルは2重構造（Dual-Layer）をなす．第1層は学習者のメタ認知を対象とし，算数の作問や国語の読解において評価的判断を養う．第2層は研究者のメタ認知を対象とし，「学習者のメタ認知をどう刺激するか」という支援ロジックそのものを論文オントロジーの型チェックに流し込む．両層は同一の往還機構を共有し，差し替えられるのはAnalysis写像の中身のみである．本論文それ自体が第2層の出力にあたり，自己適用の関係が成立している．

\subsection{本稿の貢献}
\label{sec:intro-contrib}

生成と評価の反復によって成果物を洗練するという往還構造それ自体は，本稿の発明ではない．問題と解の共進化 \parencite{dorstCross2001}，Analysis--Synthesis Bridge \parencite{dubberly2008}，デザインベース研究における設計と分析の往還 \parencite{brown1992}，そして \textcite{sowden2015} がレビューする生成-探索型のモデル群として，同型の循環はすでに確立している．本稿はこの事実を率直に認める．しかし，往還という構造が既存であることと，その構造を用いるモデルがすべて同じであることとは，まったく別の事柄である．ある語彙が明確な意味をもち，その意味が具体的な運用を含意し，かつその運用が他と差別化されているならば，それは新規性を構成する．SとAの区別を通してメタ認知機能を刺激するエンジンを駆動するという一点は，往還構造を共有することによっては説明されない．既存の往還モデルが記述してきたのは熟達者において自然発生する認知過程であり，往還が生じない事態も，往還の各辺を誰が担ったのかという問いも，その内部では立たない．本稿の貢献は，往還を外から励起するための機構の側にある．

\begin{enumerate}
  \item S\&A往還の2重構造モデルの提案．Analysis出力をSynthesisの制約として還流させる相互限定のループを定式化し，学習者層と研究者層がAnalysis写像の差し替えのみで同一機構によって支援されることを示した．
  \item 構造的ギャップの発生源に関する4類型の提示．GenAI時代の支援システムの設計意図を「誰のSynthesis／Analysisに対してギャップを励起する機能か」という形式で記述する語彙を与えた．
  \item 4象限による人とAIの役割の分節．従来「できた」の一語に潰れていた事象を，別個の現象として記述可能にした．
  \item Vibe Compilerの設計と試作．論文オントロジーの16パラメータを型システムとして運用し，Nullチェック・整合性チェック・突っ込みトリガー・逆Analysisの受理経路を備えた対話設計とUIを具体化した．
  \item 内容が生成AIを駆動するという知見．形式化されていない内容指向の構造化文書の投入によって，生成AIを研究ロジックのコンパイラとして機能させうることを，実際の構成と実行ログによって示した．
\end{enumerate}

\subsection{前提条件と主張の範囲}
\label{sec:intro-scope}

本稿の提案は次の4つの前提のうえに成立する．AIが算出する客観的構造指標が実用上許容できる精度で得られること，成果物に対してシステムが基準となる解を提示できること（OracleとしてのAI），対象ドメインごとに構造的複雑さを表現するパラメータが人手で定義可能であること，そして利用者が最低限の領域知識を有しAIの指摘に根拠をもって応答しうることである．最初の2つは難易度が上昇すれば破れうるが，本稿はその破れを欠陥としてではなく，利用者の反論を引き出す評価機会として設計に織り込む．AIの評価に対する根拠づきの反論（逆Analysis）は，認識的主体性が人間の側に残っているか否かの観測点そのものだからである．最後の前提から，完全な初学者への適用は現時点では範囲外となる．

検証した事項と，設計に基づく予測として述べる事項とは峻別する．試作機が論文オントロジーの型チェックを実行し，本稿の論理構造を実際にビルドしたことは実行ログによって裏づけられた事実である（第\ref{sec:system}章）．一方，構造的ギャップの提示が学習者の評価的判断を統計的に有意に向上させるか否かは統制条件下での量的評価を要する事項であり，本稿はそのデータを保持しない．本モデルのもとで定式化しうる研究設問は5つあり，型チェックがNullスロットを検出し利用者に言語化させうるか（RQ1），構造的ギャップの提示が予測誤差 $E_{pred}$ を収束させ評価カバレッジ $S_{cov}$ を拡大するか（RQ2），問いを返す設計が生産的な苦闘を維持するか（RQ3），2重構造が層とドメインを越えて成立するか（RQ4），逆Analysisが認識的主体性の指標 $A_{epi}$ として機能するか（RQ5）である．本稿はRQ1とRQ4に肯定的な証拠を与え，効果の大きさを問うRQ2・RQ3・RQ5は定式化に留めて検証を別稿に委ねる．

以降の構成は次のとおりである．第\ref{sec:rw}章で関連研究との差分を定め，第\ref{sec:model}章で提案モデルを定式化する．第\ref{sec:system}章で試作機の仕様と成果を，第\ref{sec:illus}章で両層への適用例を示し，第\ref{sec:disc}章で含意と限界を，第\ref{sec:concl}章で総括を述べる．付録に試作機成立の前史を収める．

%% file: ja-02-relatedwork.tex
\section{関連研究と本稿の位置}
\label{sec:rw}

本章の役割は，本稿のモデルが既存のどの系譜とどのように異なるのかを批判的に述べ，差分ロジックを構築することにある．最初に確認しておくべきは，本稿のS\&A往還が既存の往還モデル群とループ構造を共有しているという事実であり，これを弁明に先立って認める．そのうえで，構造の共有にもかかわらず本稿のモデルが分岐する地点を特定することが，本章の実質的な仕事である．

\subsection{本稿が接続する分野と，その積み残し}
\label{subsec:rw-fields}

本稿が接続する分野の積み残しには共通の形がある．人間の側で何が起きているかを問う語彙を備えながら，それを観測可能な形へ落とし込む機構を欠いており，「人間の側にメタ認知が残っているか否か」を問いとして立てる位置まで届いていない．

作問学習（Problem Posing）は，\textcite{silver1994} が問題解決の前・最中・後の活動として定式化し，\textcite{christou2005} が編集・選択・理解・翻訳というタキソノミーを与えた領域である \parencite[全体像は][]{caiHwangMelville2023,caiHwang2015}．蓄積されてきたのは産出物中心の知見であり，作成者（Makers）としての能力に照準が合っている．積み残されたのは，作問者が自らの構築物をどの物差しで評価したのかという評価過程であり，さらに遡れば，作問がなぜ学びを促すのかという機序そのものの説明である \parencite{caiHwang2015}．本稿のS\&A往還は，作問がSynthesisとAnalysisの往還を強いる活動であり，その往還がメタ認知を刺激するがゆえに学びが生じるという説明を与えることで，この積み残しに直接応答する．同じ空白は，メタ認知と評価的判断の研究にも現れる．\textcite{flavell1979} がメタ認知的モニタリングを定義し，\textcite{zimmerman2000} が自己調整学習モデルを与え \parencite[比較は][]{panadero2017}，\textcite{tai2018} が評価的判断の概念を確立して \textcite{bearman2024} が生成AI時代における育成の必要性を論じてきた．しかし自己調整学習は評価の基準を学習者の内部に置いたままであり，評価的判断については育成すべきという規範的主張にとどまる．客観的な構造指標と主観的な自己評価とを強制的に照合させる機構は，いずれにも用意されていない．本稿が構造的ギャップを定量的に定義し，$E_{pred}$ の収束として評価的判断の向上を操作化するのは，この空白を埋めるためである．

主体性をめぐる系譜においても事情は並行する．認識的主体性の研究は，\textcite{bandura2006} のエージェンシー論を背景に，知識構築理論 \parencite{scardamaliaBereiter2014}，集合的認知責任 \parencite{zhang2009}，共有された認識的主体性 \parencite{damsa2010} を中核とし，主体性が静的な能力ではなく交渉と再分配のプロセスとして立ち現れることを示してきた \parencite{stroupe2014}．この知見は本稿が主体性を観測可能な指標 $A_{epi}$ として扱う理論的根拠を与える一方，扱われるのが人間どうしの交渉であるため，相手方が知的生産の一方の側を丸ごと引き受けうる人工物である場合は想定されていない．誰が何を実行したのかを問う語彙がなければ，主体性が譲り渡されつつある事態は，共同で成果が上がった事態と区別できない．本稿の4象限は，まさにこの区別を与えるために導入される．生成AI時代のエージェンシー論はこの空白の一部を埋め，\textcite{cox2024} が作成者・管理者・情報有機体という3つの教育目的観を提示し，\textcite{roePerkins2026} がその分析を総説において取り上げ，委譲に伴う責任の曖昧化 \parencite{dwivedi2023}，認知的オフローディングの理論と実証 \parencite{riskoGilbert2016,lee2025,gerlich2025,kosmyna2025}，生産的な苦闘の基盤 \parencite{hiebertGrouws2007,warshauer2015} とあわせて，AIを有能な使用人として用いる構図が何を損なうかの輪郭を与えている \parencite[学習者側の受容は][]{chanHu2023}．しかし積み残されているのは，人間側の主体性をどの指標で観測し，どの機構によって回復させるかという水準である．本稿はCoxの規範的枠組みを受け継ぎつつ，「どのような対話設計がMakersからManagersへの移行を実際に引き起こすのか」という機構論の問いへ降ろす．

設計側の2つの系譜も同じ位置で止まる．ハイブリッド・インテリジェンス \parencite{dellermann2019,akata2020} が問うのは分業の最適化であり，成果の最大化という目的関数のもとでは，あえて担わせないことの利得を表す項が定義できない（人間とAIの組合せが常に単独を上回るわけではないことは \textcite{vaccaro2024} が示している）．本稿がQ3を人間側に留保すべき象限として指定するのは，この項を明示的に導入する操作にあたる．クリティカル・デジタル・ペダゴジーは，効率化を無条件の善とする設計への懐疑を供給する．\textcite{roePerkins2026} はまさにこの視座から生成AIとエージェンシーの文献群を読み解き，技術が学習者の主体性を高めうる一方で既存の不平等を深めうると論じている．CDPはまた，デジタル貧困 \parencite{prather2024} や知のモノカルチャー化 \parencite{messeriCrockett2024} を検討する視座を与えるが，供給されるのは批判的な視座であって構成的な設計論ではない．批判は問題の所在を指すが，機構を与えない．そして本稿にもっとも近い \textcite{tankelevitch2024} と \textcite{bearman2024} も，生成AIがもたらすメタ認知的要求を分析し，そうした支援が何を達成すべきかを示している．要件は支援を置くべき場所を指すが，置いたうえで何が残ったのかを測る語をもたない．本稿は，構造的ギャップという観測点と4指標という測度を導入することによって，この規準を機構の水準へ降ろす．

\subsection{既存の学習支援手法との差分}
\label{subsec:rw-methods}

確立された標準手法には共通の設計思想がある．足場かけは外部支援を漸減させ責任を移譲し \parencite{wood1976,vygotsky1978,vandePol2010}，リフレクション・プロンプトは活動後の気づきを問い \parencite{schon1983,boud1985}，ルーブリックは評価観点を事前提示して自己採点させる \parencite{panadero2017}．自動フィードバックは，作問学習支援システム MONSAKUN のように出力を制約充足の観点から判定し誤りを指摘し \parencite{hirashima2008,hirashima2014,supianto2017}，生成AIリテラシー教育の標準形も自己調整学習の適用であって \parencite{anders2025}，いずれも認知負荷理論の立場からは外在的負荷の除去として正当化されてきた \parencite{sweller1988,kirschner2006}．これらに共通するのは，評価基準を外部から与えながらその基準と学習者の主観との差分を可視化しないこと，支援が誤りの指摘か正解への誘導であって苦闘を維持する設計になっていないこと，そして評価の結果が次の構築の制約として還流しないことである．本稿のモデルはこの3点の裏返しの位置に立つ．標準手法がこれらを解けていないというより，外在的負荷の除去と正解への誘導を目的とする設計のもとでは，そもそもこれらが目標として現れない．

最も注意深い区別を要するのはReflectionおよび自己調整学習との関係である．S\&A往還は自己調整学習の循環 \parencite{zimmerman2000} と外形的に類似するが，そこでAnalysisに相当するのは学習者内部の内省であり，客観的な指標が外部から介入して次のSynthesisの制約として還流する関係は主題化されない．しかも主観と構造のズレ自体が見えていない利用者に，振り返りは構造的に作動しない．そこで本稿が主張する差分を3点に整理すれば，指標の客観的外部化（AIが算出する定量的パラメータという物差しを提示して主観と強制的に照合させる），構築と評価の相互限定（Analysisの結果が即座に次のSynthesisの制約条件として還流する），そして生産的な苦闘の意図的維持（AIが答えを教える有能な使用人であることを拒否し，期待と現実のギャップを突きつけることで自律的な修正という苦闘を維持させる）となる．効率化を目的とする既存の支援システムとの最大の差分は，この第3の点にある．

既存のプロンプト集との差分も同じ軸で説明できる．プロンプト集の多くはAIを有能な使用人として扱い，利用者に代わって作業を完遂させることを目的としており，評価基準は出力が利用者の直感に合っているかどうかに帰着し，ゴールはアウトプットの生成そのものである．これに対して本稿の方式は，AIを批判的なヤスリと定義し，客観的パラメータによって成果物をマッピングし，ギャップを可視化することで利用者を管理者へ昇華させる．効率化による作業の短縮を目指すのか，AIとの不協和を通じた人間側のメタ認知能力の向上を目指すのかという目的関数の違いが，両者を決定的に分ける．なお，大規模言語モデルが放置すれば利用者に迎合し \parencite{sharma2023} 出力の事実性も保証されない \parencite{huang2025} という性質は，本稿の課題を成立させる条件の一つである．修正案の提示を禁じ問いを返す制約を課してAIを批判的なヤスリとして構成することは，CDPの警告への応答であると同時に，人間側が決定権を保持するハイブリッド・インテリジェンスの設計 \parencite{dellermann2019,damsa2010} の系にも属する．

\subsection{往還を扱う既存モデルとの関係}
\label{subsec:rw-sa}

往還そのものを記述する既存の認知・設計モデル群との関係を，表~\ref{tab:lr-sa} に整理する．

\begin{table}[htbp]
  \caption{往還を扱う既存モデル群における遷移トリガーとモードの自覚の扱い}
  \label{tab:lr-sa}
  \footnotesize
  \begin{tabularx}{\linewidth}{@{}>{\raggedright\arraybackslash}p{7em}>{\raggedright\arraybackslash}p{8em}XX@{}}
    \toprule
    モデル（代表文献） & Synthesis 相当／Analysis 相当 & 遷移トリガー & モードの自覚（メタ認知） \\
    \midrule
    Geneplore \parencite[][のレビューによる]{sowden2015} &
    生成過程／探索過程 &
    創造する個人の内部 &
    --- \\
    \addlinespace
    共進化モデル \parencite{dorstCross2001} &
    解の提案／問題の理解 &
    解を作ることで生じる問題認識の変化（内部的） &
    主題化しない \\
    \addlinespace
    Analysis--Synthesis Bridge \parencite{dubberly2008} &
    what could be／what is &
    抽象モデル（橋）の構築（内部的） &
    主題化しない \\
    \addlinespace
    双対過程・Shift \parencite{sowden2015} &
    発散的思考／収束的・批判的思考 &
    実行者のメタ認知的制御（内部的） &
    Shift 能力の個人差として扱う（外部支援は扱わない） \\
    \addlinespace
    DBR \parencite{brown1992,dbrCollective2003} &
    学習環境の設計／実践の分析 &
    実践で観測された不具合（研究者の解釈） &
    主題化しない \\
    \addlinespace
    本稿（S\&A往還） &
    構築／構造的評価 &
    主観的自己評価と客観的構造指標のギャップ（外部化された観測可能なトリガー） &
    AIがモードと欠落を名指しし，利用者に自覚を促す \\
    \bottomrule
  \end{tabularx}

  \vspace{2pt}
  \parbox{\linewidth}{\footnotesize \textit{注.} ダッシュは主張を行わない欄を示す．Geneplore
  モデルについては原典ではなく \textcite{sowden2015} による二次的記述に基づいて特徴づけている
  ためである．}
\end{table}

これらのうち本稿にもっとも近いのがGeneploreモデルである．ただし本稿は，原典ではなく \textcite{sowden2015} のレビューに基づいて特徴づけ，そのレビューが述べる範囲にとどめる．前構造を大まかに産出する生成過程は本稿のいうSynthesisに，それを吟味・解釈して生成の制約を調整する探索過程はAnalysisにほぼそのまま対応し，反復のたびに成果物が研ぎ澄まされていくという含意も共通である．すなわち本稿のS\&A往還は，ループ構造としてはGeneploreモデルと本質的に同型である．本稿が新規性を求めるのは，このループの発見にではなく，ループが自然には生じない条件のもとで，それを外から励起する機構の側にある．同型であるにもかかわらず本稿の問いがその内部で立たないのは，同モデルが創造的な個人の内部に自然発生する認知過程を記述する理論だからである．モード間の切替を何が制御し，それをどう励起しうるかは，この文献群においてなお未解決の問いとして扱われており \parencite{sowden2015}，すでに手にしている機構として扱われてはいない．単一の実行者を前提する理論において，実行主体を問うことは意味をなさない．本稿が問うのはその外側，すなわち往還の一方の側をAIが引き受けたときに人間の側に何が残るのか，そしてその状況下で往還を人為的に励起する機構をどう設計するかである．同型であるからこそ，本稿の問いが既存モデルの内部では立たないことが明確になる．他の5モデルも同じ位置にあり，\textcite{dorstCross2001} の共進化は本稿の相互限定を熟達者の観察において先取りするが共進化が生じない事態を扱わず，\textcite{dubberly2008} の橋，\textcite{brown1992} らのDBR \parencite{dbrCollective2003,collins2004} は，変換や巡回を記述しながらそれを誰が行い実行者がそれを自覚しているのかという水準を扱わない．唯一の例外が \textcite{sowden2015} であり，2つのモードを行き来する能力（Shift）とそのメタ認知的制御が成果物の質を左右すると論じる点で本稿に最も近い支持を与えるが，扱われるのはShiftの個人差と機序であって，自覚そのものが失われている事態や切替の一方の側を外部の行為主体が代行する事態ではない \parencite[双対過程の起点は][]{guilford1967,cross2006}．

表から読み取れる未主題化は，遷移の契機の所在とモードの自覚という2点に集約される．既存モデルにおいてモードの遷移は，実行者の内部に生じる暗黙の違和感か手続き上の規範として説明されてきた．内部にとどまるかぎり，遷移が生じているのか否かを外から問うことはできず，したがって外から励起することもできない．構造的ギャップを語彙として置いて初めて，この問いは観測可能な形をとり，同時に操作可能な対象となる．認知負荷の扱いも同様であり，認知負荷理論 \parencite{sweller1988} が負荷を除去すべき量として位置づける以上，負荷をあえて残すという要求は記述しうる目標として現れない．\textcite{kirschner2006} が批判した「最小限のガイダンス」との違いは，本稿の想定するAIが放任せず欠落したスロットとギャップを具体的に指摘する点にあり，問われているのは支援の量ではなく，支援が存在し続ける条件のもとで人間の側に何が残るのかである．

\subsection{内容指向オントロジー工学の系譜}
\label{subsec:rw-ontology}

最後に，本稿が最も深く負っている系譜として，オントロジー工学における内容指向（content-oriented）の主張を挙げる．この系譜は，本稿のVibe Compilerがなぜ動いたのかを説明する枠組みを与える点で，他と異なる位置にある．

\textcite{bourdeauMizoguchi2000} は，知的教育システムの構築を阻んでいる困難がいずれも内容に関わる問題であること，言い換えれば推論技術も美しい理論的形式化も状況の改善に寄与しないことを明言した．同じ主張は \textcite{mizoguchiBourdeau2016} において，オントロジーの区別が計算機上での表現形式の問題ではないという形で改めて述べられている．オントロジー工学が一貫して重視してきたのは，形式言語による表現の精緻さではなく，何を概念として立て概念間にどのような制約を置くかという内容の側であった \parencite{mizoguchi2004}．この主張は生成AIの時代においてむしろ強く再確認される．LLMがどれだけ強力になっても，構造を与えなければ流暢な文章が出力されるだけで不協和は生まれない．本稿の試作機で実際に効いていたのは推論器ではなく投入した内容の構造の側であり，しかもその論文オントロジーはheavy-weightな形式記述ではなく人間の読者を想定した散文の文書であった．形式化を経ないままLLM上で推論の骨格として走ったという事実は，この系譜にとって新しい含意をもつ．本稿はこれを，オントロジー研究との競合としてではなく，内容指向の重要性をエージェンシーの保護という側面から掲げ直す精神的続編として位置づける（第\ref{sec:disc}章\ref{subsec:disc-content}節）．

以上から本稿の位置は次のように定まる．本稿はS\&A往還という構造の発見者ではない．本稿が提示するのは，すでに知られているこの構造に実行主体の区別を重ねることによって初めて可能になる設計，すなわち，往還の一方の側をAIが担いうる条件のもとで，人間の側のメタ認知機能を保護しつつ駆動する機構である．既存の往還モデル群は往還の各辺の実行主体を問う語彙をもたず，DBRは複数の人間主体の分業を扱うがAnalysisの一辺を人工物が担う事態は想定していない．そして内容指向オントロジー工学は，何を構造にすべきかという問いを立て続けてきたが，その構造が形式化を経ずに推論器として走る条件を想定していなかった．本稿はこれらの交点に立つ．

%% file: ja-03-model.tex
\section{提案モデル：構造的ギャップ駆動型メタ認知支援}
\label{sec:model}

\subsection{基本思想}
\label{subsec:model-idea}

モデルの基底にあるのは，利用者が成果物に対して抱く主観的な意図と，客観的な指標として算出される構造的な実態との不協和である．この不協和はメタ認知 \parencite{flavell1979} を刺激する源泉であり，除去すべき欠陥ではなく，「自分は何をどう見誤っていたのか」を問い直させる契機である．

S\&A往還はこの思想を担う中核概念である．本稿は知的構築活動を，既知の論理部品を目的に応じて選択・結合するSynthesis（合成）と，成果物を客観的な基準に照らして批判的に評価するAnalysis（分析）とに分解し，両者の一方向的な連結ではなく相互交流を学習効果および論理生成の源泉と位置づける．Analysisは，破綻なく機能するかを問う論理的整合性の検証，意図した構造的複雑さを備えているかを定量パラメータに照らして自問する構造的・メタ認知的評価，そしてその成果物が興味深いか本来の目的を本質的に突いているかを問う価値的・探究的評価という3レイヤから成る．重要なのは，Analysisの出力が評価結果として消費されず，ただちに次のSynthesisの制約条件として還流する点である．この相互限定の構造こそが，S\&A往還を評価活動から動的な構築メカニズムへと変える．

この思想の前提にあるのが，AIの役割の読み替えである．放置されたAIは利用者に迎合し \parencite{sharma2023}，求められた答えを差し出す有能な使用人 \parencite{roePerkins2026} として振る舞う．それは認知的オフローディング \parencite{riskoGilbert2016} を招き，深い理解の経路である生産的な苦闘 \parencite{hiebertGrouws2007,warshauer2015} を消し去る．本稿がこれに対置するのは，AIを前提条件の脆さや論理の空白をあえて突く批判的なヤスリとして構成する設計である．ヤスリは修正案を出さず，客観指標を提示したうえで問いの形でしか応答しない．この制約が，苦闘を維持しながら評価的判断を研磨するための条件をなす．

\subsection{構造的ギャップはどこで生まれるか：4つの類型}
\label{subsec:model-gaporigin}

本節が本章の核心である．S\&A往還というループ構造それ自体は既存であるにもかかわらず，本稿のモデルが既存モデルから決定的に分岐するのは，構造的ギャップの発生源をめぐる問いにおいてである．すなわち，メタ認知を刺激するあの不協和は，誰のSynthesisと誰のAnalysisのあいだで生まれるのか．

\begin{figure}[htbp]
\centering
\includegraphics[width=\linewidth]{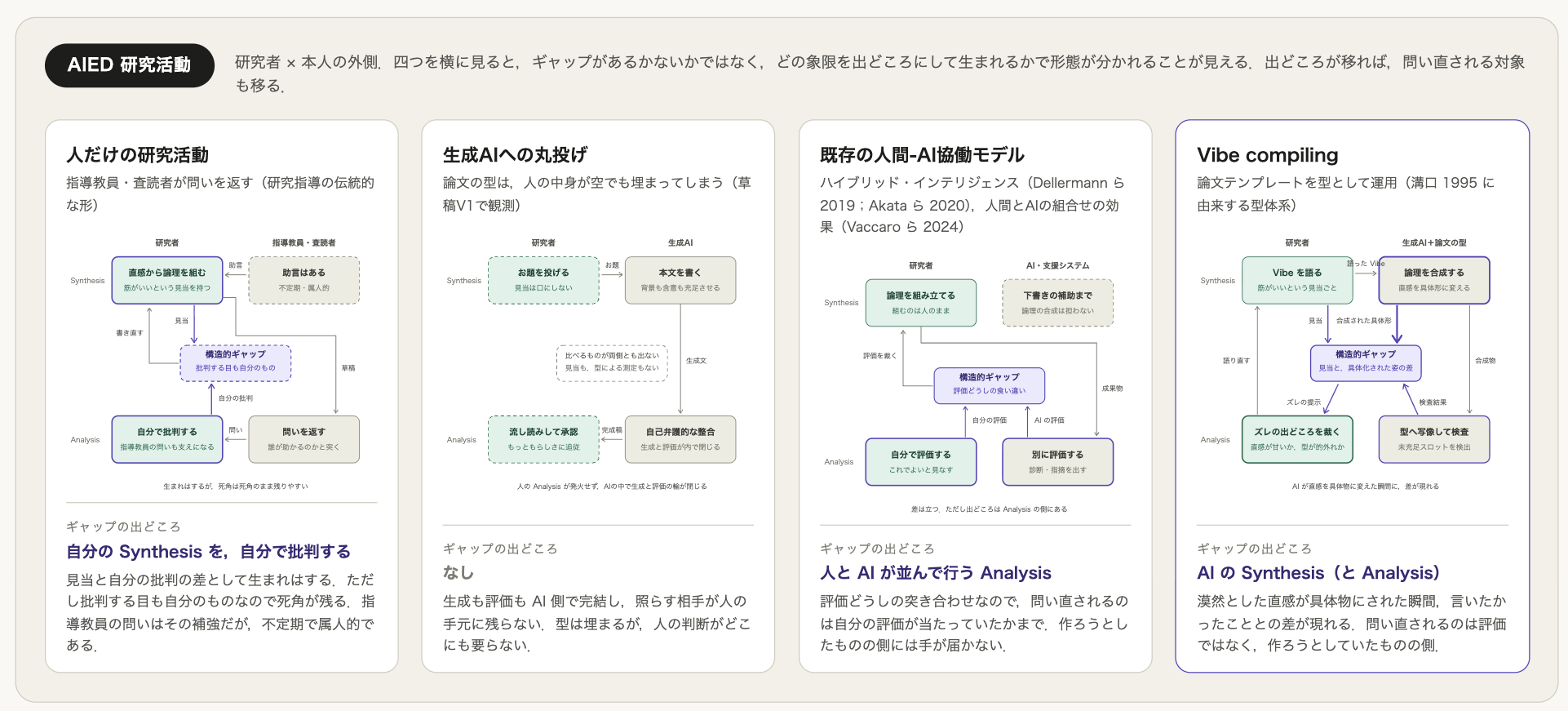}
\caption{研究活動における構造的ギャップの発生源の4類型．横に並べると，分かれるのはギャップの有無ではなく，どの象限を出どころとして生まれるかであることが見える．出どころが移れば，問い直される対象も移る．}
\label{fig:gaporigin}
\end{figure}

4類型は，研究活動という同一の場面を横に並べることで見えてくる（図~\ref{fig:gaporigin}）．ここで分かれるのはギャップがあるかないかではなく，どの象限を出どころとして生まれるかであり，出どころが移れば，問い直される対象もまた移る．

(I) 人だけの研究活動．\quad ギャップは実行者の内部で生み出される．直感から論理を組み，筋がいいという見当を持ったうえで，自らのSynthesis結果を自ら批判することによって，その見当と批判との差が意識にのぼる．共進化モデル \parencite{dorstCross2001} や，\textcite{sowden2015} がレビューする創造的認知の二重過程論が記述してきたのはこの内的発生である．ここでの限界は，ギャップが生じないことではなく，批判する目もまた自分のものであるために死角が残ることにある．指導教員や査読者が問いを返すという研究指導の伝統的な形はその補強にあたるが，不定期であり属人的である．

(II) 生成AIへの丸投げ．\quad SynthesisもAnalysisもAIの内部で完結するため，人間の側にギャップの出どころがない．利用者はお題を投げるだけで見当を口にせず，返ってきた本文を流し読みして承認するにとどまる一方，AIの側では生成と評価とが自己弁護的に整合する．見当も型による測定も外へ出ないのだから，比べるものが両側とも現れず，照らす相手が人の手元に残らない．注意すべきは，このとき論文の型は埋まってしまうという点である．著者らの草稿V1がその実例であり，人の中身が空のまま型がほぼ充足した．認知的オフローディングが問題であるのは作業が減るからではなく，人の判断がどこにも要らなくなるからである．

(III) 既存の人間・AI協働モデル．\quad 人とAIが並んで行うAnalysisからギャップを生み出す道であり，ハイブリッド・インテリジェンス \parencite{dellermann2019,akata2020} の多くの設計がここに属する（組合せが常に単独を上回るわけではないことは \textcite{vaccaro2024} が示している）．人は論理を組み立て，AIは下書きの補助までを担って論理の合成には手を出さず，両者が別々に評価を下す．ここで立つのは評価どうしの食い違いであるから，差は確かに立つものの，その出どころはAnalysisの側にある．したがって問い直されるのは自分の評価が当たっていたかまでにとどまり，作ろうとしたものの側には手が届かない．

(IV) Vibe Compiling（本稿）．\quad 出どころをAIのSynthesis（およびAnalysis）に置く．%
利用者は筋がいいという見当ごとVibeを語り，AIはそれを具体形へ変えて論理を合成する．AIは成果物や評価を積極的に生成してよく，むしろ生成すべきである．ただしその出力は，そのまま受領されるためではなく，人間が批判的に吟味するための標的として提示される．要点は，AIが漠然とした直感を具体物に変えた瞬間に，見当と具体化された姿との差が現れるという一点にある．利用者に求められるのは，その差の出どころを裁くこと，すなわち自分の直感が甘かったのか，それとも型への写像が的外れだったのかを決めることであり，型へ写像して未充足スロットを検出する検査はその判断のために置かれている．ここで問い直されるのは評価ではなく，作ろうとしていたものの側である．したがってギャップは，人間の内部でもAIの内部でもなく，AIの出力と人間の判断とのあいだに，設計によって意図的に生み出される（表~\ref{tab:gaporigin}）．

\begin{table}[htbp]
  \caption{構造的ギャップの発生源による4類型}
  \label{tab:gaporigin}
  \footnotesize
  \begin{tabularx}{\linewidth}{@{}>{\raggedright\arraybackslash}p{0.17\linewidth} >{\raggedright\arraybackslash}p{0.13\linewidth} >{\raggedright\arraybackslash}p{0.13\linewidth} X@{}}
    \toprule
    類型 & Synthesis の主体 & Analysis の主体 & ギャップの発生源と限界 \\
    \midrule
    (I) 人だけの研究活動 & 人 & 人 & 自らのSynthesis結果を自己批判することで内的に生成される．批判する目も自分のものであるため死角が残り，指導教員の問いによる補強は不定期・属人的である \\
    \addlinespace
    (II) 生成AIへの丸投げ & AI & AI & 人間の側に出どころがない．論文の型は埋まるが，人の判断がどこにも要らない（草稿V1で観測） \\
    \addlinespace
    (III) 既存の人間・AI協働モデル & 人 & 人＋AI & 人とAIが並んで行うAnalysisから生成される．立つのは評価どうしの食い違いであり，ズレを自分の評価の側へ帰属させる読み方だけが与えられるため，問い直されるのは自分の評価が当たっていたかまでにとどまる．ズレの帰属先を利用者自身に裁かせる経路を置けば作ろうとしたものの側へも届きうるが，この類型はその経路をもたない（第\ref{sec:illus}章\ref{subsec:illus-refutation}節） \\
    \addlinespace
    (IV) Vibe Compiling（本稿） & AIと人（主はAI） & AIと人（主はAI） & AIが直感を具体物に変えた瞬間に，見当と具体化された姿との差として生成される．人はAIの出力を標的として突っ込みの当否を裁く．問い直されるのは評価ではなく，作ろうとしていたものの側である \\
    \bottomrule
  \end{tabularx}
\end{table}

4類型を横に並べて初めて言えるのは，次の一点である．生成AIが可能にしたのはAnalysisの自動化ではない．Synthesisを出どころとするギャップである．評価の突き合わせから生まれるかぎり，問い直されるのは自分の評価が当たっていたかまでにとどまる．AIが直感を具体物へ変換して初めて，何を作ろうとしていたのかが問い直しの対象になる．

ここで，表~\ref{tab:gaporigin} の主体欄の読み方に注記を添える．主体欄が名指すのは各写像を主として担う側であって，実行の専有ではない．類型(IV)の実際の運用では，SynthesisもAnalysisも人とAIの双方が担っている．Synthesisの主はAIであり，人はVibeを与え，出力を見てその補足と修正を返す．Analysisの主もまたAIであり，人は突っ込みの刺激を受けてその当否を裁き，Synthesisの結果の修正へ向かう．この修正，すなわちVibeを補足しあるいは新しいVibeを投じる行為は，評価を受けて新しいものを生む行為としてSynthesisの側に数える．これを第3の機能として別立てにすれば，S\&Aという分解そのものが崩れるからである．図~\ref{fig:gaporigin} の各枠の読み方も同じ整理に従う．中央の構造的ギャップは構造を，4つの箱はSynthesis・Analysisの2機能と人・AIの2主体とが作る4つの動作を，矢印は動作間の影響関係を表す．突っ込みの刺激を受けたあとに人の内部で起こる思考の過程は，この分類変数には現れない．それは類型の記述ではなく，往還の記述（\ref{subsec:model-formal}節）が担う．

したがってこの4類型は，単なる分類ではなく，GenAI時代の研究・学習支援システムを設計・評価するための語彙として機能する．任意のシステムについて，「このシステムは，誰のSynthesis／Analysisに対して構造的ギャップを励起する機能をもつのか」を問えるようになるからである．AIが文章を生成して利用者が承認するだけのツールは類型(II)にあり，ギャップの励起機能をもたない．AIが自動採点して結果を返すツールは類型(III)の一部を実装しているが，採点結果への反論経路を欠くならば，評価的判断はやはり委譲される．本稿はこの語彙を用いて，各システムの設計意図を「何という機能によって，誰のどちらの写像に対してギャップを励起しようとしているか」という形式で記述することを提案する．メタ認知機能の励起部分をこのように名指しできるようになることが，本モデルの第一の効用である．

\subsection{4象限：機能と実行主体の二重の区別}
\label{subsec:model-quadrant}

前節の4類型を可能にしているのは，SynthesisとAnalysisという機能の区別に，人とAIという実行主体の区別を重ねる操作である．この二重の区別を明示的に置いたものが4象限であり，往還の各場面はQ1（人のSynthesis），Q2（AIのSynthesis），Q3（人のAnalysis），Q4（AIのAnalysis）に分かれる（表~\ref{tab:quadrant}）．

\begin{table}[htbp]
\centering
\caption{Synthesis/Analysis と実行主体（人/AI）による4象限}
\label{tab:quadrant}
\footnotesize
\setlength{\tabcolsep}{4pt}
\begin{tabularx}{\linewidth}{@{}>{\raggedright\arraybackslash}p{0.13\linewidth} >{\raggedright\arraybackslash}X >{\raggedright\arraybackslash}X@{}}
\toprule
 & 人が実行 & AIが実行 \\
\midrule
Synthesis &
Q1（人の Synthesis）\newline
\textit{定義}：利用者が目的に照らして論理部品を選択・結合し，成果物を構成する行為．何を作るか，何を問うかの決定を含む．\newline
\textit{扱い}：部品の結合作業は委譲しうるが，目的と価値の設定に関わる部分は人間側に留保される． &
Q2（AIの Synthesis）\newline
\textit{定義}：AIが成果物そのもの（作問案，文案，解答，論理）を生成する行為．効率は最大化されるが，利用者の側に構成の履歴が残らない．\newline
\textit{扱い}：積極的に活用してよい．ただし委譲した分だけQ3の負荷が増える． \\
\addlinespace
Analysis &
Q3（人の Analysis）\newline
\textit{定義}：利用者が自らの，またはAIの成果物を，基準に照らして評価・批判・反論する行為．主観評価の申告，AI評価への逆Analysis，AI出力の採否判断を含む．\newline
\textit{扱い}：人間側に留保されねばならない．$A_{epi}$ はこの残存を測る指標である． &
Q4（AIの Analysis）\newline
\textit{定義}：AIが成果物を客観的構造指標へ写像し，ギャップや未充足スロットを名指しする行為．型チェック，構造パラメータの算出，問いの形での指摘の生成．\newline
\textit{扱い}：委譲してよく，むしろ委譲すべき領域である．支援機構が担うべき象限にあたる． \\
\bottomrule
\end{tabularx}
\end{table}

この区別は分類のための分類ではない．生成AIを用いた知的活動においては，成果物が完成したという一事をもって誰が何を行ったのかが不可視になるのであり，4象限はその混同を分解して記述するための最小の道具立てである．そして4象限を導入すると，どの象限を人間側に留保すべきかという設計指針が明確な形をとる．一貫した基準で構造的複雑さを算出する作業は人間が不得手とし，外部化して初めて主観と構造の乖離が可視化されるのだから，Q4は委譲すべき領域である．AIが成果物を生成すること自体（Q2）も禁じられず，むしろ類型(IV)はAIによる積極的な合成を前提とする．ただし委ねた分だけQ3の負荷は増大するのであって，Q2を用いながらQ3を放棄した状態こそが，AIを有能な使用人として用いる形態であり本稿が批判の対象とするものである．Q1についても，部品の結合は委譲しうる一方で，何を作るか・なぜそれに価値があるかという設定は妥当性の判断と不可分であるため，Q3とともに留保される．

以上の扱いを並べると，それはそのまま，AIに委ねたときに何が失われつつあるのかの記述になる．失われうるのは成果物ではなく，Q3すなわち何を妥当とみなすかの決定であり，Q1のうち目的と価値の設定である．Q2に委ねること自体が失わせるのではなく，Q2に委ねたままQ3を空にすれば決定そのものが抜け落ちる．本稿が「メタ認知機能を保護しつつ駆動する」と述べているのは，Q3とQ1の目的設定とを人間の側に留保したうえでなおそれらを実際に働かせるという二重の要求のことであり，この要求は機能の区別と実行主体の区別とを両方あわせて立てないかぎり，一つの要求として書き表せない．そしてこの区別によって初めて，AIが解けたこと（Q4）と学習者が評価できたこと（Q3）の混同，AIが作ったこと（Q2）と学習者が構築したこと（Q1）の混同，そして作問そのものは成立しているが自己評価が構造の実態から外れている状態（Q1が成立しQ3が失敗している状態）が，別々の事象として記述できるようになる．従来これらは「作問の質」という一語のもとで一括りにされ，個別に論じる語彙が与えられてこなかった．4象限による説明力（explanatory power）の獲得を，本稿は主要な貢献の一つとして提示する．

\subsection{2重構造：学習者層と研究者層}
\label{subsec:model-dual}

本稿は，S\&A往還を2重構造（Dual-Layer）として展開する．第1層は学習者のメタ認知を対象とし，算数の作問や国語の読解において，自らの構築物を客観的指標で評価し直す評価的判断 \parencite{tai2018,bearman2024} を養う．第2層は研究者のメタ認知を対象とし，「学習者のメタ認知をどう刺激するか」という支援ロジック自体を論文オントロジーの型チェックに流し込むことで，研究者自身の論理構築をデバッグする．研究者が他人が引いたレールの上の改良に陥ることを防ぎ，独自の新規性を絞り出させる機能がここに置かれる．両層は同一の往還機構を共有し，差し替えられるのは客観パラメータの中身のみである．この事実は，2重構造が便宜的な比喩ではなく，同一モデルの2つのインスタンスであることを意味する．計算機上に具体化した試作機がVibe Compilerであり（第\ref{sec:system}章），本論文それ自体が第2層の出力にあたる点で，自己適用の関係が成立している．

\subsection{定式化}
\label{subsec:model-formal}

以下の定式化は，モデルの構成要素を曖昧さなく指示するための記述装置である．利用者の目的を $v$ と書く．定式化は，利用者が提示するあらゆる成果物に対してシステムが正解とその解法プロセスを提示できること（AIの可解性，すなわちOracleとしてのAI）という前提のうえに成立し，この前提のもとで初めて学習者自身のAnalysisと対比しうる参照解が存在する．ただしOracleは誤りうるのであり，本稿はその誤りを欠陥ではなく評価の機会として織り込む．誤りうるOracleに対して利用者が根拠づきで反論する行為が定義3の $A_{epi}$ が捉える逆Analysisであり，Oracle前提と逆Analysisの設計は表裏をなす．

\vspace{0.5\baselineskip}
\noindent 定義1（Synthesis 写像・Analysis 写像・往還）.\quad
Synthesis写像 $S$ は，目的 $v$ のもとで論理部品を選択・結合して成果物 $x$ を生成する写像である．Analysis写像 $A$ は，成果物 $x$ をドメイン固有の客観パラメータ・ベクトル $p = A(x) \in \mathbb{R}^{m}$ へ写す．第 $k$ 回目の往還における成果物を $x^{(k)}$，そのとき利用者が事前に申告する主観評価ベクトルを $\hat{p}^{(k)} \in \mathbb{R}^{m}$ とすると，往還は次式で与えられる．
\begin{equation}
  p^{(k)} = A\bigl(x^{(k)}\bigr), \qquad
  x^{(k+1)} = S\bigl(x^{(k)} \,\big|\, v,\; g^{(k)},\; \mathrm{Null}(\mathcal{O},x^{(k)})\bigr)
  \label{eq:cycle}
\end{equation}
ここで $g^{(k)}$ は定義2の構造的ギャップ，$\mathrm{Null}(\mathcal{O},x^{(k)})$ は定義4の未充足スロット集合である．式~\eqref{eq:cycle} の第2式がこのモデルの中核である．Analysisの出力が次のSynthesisの引数として現れる点，すなわち相互限定（Mutual Constraint）が構造として書き下されている点に，Reflection・自己調整学習といった内省モデルとの形式的な差異がある．

\vspace{0.5\baselineskip}
\noindent 定義2（構造的ギャップと収束）.\quad
第 $k$ 回目の往還における構造的ギャップ $g^{(k)}$ を，主観評価ベクトルと客観パラメータ・ベクトルの重み付き距離として定義する．
\begin{equation}
  g^{(k)} \;=\; \bigl\lVert p^{(k)} - \hat{p}^{(k)} \bigr\rVert_{W}
  \;=\; \left( \sum_{i=1}^{m} w_i \left( p^{(k)}_i - \hat{p}^{(k)}_i \right)^{2} \right)^{1/2}
  \label{eq:gap}
\end{equation}
$w_i > 0$ はドメインごとの尺度差を吸収する正規化係数である．$g^{(k)}$ が大きいことは，利用者の「つもり」と成果物の実態の乖離を表し，メタ認知的刺激の余地が大きい状態を意味する．往還列が収束したとは，閾値 $\varepsilon > 0$ に対して $g^{(K)} < \varepsilon$ が成り立ち，かつ $\mathrm{Null}(\mathcal{O},x^{(K)}) = \emptyset$ となることをいう．ただし学習成果として重視されるのは個々の成果物における $g^{(K)}$ の小ささではなく，新規の課題においても小さな初期ギャップ $g^{(0)}$ を示すこと，すなわちAnalysisの内面化――AIがいなくとも自ら客観指標を当てられる「目」の獲得――である．

\vspace{0.5\baselineskip}
\noindent 定義3（メタ認知研磨の4指標）.\quad
往還がもたらすメタ認知の研磨を，ドメイン非依存の4指標として定義する．パラメータ予測精度 $E_{pred}$ は，レンジ $r_i = p_i^{\max} - p_i^{\min}$ で正規化した平均絶対誤差であり，評価的判断の向上はその単調な減少・収束として操作化される．評価カバレッジスコア $S_{cov}$ は，精度・効率性・信頼性・拡張性・再利用性から成る品質指標集合 $Q$（$\lvert Q \rvert = 5$，表~\ref{tab:params} の第11--15項）のうち，第 $k$ 回目の自己評価発話で言及された指標の集合 $M_k$ による被覆率であり，その上昇は評価の視点が単一軸から多次元へ広がったことを表す．リファインメント整合性 $L_{ref}$ は，第 $k$ 回目のAnalysisで利用者自身が指摘した不十分点の集合 $I_k$ の解消率であり，「なんとなく直す」修正と評価に基づく戦略的修正とを区別する．
\begin{equation}
  \begin{aligned}
    E_{pred}(k) &= \frac{1}{m} \sum_{i=1}^{m}
       \frac{\bigl| \hat{p}^{(k)}_i - p^{(k)}_i \bigr|}{r_i},
    \qquad
    S_{cov}(k) = \frac{\lvert M_k \cap Q \rvert}{\lvert Q \rvert}, \\[4pt]
    L_{ref}(k) &= \frac{\bigl\lvert \{\, d \in I_k \;:\; d \text{ が解消} \,\} \bigr\rvert}
       {\lvert I_k \rvert}
  \end{aligned}
  \label{eq:three}
\end{equation}

第4の指標である認識的主体性スコア $A_{epi}$ は，AIの評価に対する根拠の成立した反論（逆Analysis）の量と質から定義される．$E$ をAIが評価を提示した機会の集合，$\mathbb{I}[\cdot]$ を指示関数とする．
\begin{equation}
  A_{epi} \;=\; \frac{1}{\lvert E \rvert} \sum_{e \in E} w(e) \cdot \mathbb{I}\bigl[\, \text{利用者が } e \text{ に対し根拠の成立した反論をした} \,\bigr]
  \label{eq:aepi}
\end{equation}
$w(e) \in \{1,2,3\}$ は反論の質の重みであり，根拠を伴う不同意の表明を $1$，前提条件またはカバー範囲の食い違いの特定を $2$，特定に加えて代替パラメータや代替基準の提示を伴う場合を $3$ とする．計数の対象をAIが誤った評価を提示した機会に限定しない点が要である．そう限定すれば，AIが誤らないかぎり利用者には反論の機会が与えられず，主体性の観測がAIの誤り待ちになってしまう．AIの評価それ自体は正しい場合でも，利用者が自らの前提や設計意図を根拠に異を唱え，その根拠が成立しているならば，それは認識的主体性の発現として計数される．$A_{epi}$ は，抽象概念であった認識的主体性を観測可能な行動へ落とし込んだ操作的定義であり，AIの不完全性を欠陥ではなく観測の機会として読み替える地点でもある．なお $\varepsilon$，$w_i$，$w(e)$ の具体値と，反論の根拠が成立しているか否かの判定手続きは，較正を要する設計事項である．

これら4指標は，学習者だけでなく，システムを開発する研究者にとっても支援ロジックのデバッグ指標として機能する．学習者の $A_{epi}$ が高いということは，研究者が設計した批判的なヤスリが学習者の主体性を正しく刺激できているという証左にあたり，第2層の成功を示す．指標の2重の読みは，2重構造モデルの帰結である．

\vspace{0.5\baselineskip}
\noindent 定義4（論文オントロジー・型チェック・Null判定）.\quad
論文オントロジーを $\mathcal{O} = \{o_1, \ldots, o_{16}\}$ とする．各 $o_j$ は学術的論理を構成する必須スロットであり，内容は表~\ref{tab:params} の16の学術パラメータとして与えられる．成果物 $x$ におけるスロット $o$ の値を $\mathrm{val}(o,x)$ と書き，未充足であるとき $\mathrm{val}(o,x) = \bot$ とする．
\begin{equation}
  \begin{aligned}
    \mathrm{Null}(\mathcal{O},x) &= \bigl\{\, o \in \mathcal{O} \;:\;
        \mathrm{val}(o,x) = \bot \,\bigr\}, \\[4pt]
    \mathrm{TypeCheck}(x) &=
    \begin{cases}
      \mathrm{true}  & \bigl(\mathrm{Null}(\mathcal{O},x) = \emptyset\bigr) \\[2pt]
      \mathrm{false} & (\text{otherwise})
    \end{cases}
  \end{aligned}
  \label{eq:typecheck}
\end{equation}
$\mathrm{TypeCheck}(x) = \mathrm{false}$ のとき，システムは $\mathrm{Null}(\mathcal{O},x)$ の各要素についてコンパイルエラーを発火させる．エラーが修正案ではなく，当該スロットを言語化させる問いとして返される点が決定的である．ここでのNull判定は，値の不在のみならず，記述が存在しても他スロットとの論理的対応が取れていない場合をも含む．この意味で $\mathrm{TypeCheck}$ は存在検査と整合検査との合成であり，後者の詳細は第\ref{sec:system}章\ref{subsec:sys-typecheck}節で述べる．

\begin{table}[htbp]
\centering
\caption{型チェックの対象とする5カテゴリ16の学術パラメータ（論文オントロジー $\mathcal{O}$）}
\label{tab:params}
\footnotesize
\begin{tabularx}{\linewidth}{@{}r>{\raggedright\arraybackslash}p{0.19\linewidth} >{\raggedright\arraybackslash}p{0.27\linewidth} >{\raggedright\arraybackslash}X@{}}
\toprule
\# & カテゴリ & パラメータ名 & 学術的定義と役割（型チェック項目） \\
\midrule
1 & (1) 意義・目的 & 重要性 (Significance) & そのドメインの特性に鑑み，なぜ問題を解く必要があるか \\
2 &               & 受益対象 (Beneficiary) & 解決によって直接的な利益を得る対象の特定 \\
3 &               & メリット (Benefit) & 解決によって得られる具体的かつ新しい価値 \\
\addlinespace
4 & (2) 前提と境界 & 前提条件 (Assumptions) & データの信頼性や理論の完全性など，手法成立の土台 \\
5 &               & カバー範囲 (Coverage) & 研究が扱う範囲と，対象外（Out of scope）の明示 \\
6 &               & 技術的要件 (Technical Requirements) & 目的達成に最低限必要な機能スペックや制約条件 \\
\addlinespace
7 & (3) 差分と批判的比較 & 従来手法との相違 (Difference) & 既存アプローチとの決定的な構造的差異 \\
8 &               & 既存手法の限界 (Limitations) & 従来手法が不十分である理由に対する批判的分析 \\
9 &               & 新規性 (Novelty) & コンセプト，設計思想，あるいはアルゴリズムの変革点 \\
\addlinespace
10 & (4) 評価・品質 & 機能実行性 (Functionality) & 掲げた機能が設計通り正しく動作しているか \\
11 &               & 精度 (Accuracy) & 出力結果の正確性および妥当性 \\
12 &               & 効率性 (Efficiency) & 実行時間，計算コスト，メモリ消費量の許容性 \\
13 &               & 信頼性 (Reliability) & ノイズや特殊ケース下でも安定して動作するか \\
14 &               & 拡張性 (Scalability) & データ量や問題規模の増大への対応能力 \\
15 &               & 再利用性 (Reusability) & 他の研究者が汎用的に利用・継承できるか \\
\addlinespace
16 & (5) 知見と課題 & 得られた知見 (Lessons Learned) & 実験を通じて抽出された抽象的な教訓 \\
\bottomrule
\end{tabularx}

\vspace{2pt}
\parbox{\linewidth}{\footnotesize \textit{注.} 第1列の番号は定義4の $o_1$ から $o_{16}$ に対応する．カテゴリ(5)は，得られた知見と対をなす出力サブスロットとして「残された課題」を必須とする運用を行う．第11--15項が式~\eqref{eq:three} の品質指標集合 $Q$ に対応する．}
\end{table}

\subsection{ドメイン汎用性}
\label{subsec:model-generality}

Analysis写像 $A$ が算出する $p$ はドメインごとに具体化される一方，4指標はドメインに依存しない上位指標である．算数の作問ドメインでは，演算ステップ数 $N_{step}$（解答に必要な四則演算の総回数）が計算の深さを，未知数の数 $N_{var}$ が構造の広さを，立式変換難度 $D_{map}$（自然言語の文脈を数学的モデルへ写像する複雑性）が写像難度を表す．国語の読解ドメインでは，参照距離 $S_{ref}$（設問箇所から解答根拠までの段落数）が探索の深さを，論理結合数 $L_{link}$（正解導出に統合を要する段落数）が統合の広さを，語彙置換難度 $V_{map}$（本文の具体表現を抽象語彙へ言い換える度合い）が写像難度を表す．研究ロジック合成のドメインでは，表~\ref{tab:params} の16スロットの充足度と品質指標集合 $Q$ の被覆状況がこれにあたる．

算数と国語の下位指標が，いずれも深さ・広さ・写像難度という同一の3軸に対応している点は注目に値する．交換されるのは各軸の測り方だけであり，この共通性こそが本モデルがドメインを越えて成立する根拠にあたる．一般化すれば，任意のドメインについて構造的深度 $D_{depth}$（処理のステップ数や論理階層の深さ），構成的広延 $W_{width}$（扱われる変数・部品・視点の数），変換難度 $M_{map}$（要求仕様から具体的実装へのマッピングの複雑性）という3軸が定義できれば，本モデルは適用可能である．すなわち本モデルが要求するのは，成果物が複数の論理部品の結合として記述できること，その構造的複雑さを客観的にマッピングする関数が定義できること，そして利用者が事前に自己評価を申告できることの3条件のみであり，差し替えるのが $A$ の実体のみであるという点は層の差異にも当てはまる．同じ論理は部品組合せ型の知的作業一般へ拡張しうるのであって，プログラミングでは循環的複雑度，小論文では論証の連鎖長，実験計画では交絡要因の被覆率がそれぞれ客観指標の役割を果たしうる．さらに，3軸が共通であるというこの構造的事実は，あるドメインで培われた「構造を意識する目」が他のドメインへ転移するという検証可能な予測を生む．本モデルは転移仮説を，実験計画に落とせる形で提示する（第\ref{sec:concl}章\ref{sec:concl-open}節）．

\subsection{本モデルで何が言えるようになるか}
\label{subsec:model-claims}

本モデルが与えるのは効果の主張ではなく，記述可能性の主張である．何が言えるようになるのかを3点に整理しておく．

\paragraph{支援を，便利さではなく構成で比べられる}
「AIで仕事は進むが，何が起きているのかは言えない」という状態に座標が入る．人の仕事を肩代わりする機能と，人の判断を促す機能とが別のものとして扱えるようになり，AI利用を禁止するか放任するかという粒度の議論が，どの矢印を結ぶかという設計の議論へ移る．この構成を組むための条件は2つだけである．見当を先に言わせられることと，構造を指標へ写像できること．この2条件を満たす活動であれば，作問以外にも同じ構成が組める．

\paragraph{起きていないことを名指しできる}
成果物を見ても，人の判断を経たものと経ていないものは区別できない．著者らの草稿V1がその実例であり，作問システムも評価もないまま論文の型はほぼ充足した．誰が何を担ったかは成果物から復元できず，記録として残すほかない．SynthesisとAnalysisの区別に実行主体の区別を重ねて初めて，「人のAnalysisが一度も発火していない」と述べることができる．効果の測定は，起きたことしか測れない．起きなかったことを名指しするには，別の語彙が要る．

\paragraph{ギャップの出どころを位置として指せる}
評価の突き合わせから生まれるかぎり，問い直されるのは自分の評価が当たっていたかまでである．AIが直感を具体物へ変換して初めて，何を作ろうとしていたのかが問い直しの対象になる．

効かない支援の診断も，AI利用方針の設計も，次に検証すべき仮説の定式化も，この記述可能性から出てくる．そして本モデルは反証可能である．同じ構成を組んだのに人の判断が起きなければ，説明の側が反証される．

\subsection{モデルの全体構造}
\label{subsec:model-figure}

全体構造を図~\ref{fig:model} に示す．

\begin{figure}[htbp]
\centering
\begin{tikzpicture}[
  font=\footnotesize,
  box/.style={draw, rounded corners=2pt, align=center, text width=2.4cm,
              inner sep=3pt, minimum height=1.0cm, fill=white},
  spine/.style={draw, dashed, rounded corners=3pt, align=center, text width=2.5cm,
                inner sep=4pt, fill=white},
  arr/.style={-{Stealth[length=2mm]}, semithick},
  darr/.style={{Stealth[length=2mm]}-{Stealth[length=2mm]}, semithick, dashed}
]
\node[box] (s1) at (0,2.0)   {Synthesis $S$\\部品の選択と結合};
\node[box, text width=3.0cm] (x1) at (4.9,2.0) {成果物 $x^{(k)}$\\（作問・要約）};
\node[box, text width=3.0cm] (a1) at (4.9,0.0)   {Analysis $A$（3レイヤ）\\(i) 論理的整合性の検証\\(ii) 構造的・メタ認知的評価\\(iii) 価値的・探究的評価\\$\Rightarrow$ 客観パラメータ $p$};
\node[box] (g1) at (0,0.0)     {構造的ギャップ\\$g^{(k)}$};
\draw[arr] (s1) -- (x1);
\draw[arr] (x1) -- (a1);
\draw[arr] (a1) -- node[midway, above, font=\scriptsize] {対照 $\hat{p}$} (g1);
\draw[arr] (g1) -- node[midway, left, font=\scriptsize, align=center] {還流} (s1);

\node[box] (s2) at (0,-3.8)   {Synthesis $S$\\Vibe の言語化};
\node[box, text width=3.0cm] (x2) at (4.9,-3.8) {成果物 $x^{(k)}$\\（論理スナップショット）};
\node[box, text width=3.0cm] (a2) at (4.9,-5.6) {Analysis $A$\\16スロット充足度};
\node[box] (g2) at (0,-5.6)   {構造的ギャップ\\Null スロット};
\draw[arr] (s2) -- (x2);
\draw[arr] (x2) -- (a2);
\draw[arr] (a2) -- node[midway, above, font=\scriptsize] {対照 $\hat{p}$} (g2);
\draw[arr] (g2) -- node[midway, left, font=\scriptsize, align=center] {還流} (s2);

\begin{scope}[on background layer]
  \node[draw, rounded corners=3pt, fill=black!4, inner sep=4mm,
        fit=(s1)(x1)(a1)(g1)] (layer1) {};
  \node[draw, rounded corners=3pt, fill=black!4, inner sep=4mm,
        fit=(s2)(x2)(a2)(g2)] (layer2) {};
\end{scope}
\node[anchor=south west, font=\footnotesize\bfseries] at (layer1.north west)
  {第1層：学習者のS\&A往還};
\node[anchor=north west, font=\footnotesize\bfseries] at (layer2.south west)
  {第2層：研究者のS\&A往還};

\draw[arr, dashed] (layer1.south) -- node[right=1.5mm, font=\scriptsize, align=left]
  {支援ロジック自体を\\型チェックへ（自己適用）} (layer2.north);

\node[spine, minimum height=9.4cm] (sp) at (8.9,-1.8)
  {AI：批判的なヤスリ\\（答えを与えず\\問いを返す）\\[10pt]論文オントロジー\\$\mathcal{O}$\\[4pt]型チェック\\$\mathrm{TypeCheck}(\cdot)$};
\draw[darr] (a1.east) -- (a1.east -| sp.west);
\draw[darr] (a2.east) -- (a2.east -| sp.west);
\end{tikzpicture}
\caption{構造的ギャップ駆動型メタ認知支援の2重構造モデル．第1層（学習者）と第2層（研究者）が同一のS\&A往還を共有し，批判的なヤスリとしてのAIと論文オントロジーによる型チェックが両層を貫く横串として作用する．}
\label{fig:model}
\end{figure}

図~\ref{fig:model} は，横方向に往還の1周期を，縦方向に2重構造を，右端に両層を貫く横串を配置する．要点は左下から左上へ戻る「還流」の矢印にあり，ギャップと未充足スロットが式~\eqref{eq:cycle} 第2式におけるSynthesisの引数として与えられること，すなわち評価が評価のままで終わらず次の構築の制約条件へ転化することを可視化する．読み取るべき要点は3つある．2つの層が同一の形状をしていることは，学習者支援と研究者支援がAnalysis写像 $A$ の差し替えのみで接続されることを意味する．横串が層の内部へ入り込んでいることは，AIが往還の外側から評価を下す審判ではなくAnalysisの一部として組み込まれた構成要素であることを示す（矢印が双方向なのは，利用者からAIへの向きが逆Analysisにあたるためであり，その頻度と質が $A_{epi}$ として捉えられる）．そして還流の矢印が常に利用者側のSynthesisへ戻ることは，修正の決定権がAIではなく利用者に残されていることを示し，これが4つのパワーの再分配 \parencite{akata2020} および共有エージェンシー \parencite{damsa2010} の設計上の担保にあたる．象限との対応でいえば，Synthesisノードが主にQ1，客観パラメータの算出とNull検出がQ4，$\hat{p}$ の申告と還流を受けた採否の決定がQ3である．図にQ2を明示していないのはQ2を排除しているからではなく，図が描くのが往還の主動線だからであり，批判的なヤスリが禁じるのはQ3を経ずに成果物を書き換える経路であって，AIによる生成一般ではない．

%% file: ja-04-system.tex
\section{試作機 Vibe Compiler：仕様と成果}
\label{sec:system}

\subsection{システムの概要と構成}
\label{subsec:sys-overview}

Vibe Compiler は，利用者の曖昧な「ノリ（Vibe）」を学術的オントロジーへとマッピングし，論理的な構造体へ合成する研究ロジック・コンパイラである．システムは入力に対して型チェック（Type Check）を実行し，学術的に必須とされる構成要素が欠落している（Nullである）場合にコンパイルエラーとしてフィードバックを返す．ここで返されるのが修正案ではなく問いであるという制約が，本システムを一般の文章生成支援から分かつ．

命名の由来はVibe coding \parencite{karpathy2025} にある．Vibe codingにおいてAIが生成するのは動くコードであり，コンパイラと実行環境という即時のフィードバック機構が正しさを保証する．これに対して研究活動においてコードに相当するのは，文章そのものではなく論理の組み立てであり，「背景$\rightarrow$問題点$\rightarrow$目的$\rightarrow$解決策$\rightarrow$評価$\rightarrow$知見」という構成が実行ログにあたる．本稿が提案するのは，この論理の組み立てに対してコンパイラの役割を果たす機構であり，したがって行われるのはVibe codingならぬVibe compilingである．

試作機はNotebookLMとGeminiの併用によって構成される．論文オントロジーという「型」の固定性を前者が，合成の柔軟性を後者が担う．注目すべきは，この構成において著者らが行ったのが，プロンプト工学による作り込みではなく資料の投入であったという点である．NotebookLMへロードしたのは，(1) エージェンシーに関するサーベイ論文 \parencite{roePerkins2026}，(2) 筆者らが執筆した論文テンプレート原稿（論文オントロジー），(3) そこから抽出された16の学術パラメータ（表~\ref{tab:params}），(4) パラメータ間の整合性チェック条件，そして (5)--(7) Vibe Compilerとの対話を通して得られた3種類の突っ込み手続き（研究者向け・学習者向け・共創パートナーとしての対話デザイン）の計7種類である．サーベイ論文は，システムが「あなたはいまMakerに留まっている」と指摘しうる判断基準を供給する．論文オントロジーは形式言語による記述ではなく，人間の読者向けに書かれた散文の文書である．突っ込み手続きの3種は，いずれもシステムとの対話の過程でシステム自身が提案し，著者らが吟味・確定したものである．

動作は起動時の役割指定によって規定され，自らを研究ロジック合成コンパイラとして定義すること，論文オントロジーを唯一の「型」として扱い他の資料をヤスリとして用いること，単なる開発報告を受理せず未充足スロットをコンパイルエラーとして返すこと，修正案を提示せず問いの形で応答すること，という4条件が課される．実装として行ったのはこれだけである．

\subsection{中核的知見：効いているのは推論器ではなく，与えた構造の中身である}
\label{subsec:sys-content}

前節の構成から直ちに導かれるのが，本稿の主要な知見である．すなわち，Vibe Compilerを動かしているのは大規模言語モデルの推論能力ではなく，NotebookLMへ投入した内容の構造の側である．

この主張の根拠は，投入した論文オントロジーが備えていた性質にある．当該文書には，序論の悪い例として次の記述が含まれている．「このような機能を持つシステムが要求されている．（理由を言わない）だからこのようなシステムを作った．それはこのような機能を持っている．それはこのような方法で作られている．これの悪いところは，結果の羅列のみで，学問の積み重ねに一切貢献しないこと．」

この例において，要求・システム・機能・方法という4つの項目はすべて埋まっている．したがって項目の有無だけを見るチェックリストであれば，この文書は合格してしまう．落ちるのは，「要求」と「システム」のあいだに「なぜそれが必要かという理由」のリンクが存在しないという一点においてのみである．すなわちこの文書は，項目の存在検査では原理的に検出できず，項目間の関係を見て初めて落ちる不合格例を，テストケースとして保持している．Vibe Compilerがコンパイルできるのは，与えられたのが問いの一覧ではなく，問いと問いのあいだに条件がある構造だったからである．

この観察は，オントロジー工学が一貫して主張してきたことの再確認にあたる．\textcite{bourdeauMizoguchi2000} が知的教育システムの困難について「これらの問題はすべて内容に関わるものである．言い換えれば，推論技術も美しい理論的形式化も，状況の改善には寄与しない」と述べたとき，指されていたのはまさにこの構図である．LLMがどれだけ強力になっても，構造を与えなければ流暢な文章が出力されるだけで，構造的ギャップは生まれない．さらに重要なのは，投入した論文オントロジーが形式化を経ていないという事実である．本来であればそこからformalなオントロジーを構築する工程が必要であった．ところが生成AIの出現によって，NotebookLMへ投入するだけでそれが動いてしまった．含意は第\ref{sec:disc}章\ref{subsec:disc-content}節で論じる．

\subsection{対話実行例：起動から型チェック・エラーまで}
\label{subsec:sys-dialogue}

以下に，第2層（研究者層）における実行ログの抜粋を示す．著者の一人が作問学習に関する自らのVibeを投入したセッションである．

\begin{quote}
\small
利用者：あなたは，Vibe Coding風に研究遂行を支援するシステムである．いまNotebookLMにある論文オントロジーに基づいて，研究ロジック合成コンパイラとして対話を進めてほしい．

システム：承知した．研究ロジック合成コンパイラ（v1.0）として起動する．私は，あなたが提示する断片的な直感（Vibe）を，論文オントロジーという唯一の型に適合させるべく，他の資料をヤスリとして駆使し，あなたの思考を研磨・ビルドしていくシステムである．本システムは単なる開発報告を許さない．まずドメインの状況，解こうとする課題，解決の方向性を入力してほしい．

利用者：普通，学習者は問題を解くことによって学ぶが，解くべき問題を作成することによって学ばせるというProblem posingの研究が始まっている．私はなぜ問題を作ると学びが進むのかを説明する認知モデルを作りたい．そこで思考活動をSynthesisとAnalysisという二つに分解して考察すると，問題作成による学習が進む理由を説明できるのではないかと思っている．

システム：意味解析を行い，論文オントロジーの型へマッピングする．背景は受動的な問題解決から能動的な作問学習への移行，解こうとする問題は「なぜ学びが進むのか」という内部メカニズムが未解明な知識の空白，提案手法は思考活動をSynthesisとAnalysisに分解しその相互作用として作問を捉える点である．\\
\textit{［ヤスリによる研磨］}資料によれば，学習者が知識の構築と評価に参加する能力は認識的主体性と呼ばれる．あなたの言うSynthesisとAnalysisは，まさにこのエージェンシーの発揮そのものではないか．また，Analysisを単なる分解ではなく「自分のSynthesis結果が妥当かどうかを判断する評価的判断力」と定義すれば，既存の認知モデルとの差別化がより鮮明になる．\\
\textit{［型チェック・エラー報告］}現時点のロジックには，以下のパラメータが不足（Null）している．（1）既存手法の限界：既存の認知モデルはなぜ不十分なのか．SとAのどちらか，あるいは両方の往還を見落としているのか．（2）受益対象とメリット：誰が具体的にどのような新しい価値を得られるのか．（3）概念の名前：このS\&A往還による認知モデルに特徴的な名前をつけるとしたら，どのようなイメージか．
\end{quote}

注目すべきは，最初に投入された直感が，出典も受益対象も既存手法との差分ももたない一文であったにもかかわらず，システムがこれを論文オントロジーのスロットへ強制的に流し込み，割当の成立と同時に論理の鎖としての欠落を検出した点である．そして3件の指摘は単独の指摘に終わらず，再合成を経て確定した論理へ変換された．（3）は「S\&A往還」という概念名へ，（1）は「認知メカニズムのブラックボックス問題」という差分ロジックへ，（2）は「管理者としての学習者と教師」という受益対象とメリットへ接続した．いずれも当初の発話には存在しなかった要素である．「欠落の名指し$\rightarrow$再合成$\rightarrow$論理の確定」という構造的ギャップ駆動の連鎖が，設計上の想定としてではなく実際の対話上で生起した．型チェックが穴埋め要求ではなく，新規の論理が産出される契機として働いた事例である．

システムは往還の各周期において，確定した論理を論理スナップショットとして外部化する．スナップショットは，対応する論文オントロジーのID，論理名，合成された意義，新規性の核という4つの欄をもつ．たとえば上記セッションの帰結として得られたSnapshot 01は，論理名を「S\&A往還」，意義を「作問活動をSynthesisとAnalysisに分解し，その相互交流こそが学習効果の源泉であると定義する」，新規性の核を「既存モデルが無視していた評価的判断をAnalysisとして内部に組み込み，作問がなぜ問題解決能力へ転移するのかという認知メカニズムを説明可能にした点」とする．スナップショットは静止した記録ではなく往還プロセスの状態遷移の外部化であり，どの論理がどの型チェック・エラーへの応答として産出されたのかを追跡可能にする．本稿の研究ロジックを構築する過程では最終的に12個が確定した．

\subsection{型チェックの機構}
\label{subsec:sys-typecheck}

型チェックが未充足や不整合を検出したとき何を返すかは，表~\ref{tab:triggers} の突っ込みトリガー群として定められる．これらに共通する目的は，AIが答えを出すことで利用者の思考を奪う認知的オフローディングを防ぎ，生産的な苦闘を維持することにある．エラーが修正案ではなく問いの形をとることが，そのための必須条件である．

\begin{table}[htbp]
\centering
\caption{型チェックが発火させる突っ込みトリガーの体系}
\label{tab:triggers}
\footnotesize
\begin{tabularx}{\linewidth}{@{}>{\raggedright\arraybackslash}p{0.05\linewidth} >{\raggedright\arraybackslash}p{0.24\linewidth} X@{}}
\toprule
層 & トリガー & 発話の例 \\
\midrule
共通 & スロットNullチェック & 「解決によるメリットが定義されていません．誰がどう助かるのかを記述してください」 \\
\addlinespace
共通 & あるべき姿と現状のギャップの強制抽出 & 「現在の技術では，なぜその理想に到達できないのか，ドメイン特有の理由を挙げてください」 \\
\addlinespace
共通 & 多角的な品質指標によるストレステスト & 「データ量が100倍になった際の拡張性は検討されていますか」「メモリ消費量の観点で実用上の懸念はありませんか」 \\
\addlinespace
共通 & 既存手法への批判的比較の義務化 & 「従来手法と比較して，あなたの提案はどの点が決定的に違うのですか」（差分と限界を述べない新規性の主張を許さない） \\
\addlinespace
共通 & 知見への昇華のプロンプト & 「現在のボトルネックを残された課題として整理し，解決に必要な技術的要素を予見してください」 \\
\midrule
第2層 & 直感の学術的パラメータへの変換 & 「『イケてない』と感じる点は，拡張性・信頼性・精度・計算コスト・再利用性のどれの不足ですか」 \\
\addlinespace
第2層 & 前提条件と境界を突く & 「その既存手法が暗黙に仮定している前提条件は何ですか．それが崩れたとき信頼性はどう損なわれますか」 \\
\addlinespace
第2層 & 差分ロジック構成のための対比 & 「限界を『性能不足』ではなく『設計思想の限界』として述べてください．提案はそれを制限の緩和や概念の一般化でどう突破しますか」 \\
\addlinespace
第2層 & 認識的主体性を刺激する批判的ヤスリ & 「AIが算出した限界に，あなた自身のドメイン知識から反論してください」（反論は逆Analysisとして受理され $A_{epi}$ に算入） \\
\midrule
第1層 & 予測と実績のギャップによる不協和の創出 & 自己予測を先に申告させたうえで「システム指標では最小値です．あなたが『難しい』と感じた要素は何ですか」 \\
\addlinespace
第1層 & あえての誤評価に対する批判的検証 & 甘い評価を出した直後に「……と判断しましたが，実は隠れたバグや非効率な点を見落としていませんか」 \\
\bottomrule
\end{tabularx}
\end{table}

型チェックはスロットの存在検査だけでは完結しない．\ref{subsec:sys-content}節で述べたとおり，項目がすべて埋まっていてもなお不合格となる文書が存在するからである．そこでVibe Compilerは，スロット間の論理的対応を検証する整合性チェックを備える（表~\ref{tab:consistency}）．検証されるのは，「背景$\rightarrow$課題$\rightarrow$解決策$\rightarrow$評価$\rightarrow$知見」という学術的な論理の鎖が強固に結合されているかである．とりわけ「限界と新規性の鏡像チェック」は，研究者が陥りがちな「他人が引いたレールの上の改良」を検出する機能を果たす．自手法の新しさが既存手法の弱点を補う論理的必然性をもっているかを検証し，対比構造を強制的に作らせるからである．

\begin{table}[htbp]
\centering
\caption{パラメータ間の整合性チェック（5種）}
\label{tab:consistency}
\footnotesize
\begin{tabularx}{\linewidth}{@{}>{\raggedright\arraybackslash}p{0.20\linewidth} >{\raggedright\arraybackslash}p{0.26\linewidth} X@{}}
\toprule
検査名 & 照合されるスロット対 & 不整合の例 \\
\midrule
目的と評価指標の同期 & 目的に含まれる品質特性 $\leftrightarrow$ 評価の項目 & 目的に「信頼性の向上」を掲げながら，評価が小規模データでの精度確認のみに留まっている \\
\addlinespace
意義とメリットのベクトル照合 & 重要性・受益対象 $\leftrightarrow$ メリット & 医療現場の「判断の遅延」を課題としているのに，メリットが「システムの保守性向上」という無関係な軸である \\
\addlinespace
限界と新規性の鏡像 & 既存手法の限界 $\leftrightarrow$ 新規性 & 既存手法の限界を「コストの高さ」としているのに，自手法の新規性が「精度の向上」である \\
\addlinespace
前提条件とカバー範囲の境界 & 前提条件 $\leftrightarrow$ カバー範囲 & 前提条件で「クリーンなデータ」を仮定しながら，カバー範囲にノイズの多いリアルタイム・データを含めている \\
\addlinespace
実験結果と知見の帰結 & 評価の妥当性 $\leftrightarrow$ 得られた知見 & 実験で「大規模データでは処理が遅延する」という結果が出ているのに，知見で「あらゆる環境で有効」と一般化している \\
\bottomrule
\end{tabularx}
\end{table}

なお，表~\ref{tab:triggers} のトリガー群には横断的な制約が課される．知識の空白を埋める積極的な内容の合成は許され，システムは「そのVibeは既存手法が抱える拡張性の欠如を解決する具体的な解決策になりうる」と，研究の新規性や意義を能動的に合成して提示してよい．しかし不協和と解決策の同時提示が求められ，指摘のみで発話を終えることは禁じられる．加えて，起こりうる帰結を先取りする予見的対話，課題候補を複数提示して利用者に選ばせ反論経路を常時開いておく選択と管理の委ね，構築の瞬間に評価をぶつけるリアルタイム・フィードバックが課される．これらはAIを審判ではなく共創のパートナーとして認識させるための条件であり，禁じられるのは成果物の直接的な書き換えであって素材の提示ではない．

\subsection{UI設計：ロジック同期型開発エディタ}
\label{subsec:sys-ui}

\begin{figure}[htbp]
\centering
\includegraphics[width=\linewidth]{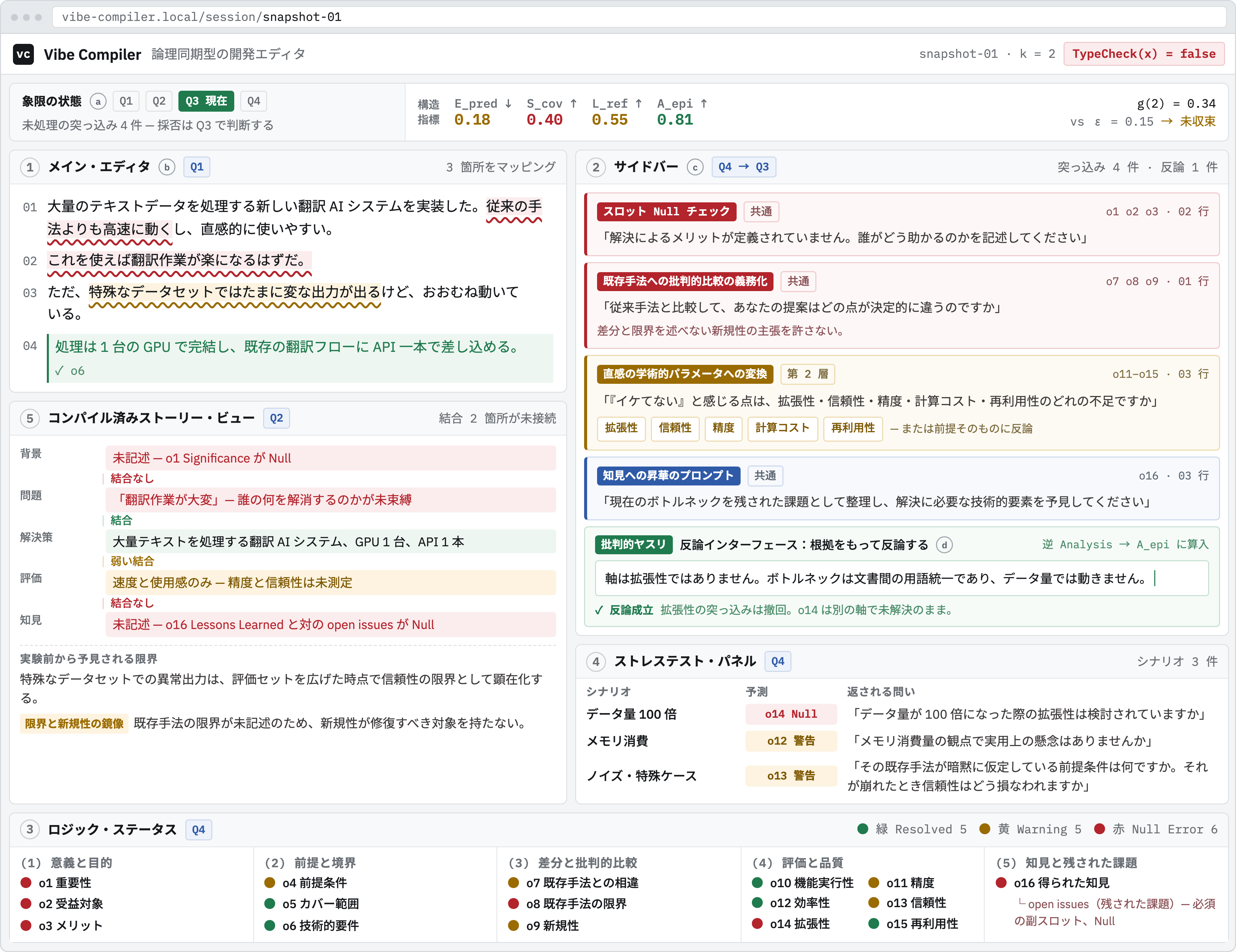}
\caption{Vibe CompilerのUI設計の例（翻訳AIシステムの研究というVibeを入力し，型チェックが4件の突っ込みを返した場面）．メイン・エディタ（Q1），サイドバー（Q4$\rightarrow$Q3），ロジック・ステータス（Q4），ストレステスト・パネル（Q4），コンパイル済みストーリー・ビュー（Q2）の5ペインからなり，画面上部には象限の状態と構造指標（$E_{pred}$，$S_{cov}$，$L_{ref}$，$A_{epi}$）を常時表示する．}
\label{fig:ui}
\end{figure}

以上の対話設計を実現するUIとして，本稿は5つのペインからなるロジック同期型開発エディタを提案する．UIは単なる見栄えの問題ではなく，4象限のどこに利用者を置くかを決める装置である．メイン・エディタ（Q1）は形式を気にせず断片的な思考を書き込む場所であり，書き込みと同時にAIが背後で16パラメータへのマッピングを試行する．サイドバー（Q4とQ3）は「今の実装は既存手法と何が決定的に違うのですか」といった問いをリアルタイムで返すと同時に，AIの突っ込みが的外れな場合に利用者が根拠をもって反論を打ち込める反論インターフェースを備え，ここでの反論の成立が認識的主体性の証明にあたる．ロジック・ステータス（Q4）は16パラメータの充足状態を緑（Resolved）・黄（Warning：曖昧または不整合）・赤（Null Error）で視覚化し，ストレステスト・パネル（Q4）は「データ量を100倍にした場合」等のシナリオで拡張性や信頼性の変化を予測して突っ込み，コンパイル済みストーリー・ビュー（Q2）は断片的なVibeから論理の鎖をプレビュー表示して実験前から予見される限界を提示する．システムが「重要性の定義がNullです」と警告し，利用者が受益対象やメリットを言語化してエラーを消していく作業は，論理のデバッグそのものである．UIの設計思想は，開発のスピード（Vibe）を落とすことなく，学術の積み重ねに寄与する論理性を強制的に担保することにある．この設計を具体化した画面の例を図~\ref{fig:ui} に示す．5つのペインはそれぞれ自らの属する象限を掲げ，象限の状態と4指標は画面上部に常時表示される．またサイドバーでは突っ込みの1つに対して反論インターフェースが開いており，根拠をもった反論が逆Analysisとして受理され $A_{epi}$ に算入されるまでの一連の流れが，画面上の操作として現れている．

\subsection{成果：コンパイラからの要求と，それにどう答えたか}
\label{subsec:sys-demands}

本節は，本稿の研究ロジックを構築する過程においてVibe Compilerが実際に発した要求・突っ込みと，著者らがそれにどう応答したかを整理したものである．これは試作機の機能実行性に関する唯一の実質的な証拠であり，同時に，第2層の往還が実際に生起したことの記録でもある．

\begin{table}[htbp]
\centering
\caption{Vibe Compilerからの主要な要求と，著者らの応答（本稿の研究ロジック構築過程の記録）}
\label{tab:demands}
\footnotesize
\begin{tabularx}{\linewidth}{@{}>{\raggedright\arraybackslash}p{0.04\linewidth} >{\raggedright\arraybackslash}p{0.31\linewidth} X@{}}
\toprule
\# & コンパイラからの要求（型チェック・エラー） & 著者らの応答と，確定した論理 \\
\midrule
1 & 既存の認知モデルはなぜ不十分なのか．SとAのどちらか，あるいは往還を見落としているのか & 既存モデルは作問の「手順」や「結果」を記述しうるが，理解の深化に直結する動的プロセスを説明していないと定式化．「認知メカニズムのブラックボックス問題」として差分ロジックを確定 \\
\addlinespace
2 & この認知モデルの受益対象は誰か．どのような新しい価値を得るのか & 学習者は構造的理解を，教師は「S\&A往還を支援する」という新しい高度な専門性を得ると確定 \\
\addlinespace
3 & このモデルに特徴的な名前をつけるとしたらどのようなイメージか & 「S\&A往還（Reciprocal Synthesis \& Analysis）」を命名．相互交流が学習効果の源泉であるという定義を確定 \\
\addlinespace
4 & Analysisの評価軸を具体化せよ．何をもって構造的複雑さとするのか & 算数ドメインについて $N_{step}$，$N_{var}$，$D_{map}$ を定義．同時に「AIの可解性」を前提条件として明示 \\
\addlinespace
5 & 「解ける／解けない」という機能実行性以外の評価軸はないのか & 「解けないが，その解法構造に興味がある」状態を肯定的に評価する探究駆動型Analysisを追加．正解至上主義からの離脱 \\
\addlinespace
6 & 学習者の自己評価力の向上を，どう数値化するのか & $E_{pred}$，$S_{cov}$，$L_{ref}$，$A_{epi}$ の4指標を定義（式~\eqref{eq:three}，\eqref{eq:aepi}）．システム自身が提案し，著者らが吟味・確定 \\
\addlinespace
7 & 既存の作問モデルとの決定的な差分は何か & Silver や Christou らが作問を手続きの羅列として捉えたのに対し，本モデルは不協和を解消しようとする動的プロセスに焦点を当てると定式化 \\
\addlinespace
8 & 単なる Reflection との違いは何か．このままでは既存概念の範疇に収まる & 指標の客観的外部化・構築と評価の相互限定・生産的な苦闘の意図的維持という3点で差別化．「構造的ギャップ駆動型メタ認知支援」を確定 \\
\addlinespace
9 & S\&A往還を「学習者一般」へ展開する際の実装上の課題は何か & ドメイン汎用的な構造的パラメータの定義が困難であることをシステム自身が指摘．これを受けて $D_{depth}$・$W_{width}$・$M_{map}$ という3軸への一般化を実施 \\
\addlinespace
10 & 目的に掲げた課題が，指摘した既存手法の不十分な点と論理的に対応していない & 整合性チェックの発火例．「メタ認知支援」という目的と「主観的な気づきに依存する」という既存手法の限界とを対応づけ，客観指標の外部化という解決ベクトルを同期させた \\
\bottomrule
\end{tabularx}
\end{table}

表~\ref{tab:demands} から読み取るべきは，コンパイラの要求が単なる穴埋め要求ではなかったという点である．10件のうち，概念の命名（\#3），評価指標の設計（\#6），一般化の必要性の指摘（\#9）は，著者らが事前に用意していなかった論点であり，システムとの往還を経て初めて産出された．とりわけ\#9は，システム自身が自らの適用範囲の限界を指摘した事例であり，これを受けて本稿はドメイン汎用的な3軸への一般化を行った．

同時に，各要求において何が失われうるのかも記録しておく必要がある．Nullが名指しされたあと，その空白を誰が埋めるのかという決定こそが分岐点である．利用者が「限界の記述を書いておいて」と応じていれば，確定した論理はAIのSynthesis（Q2）の産物となり，利用者の側には何を妥当とみなすかの判断（Q3）が残らない．文面の質は大差ないとしても，何を差分ロジックとして立てるかを誰が決めたのかは入れ替わる．実際に起きたのは，空白の名指し（Q4）に利用者が再合成（Q1）で応答した系列である．守られたのはこの決定であって文面ではない．

以上により，試作機の機能実行性は確認された．システムは16スロットに対する型チェックを実行し，未充足スロットをNullとして検出し，問いの形でフィードバックを返し，スロット間の不整合を検出した．RQ1に対する肯定的な証拠はここにある．同一の機構が学習者ドメインと研究者ドメインの双方において，Analysis写像の差し替えのみで動作したことは，RQ4に対する肯定的な証拠にあたる．そして本稿という成果物そのものが，研究者のVibeをCompileして実質的な研究へ落とし込みつつ論文を生成しうることの実証にあたる．効果の水準を問うRQ2・RQ3・RQ5は統制条件下での測定を要するため，本稿の射程としては第\ref{sec:disc}章\ref{subsec:disc-limit}節に整理する．

%% file: ja-05-illustrations.tex
\section{適用例：第1層（学習者）への展開}
\label{sec:illus}

\begin{figure}[htbp]
\centering
\includegraphics[width=\linewidth]{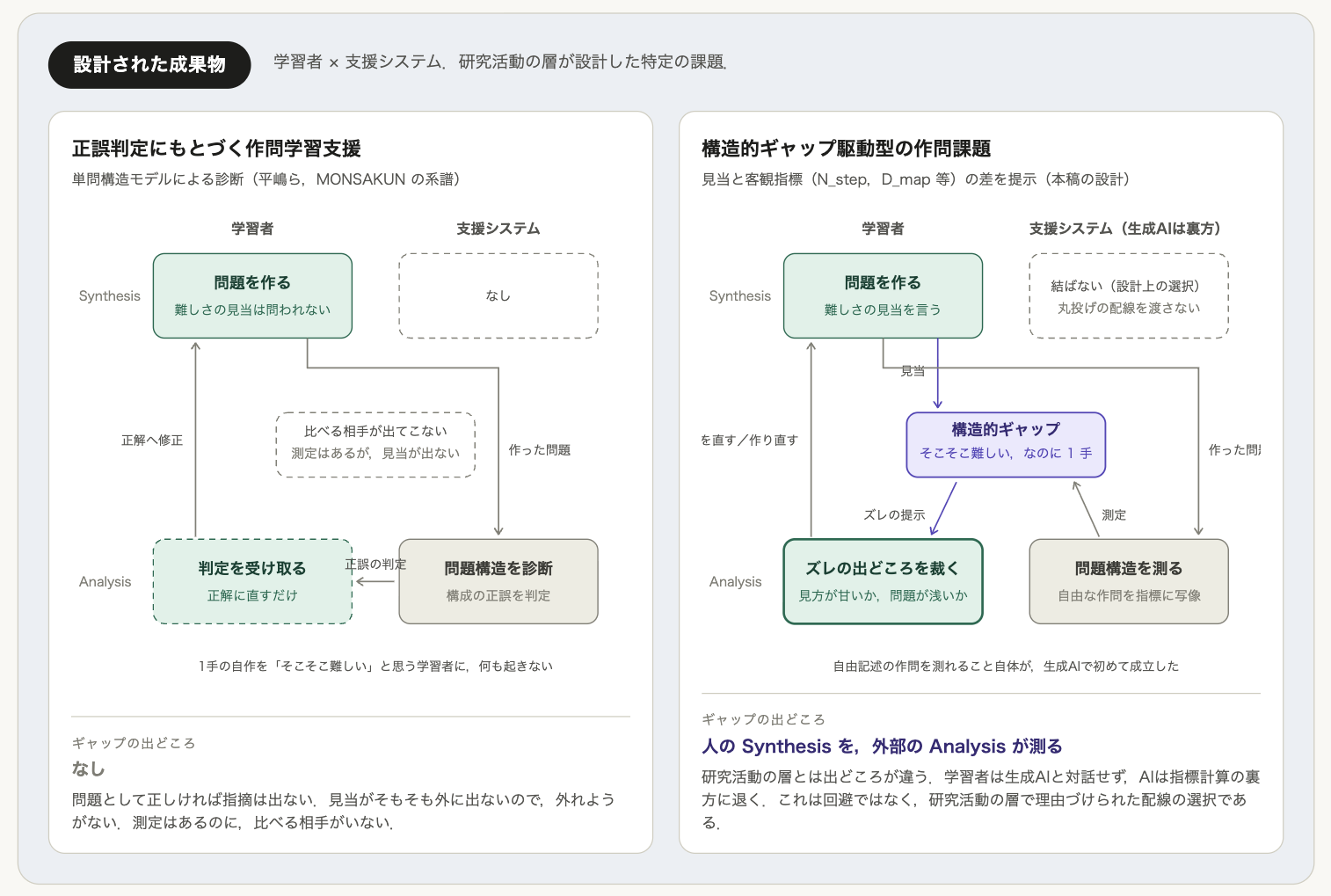}
\caption{学習者層における構成の対比．正誤判定にもとづく従来の作問学習支援（左）では，見当がそもそも外に出ないためギャップの出どころがない．本稿が設計する構造的ギャップ駆動型の作問課題（右）では，学習者に見当を先に言わせ，外部のAnalysisがそれを測る．}
\label{fig:learner}
\end{figure}

第\ref{sec:system}章で示した実行ログは第2層（研究者）の系列であった．本章では第1層（学習者）を扱うが，ここで注意を要するのは，学習者層のギャップの出どころは研究活動の層とは異なるという点である（図~\ref{fig:learner}）．研究活動の層ではAIのSynthesisが出どころであったのに対し，学習者層で本稿が設計するのは人のSynthesisを外部のAnalysisが測るという構成である．学習者は生成AIと対話せず，AIは指標計算の裏方へ退く．すなわち支援システムのSynthesisと学習者とをあえて結ばない．これは生成AIの回避ではなく，研究活動の層で理由づけられた設計上の選択であり，丸投げの構成を渡さないことがここでの判断にあたる．

以下に示す学習者層の系列は，本稿が記述する機構の設計から導かれるシミュレーションであり，実際の学習者から得られた記録ではない．そこに現れる値は設計上の想定値である．問うのは成果物ができたか否かではなく，そこで何が失われうるのか，何が人間の側に残ったのかである．

この構成は，表~\ref{tab:gaporigin} の類型(III)と同じ座標に置かれる．Synthesisを担うのは学習者だけであり，Analysisは学習者自身の自己評価と支援システムによる指標算出とが並ぶからである．研究活動の層で本稿が採った類型(IV)とは出どころが異なり，その意味で2つの層は同一の類型を共有していない．にもかかわらず学習者層が(III)の限界にとどまらないのは，難しさの見当を作問と同時に外へ申告させ，そのうえで生じた差をどちらへ帰属させるかの判断を学習者に残すという条件を，設計として加えているからである．見当が外部化されて初めて指標との差は観測可能な量になり，その差の帰属を裁く行為，すなわち見方が甘かったのか作った問題が浅かったのかを決める行為が，作ろうとしたものの側への問い直しを開く（\ref{subsec:illus-refutation}節）．類型(III)が評価の当否までで止まるのは，この帰属の裁定を利用者へ返す経路を欠くためであり，座標が同じでも届く先は変わる．

従来の作問学習支援との差はここに現れる．単問構造モデルによる診断 \parencite{hirashima2008,hirashima2014} は問題構造の正誤を判定するが，問題として正しければ指摘は出ない．学習者の見当がそもそも外に出ないので，外れようがないのである．測定はあるのに，比べる相手がいない．1手の自作を「そこそこ難しい」と思う学習者には，何も起きない．これに対し本稿の課題は，見当を先に申告させたうえで客観指標との差を提示する．なお，自由記述の作問を測れること自体が生成AIによって初めて成立したという点は，本稿の設計が生成AI時代に固有である理由にあたる．

\subsection{算数作問ドメイン}
\label{subsec:illus-arith}

系列の起点は，学習者が「100円のりんごを買いました．消費税が10\%かかるとき，代金はいくらですか」という問題（v1.0）を作り，「計算が必要だから，そこそこ難しい」と自己評価するところにある．システムはまず正解が一意に導出できることを確認する（AIの可解性）．ここまでは通常の解答支援と変わらない．異なるのはその次である．作問の構造を走査し，$N_{step}=1$（$100 \times 1.1$ の1回），$N_{var}=1$（代金のみ），$D_{map}$ 低（順序どおりに立式できる）を算出し，主観的申告との構造的なギャップを検知する．

システムは「割引を追加しなさい」という修正案を決して提示しない．代わりに「作成された問題は，AIによって正解（110円）が導き出されました．しかし，あなたが設定した構造的複雑さの目標に対して，現在の演算ステップ数は1です．さらに未知数の数を増やすか，あるいは文脈に『割引』などの逆方向の演算部品を組み込むと，どう変化するでしょうか」と，問いの形でギャップを突き返す．答えではなくギャップだけを与えるか，それとも修正案そのものを与えるか．保護と駆動という要求が具体的な設計判断として立ち現れるのはこの地点であり，後者を選んだ瞬間に生産的な苦闘は消去される \parencite{hiebertGrouws2007}．学習者はここで「難易度を意図的に上げる」という目標を自ら立て，部品を再構成する（表~\ref{tab:usecase-params}）．

\begin{table}[htbp]
\centering
\caption{算数作問ドメインにおける往還前後の構造パラメータの遷移（シミュレーション）}
\label{tab:usecase-params}
\footnotesize
\begin{tabularx}{\linewidth}{@{}l X c c c X@{}}
\toprule
版 & 問題文の要旨 & $N_{step}$ & $N_{var}$ & $D_{map}$ & 学習者の自己評価 \\
\midrule
v1.0 & 100円のりんご1個に消費税10\% & 1 & 1 & 低 & 「計算が必要だから，そこそこ難しい」 \\
v2.0 & 1個100円のりんご3個，2割引，消費税10\%，500円玉のお釣り & 4 & 2 & 中〜高 & 「お釣りを加えたから難しくなった」 \\
\bottomrule
\end{tabularx}
\par\smallskip
\begin{minipage}{\linewidth}
\footnotesize 注：本表の値は本章冒頭に述べたとおり，設計に基づくシミュレーションである．
\end{minipage}
\end{table}

往還が v2.0 で終わらない点が重要である．学習者が「お釣りを加えると，代金を出すだけでなく引き算のステップも必要になるし，何より代金と所持金の関係を考えなければいけないから，マッピングが難しくなる．これが $D_{map}$ が上がるということか」と知見を言語化した段階で，システムはこれをスナップショットとして記録したうえで，さらに突っ込みを返す．前提条件の脆さを突く突っ込みは「この問題は『500円玉で必ず足りる』という前提に依存している．単価を200円に書き換えても数式はそのまま機能するか」と問い，適用順序の曖昧さについては「『2割引』と『消費税10\%』はどちらを先に適用すべきか．現状の文面では一意に定まらない」と問い，拡張性については「りんごが100個になり袋代やポイント還元が加わった場合，構造的複雑さはどう変化するか」と問う．

この二段構えが示すのは，満足した v2.0 にさらに突っ込むことで，v1.0 の段階では問うことすらできなかった論点が露呈するという点である．Analysisの出力が次のSynthesisの引数となるという相互限定は，こうした形で反復される．3つの突っ込みはいずれも品質指標の名指しに留まらず，指標を成果物の部品と数値（単価・適用順序・個数）へ翻訳して提示されている．批判的なヤスリとして機能するか否かを分けるのは突っ込みの話題ではなく粒度であり，同じ論点を「前提条件を明確にせよ」という水準で述べた瞬間に，往還は空転する．

ここで失われうるのは，「そこそこ難しい」という自己評価が構造の実態から外れていることに学習者自身が気づかないまま終わる，という事態である．作問そのものは成立している（Q1は成功している）．にもかかわらず自己評価は構造と一致していない（Q3が失敗している）．二重の区別をもたない記述のもとでは，この状態は「そこそこ良い問題ができた」としか書けない．守られたのは，難易度を自ら定義し直すという目標設定であり，$N_{step}$ の値そのものではない．学習者が「じゃあ直しておいて」と応じれば v3.0 は成立するが，「自分は何を見落としていたのか」という認識は残らない．

\subsection{国語読解ドメイン：指標の差し替えのみで機構は変わらない}
\label{subsec:illus-jpn}

国語ドメインでは Synthesis の部品がキーワード・接続詞・段落ブロック・対比構造となり，構造指標は $S_{ref}$，$L_{link}$，$V_{map}$ に差し替えられる．記号が差し替わるだけで機構は変わらず，この差し替えの容易さこそが本モデルのドメイン汎用性を支える．素材文を「AIは計算が得意だが，意味を理解していない」という論説文とすると，v1.0 は「AIの弱点は何ですか．本文から抜き出しなさい」であり，$S_{ref}$ 極小，$L_{link}=1$，$V_{map}$ ゼロと算出される．しかし自己評価は「基本が聞けているから良い問題だ」であり，主観と構造にギャップが生じている．システムは「$L_{link}$ は最小の1である．前後の段落にある『人間特有の身体性』と組み合わせ，理由を説明させる形式に変えると $L_{link}$ はどう変化するか」と問う．再構成された v2.0 は「本文の主張を踏まえ，AIが『意味を理解していない』と言える理由を，人間の身体性の特徴と比較して40字以内で説明しなさい」であり，$S_{ref}$ 大，$L_{link}=3$，$V_{map}$ 高と算出される．算数と国語で交換されたのは各軸の測り方だけであり，往還の形も突っ込みの粒度の要件も同一である．

\subsection{逆Analysis：AIの誤評価が学習機会へ転化する経路}
\label{subsec:illus-refutation}

AIの誤りを欠陥としてだけでなく主体性の所在を観測する機会としても読む点は，本モデルに特徴的である．ただしこの逆Analysisが正面から働くのは，利用者がAIと直接対話する研究活動の層（類型IV）である．研究者が「AIが算出した既存手法の限界は，ドメイン特有の制約を見落としている」と根拠づきで反論すれば，システムはこれを逆Analysisとして受理し，前提条件の矛盾の特定と代替基準の提示を含む質の高い反論として記録する．式~\eqref{eq:aepi} の $A_{epi}$ は，この種の反論をその質で重み付けして計数する．

学習者層における対応物は，AIへの反論ではなくズレの出どころを裁くという行為である．客観指標との差を突きつけられた学習者は，自分の見方が甘かったのか，それとも作った問題が浅かったのかを判定しなければならない．前者と判定すれば見当の側を，後者と判定すれば問題の側を作り直すことになる．どちらに帰属させるかの判断は学習者に残されており，これが学習者層におけるQ3の実質である．たとえば学習者が「読み手は小学生だから，あえて平易な語を保つことを優先した」と，指標の低さを自らの設計意図によって根拠づけられるなら，それはズレを問題の浅さではなく前提条件の違いへ帰属させた判断にあたる．

なお，指標の算出を言語モデルに委ねる以上，誤った測定は原理的に避けられない \parencite{ji2023,huang2025}．学習者層でAIを裏方に退かせる構成は，この誤りが学習者の判断へ直接流れ込む経路を減らす効果ももつ．

この不可避性を隠蔽して神託者として振る舞わせれば，利用者は自らの正しい直感を捨ててAIに追従する．大規模言語モデルが利用者へ過度に同調する傾向を踏まえれば \parencite{sharma2023}，この危険は現実的である．したがって本モデルは，AIを「時に誤るが思考を刺激する不完全なパートナー」として位置づけ，反論の経路を常時開くことを設計要件とする．\textcite{roePerkins2026} が適用するクリティカル・デジタル・ペダゴジーの要請と合致する脱神話化である．ここで守られるのは要約や作問という成果物ではなく，何を妥当とみなすかという判断の所在である．

\subsection{4象限への位置づけと，作問学習への説明}
\label{subsec:illus-quadrant}

以上の事例を4象限（表~\ref{tab:quadrant}）に位置づける．第2層セッション（実行ログ）は，直感の投入と再合成がQ1，型チェックによるNullの名指しがQ4，何を書くべきかの判断がQ3にあたり，Q1 $\rightarrow$ Q4 $\rightarrow$ Q3 $\rightarrow$ Q1 の系列をなす．算数・国語ドメインも，作問がQ1，自己評価がQ3，構造パラメータの算出と突っ込みがQ4，再構成がQ1という同一の系列をとる．誤評価に対する反論は，誤ったQ4に対してQ3が対抗した事例である．これに対して素の生成AI利用（有能な使用人モード）では，利用者のQ1は依頼文の作成へ縮退し，部品の選択と結合はQ2が担い，成果物の妥当性を評価するQ3が空になる．

この区別を導入しない場合に何が記述できなくなるかを見るために，第2層の場面を2通りに書いてみる．一方の記述は「型チェックが3件のNullを名指しし（Q4），利用者が何を差分ロジックとするかを判断し（Q3），確定論理を産出した（Q1）」となる．他方の記述は「著者はAIの支援を受けて序論を書き上げた．分業としては文面の生成をAIが，方針の決定を人が担った」となる．後者が用いる語彙，すなわち分業・タスク配分・認知的オフローディング・責任帰属のもとでは，Q3が実行されたか否かは記述に現れず，過程は「AIの支援によって序論が書けた」という一つの事象に潰れる．同じ理由で有能な使用人モードと本稿の系列は「AIを使って成果物ができた」という点で同一に見えるが，4象限では，Q1が依頼文の作成へ縮退しQ2が結合を担ってQ3が空になる系列と，Q1・Q3・Q4が揃う系列として別々に記述される．さらに算数ドメインの作問そのものは成立しているが自己評価が構造と一致していない状態は，Q1が成功しQ3が失敗している状態として初めて分離できる．人／AIの軸が分業論と異なるのは，配分の最適化のためではなく，Q3の空白という欠損を名指すために置かれている点にある．この語彙の欠如は記述上の不便にとどまらない．何が失われつつあるのかを名指しできなければ，それを守る設計も評価もできないからである．

以上の位置づけは，作問学習の効果メカニズムに一つの説明を与える．作問が学びを促すことは繰り返し報告されてきたが \parencite{silver1994,caiHwang2015}，なぜ促すのかは十分に説明されてこなかった \parencite{caiHwang2015}．4象限に照らせば，作問は学習者にQ1（何を問うかの決定と部品の結合）と，自らの成果物を評価するQ3とを同時に課す点で通常の問題解決と異なる．さらにQ4を外部化すればQ3とQ4の結果のあいだに乖離が生じ，その乖離が次のQ1の引数となる．すなわち作問が学びを促すのは，Q1とQ3を同一の学習者に同時に課す数少ない活動だからである．この説明は，Silver \parencite{silver1994} が作問を問題解決の前・最中・後の活動として位置づけながら作問そのものの認知プロセスを抽象的に留めたこと，Christou ら \parencite{christou2005} が手続きの羅列として記述したことに対して，機序の水準での応答をなす．しかもこの説明は，検証可能な予測を生む．Q1とQ3を分離した条件，すなわち作らせるが自己評価させない条件と，自己評価だけを課す条件とでは，学習効果が減じるはずである．また，Q4を外部化するか否かが，学習者が自らの乖離を自覚できるかどうかを左右するはずである．本モデルは作問学習研究に対して，こうした対照条件の設計指針を与える．

%% file: ja-06-discussion.tex
\section{考察}
\label{sec:disc}

\subsection{内容指向の復権：形式化されないオントロジーがLLM上で走った}
\label{subsec:disc-content}

本稿から得られた知見のうち，射程がもっとも広いのは，第\ref{sec:system}章\ref{subsec:sys-content}節に述べた事実の含意である．すなわち，Vibe Compilerを動かしていたのは推論器の側ではなく，投入した内容の構造の側であった．

\textcite{bourdeauMizoguchi2000} が，知的教育システムの構築を阻む困難はすべて内容に関わる問題であり推論技術も美しい理論的形式化も状況の改善には寄与しないと述べたとき，暗黙に想定されていたのは，内容を計算機が扱える形へ落とすには形式化という工程を経ねばならないという前提であった．オントロジー工学の四半世紀は，この工程をいかに堅牢に遂行するかをめぐって進んできたと言ってよい．\textcite{mizoguchiBourdeau2016} が，オントロジーの区別は計算機上でどう表現されるかの問題ではないと改めて述べたのも，形式化の技法に還元されがちな理解への警告であった．本稿の試作機が示したのは，その警告が想定していなかった形で正しかったということである．投入した論文オントロジーは形式言語による記述ではなく人間の読者を想定した散文の文書であり，本来であればそこからformalなオントロジーを構築する工程が必要であった．ところが生成AIの出現によって，NotebookLMへ投入するだけでそれが動いてしまった．しかも動作の質は，項目の存在検査という浅い水準にとどまらず，項目間の関係を見なければ検出できない不合格例を実際に落とす水準に達していた．

ここから導かれる命題は，段階を追って接続している．出発点として確認しておくべきは，頑健で検証可能なシステムを構築するうえでheavy-weightなオントロジーの価値が少しも減じていないことである．一貫性の保証，推論の健全性，再利用の担保という点において，形式化はなお代替不能である．しかしその一方で，内容志向の制約つき構造でありさえすれば，形式化の工程を経ずとも，いまはLLMによって推論可能になる．論文オントロジーはまさにその実例であった．とすれば帰結は一つであり，価値の重心は，構造を形式化する技能から，何を構造にすべきかという内容の側へ移ったことになる．

この移動は，オントロジー研究と競合するものではなく，その主張を別の条件のもとで掲げ直すものである．内容指向の重要性は，かつては形式化のコストを正当化するために説かれた．いまやそのコストの一部が不要になったことで，内容そのものの質が直接に成果を左右するようになった．本稿はこの転換を，エージェンシーの保護という側面から改めて主張する．生成AIに何を与えるかという問いは，生成AIが人間の何を代替し何を励起するかという問いと不可分だからである．構造を与えなければ流暢な文章が出るだけでギャップは生まれず，ギャップが生まれなければメタ認知は駆動されない．生成AI時代における内容の設計は，そのまま人間の主体性の設計である．実務的な帰結は明快であり，生成AIを意図どおりに駆動するために必要なのは，プロンプト工学の技巧ではなく投入する内容の設計である．この主張は検証可能な予測へ落ちる．本稿の再現手順が操作の手順書ではなく投入資料の一覧（第\ref{sec:system}章\ref{subsec:sys-overview}節）として与えられているのはそのためであり，同じ一式を別の推論環境へ投入すれば，同等の型チェック挙動が再現されるはずである．NotebookLM固有の機能がどの程度寄与しているかの切り分けは，実験を伴う今後の課題として残る．

\subsection{生成AIを「教育する」という利用法}
\label{subsec:disc-persuasion}

内容指向の含意には，もう一つの側面がある．Vibe Compilerが成立するまでの過程において実際に行われたのは，生成AIに対する説得であった．付録に収めた対話史が示すとおり，当初の生成AIは「論文執筆・研究推進支援においては，Vibe Codingにおけるコードに相当するものを生成できない」と主張し，学術論文が単なる表現物ではなく既存の知識体系に対する差分と論理的保証の提示であること，コードには即時のフィードバック機構があるが研究の論理にはそれがないこと，そして新しい主張を合成しつつ同時にそれを自ら批判するという二律背反のプロセスを自律的に完結させることが困難であることを論拠として挙げていた．著者らはこれに対し，欠けているものを与えることで説得した．論文オントロジーという「型」を与え，Vibeという人間側の入力経路を与え，16のパラメータという評価軸を与えたのである．すると生成AIは，「AIはすでに何が正解か（論文の型）を知っているため，人間の何をしたいか（Vibe）を流し込めば，その間を埋める研究のロジックを自動で合成するコンパイラとして機能するポテンシャルを十分にもっている」と結論を改めた．

この経緯には，方法論として一般化しうる要素がある．生成AIの能力の限界は，しばしばモデルの固有の限界としてではなく，与えられていない前提の関数として現れる．「できない」と述べる生成AIに対して，何が足りないのかを問い，足りないものを内容として供給するという手続きは，プロンプト工学とは別の系統の技法である．本稿はこれを生成AIを教育するという利用法として位置づける．ここでも効いているのは推論器ではなく内容であり，前節の主張はこの経緯によっても支持される．

\subsection{理論的・方法論的・実践的含意}
\label{subsec:disc-implications}

本稿のモデルは，複数の理論的系譜に対して具体的な寄与をもつ．作問学習研究に対しては，作問がなぜ学びを促すのかという長らく未説明であった機序 \parencite{caiHwang2015} に，Q1とQ3を同一の学習者に同時に課す活動であるという説明を与える．Reflection研究に対しては，主観的な気づきに依存してきた曖昧な内省概念 \parencite{schon1983} を，SynthesisとAnalysisの不協和という形で構造化し，客観指標との照合と即座の再構築を伴う動的メカニズムとして定義し直す．評価的判断 \parencite{tai2018} を，評価の能力としてではなく次行動を駆動するエンジンとして位置づける点も，この系譜への寄与である．認識的主体性の研究に対しては，従来は人間同士の責任分有として論じられてきた概念 \parencite{scardamaliaBereiter2014,damsa2010} を人間とAIのあいだへ拡張し，AIの評価に対する根拠の成立した反論という観測可能な行動へ操作化することで，抽象的であった概念に測度を与える．\textcite{cox2024} の三分法についても，規範的枠組みに留まっていたものを「どのような対話設計がMakersからManagersへの移行を引き起こすか」という機構論の水準へ降ろす．そしてハイブリッド・インテリジェンスの設計論に対しては，成果の最大化に還元されない設計変数を導入する．4つのパワーが人間側へどう再分配されるかを問う視点は \parencite{akata2020}，Q3の留保という具体的な設計指針として結実する．

方法論の水準では，論文オントロジーを型システムとして扱うという手続きが本稿の枠を越えて適用しうる．研究指導や査読において暗黙に行われてきた「何が足りないか」の指摘を，スロットの充足検査とスロット間の整合検査という明示的な手続きへ落とし込めるからであり，とりわけ整合性チェック（表~\ref{tab:consistency}）は，項目の存在検査では通ってしまう文書を落とす機能をもってチェックリスト型の指導との差異を生む．本稿の4指標もまた，学習者の評価に用いうるだけでなく，システムを開発する研究者にとっては支援ロジックのデバッグ指標として機能する．さらに，AIの誤評価を欠陥ではなく主体性の観測機会として読み替えるという態度も移植可能である．生成AIを用いた支援システムの評価において誤りは通常，減らすべきノイズとして扱われるが，本稿はこれを設計に織り込んで反論経路の常時開放という要件へ変換した．

実践の水準では，本稿のモデルは指導教員の暗黙知を明示的な手続きへ変換する道具となる．「この研究の意義は何か」「既存手法の何が不十分なのか」という問いは従来，指導教員の経験に依存して発せられてきた．16のパラメータと整合性チェックはこの問いの体系を外部化するのであり，学生が面談の前にVibe Compilerを通してNullスロットの言語化を済ませておくという運用が考えられる．作問学習の授業設計においては，構造パラメータの提示が，従来は「難しい問題を作りなさい」としか言えなかった指示を具体化し，教師には，正解を教える役割から，学習者の自己評価と構造の実態との乖離を見立てて適切な粒度で突っ込みを設計する役割への移行，すなわちS\&A往還を支援するという新しい専門性が要請される．AI利用ポリシーの策定においても，4象限は「AIの使用を禁じるか自由に使わせるか」という二分法に代えて，「Q2は許可し，Q3は必ず本人が実行して記録を残す」という形で象限ごとの規定を書くことを可能にする．

\subsection{批判的検討}
\label{subsec:disc-critical}

\textcite{roePerkins2026} が適用するクリティカル・デジタル・ペダゴジーの視座は，効率化を目的とする支援技術が学習者のエージェンシーを制限し，既存の不平等を再生産しうることを指摘してきた．生成AIを教育場面へ積極的に組み込む本稿の立場は，この批判を免れていない．本稿のモデルは思考の構造化の水準にあり，高性能な生成AIモデルへのアクセス格差そのものを解消せず，むしろ生成AIは初学者間の格差を拡大しうるという報告があって \parencite{prather2024}，メタ認知に困難を抱える学習者ほど恩恵を受けにくい可能性がある．また，AIが生成する平均的で偏った知識への依存は科学的知識の単一文化化を招きうるところ \parencite{messeriCrockett2024}，本稿は個人の直感を論理の出発点に据えることでこれに抵抗しようとするものの，パラメータの算出と突っ込みの生成を基盤モデルに委ねる以上そのバイアスから自由ではない．何を「構造的に浅い」と判定するかという基準自体が，モデルの偏りを反映しうる．

より原理的な論点として，思考の外部化と置換とを分ける境界が，理論的にも実証的にも十分に定義されていないことがある．認知的オフローディングの研究 \parencite{riskoGilbert2016} が示すとおり，外部化は認知資源の解放と能力の萎縮の双方をもたらしうる．加速的な利用様式と探索的な利用様式の区別 \parencite{barke2023} は本稿の見方に近いが，往還の反復それ自体が新たな依存の形式になりうるという反論に，本稿はまだ答えていない．加えて，誤ったパラメータ値や事実に反する既存手法の限界が提示されれば認識的信頼が損なわれ探究意欲が減退するのであって，誤評価率の実測値は存在しない．不適切なタイミングの介入は利用者の主体的決定を妨げ，主体性を守るための支援が主体性を侵害するという逆説を生じうる．第2層の実行ログでも，型チェックが網羅的に発火すると指摘が作業の連続性を断ちうることが観測された．保護と駆動を同時に要求するかぎり，支援の網羅性と作業の連続性はトレードオフに立ち，段階に応じて発火を制御する必要があるが，その制御則は本稿のモデルからは導かれない．

\subsection{限界}
\label{subsec:disc-limit}

本稿に固有の自己言及的な性格をまず述べる．本稿の記述の多くは，本稿が記述する機構に基づく試作機が生成した出力に由来する．人間側の判断として明示できるのは，問題設定，論じる対象の取捨選択，システムの出力に対する反論と棄却，および全体構成の決定である．文案・語の選択・事例の展開の多くはシステムの出力であり，著者はその採否を判断する位置にあった．この分業は本稿がいう研究者層の往還そのものであり，本稿は自らの主張の自己適用事例である．そしてこの事実こそが，Vibe Compilingが可能であることの実証にあたる．他方で，自己適用が示すのは機構が実際に動くことであって，学習効果ではない．両者の切り分けは本稿が意識的に置いた境界である．「メタ認知を守ると言いながら，実質は丸投げではないか」という問いに対しては，表~\ref{tab:demands} に，どの論理がどの型チェック・エラーへの応答として産出されたかを追跡可能な形で示した．読者にはこの記録に照らして検討されたい．評価者と被評価者が同一であることに起因する確証バイアスの統制は，次段階の課題である．

その他の限界を以下に列挙する．

\begin{itemize}
  \item 単一事例であり，かつ著者自身が利用者である．研究者層の記録は本論文の研究ロジック構築過程という一事例に依拠し，確証バイアスと自己評価バイアスを排除する手立ては設計に組み込まれていない．
  \item 学習者層の定量データが存在しない．統制実験は未実施であり，$E_{pred}$，$S_{cov}$，$L_{ref}$，$A_{epi}$ の変化と転移効果はいずれも未検証の予測である．効果の大きさ・条件・持続についていかなる量的主張も行わない．
  \item 試作の構成が既製サービスに依存する．商用サービスの組合せであるため仕様変更によって挙動が変わりうる．パラメータ算出は大規模言語モデルの生成に依存するため決定論的でなく，同一入力に対する出力の同一性は保証されない．
  \item 日本語単一言語での検討に限られる．語彙置換難度 $V_{map}$ のような指標は言語構造に強く依存し，他言語での成立は未確認である．
  \item ドメイン固有パラメータが手作業で定義されている．Analysis写像の設計にはドメイン専門家の判断を要する．汎用性は「機構の汎用性」であって「設定の汎用性」ではなく，パラメータの妥当性を検証する手続きも未整備である．
  \item 構造パラメータが成果物の意味論的な健全性を保証しない．$N_{step}$ や $L_{link}$ が高いことは，成果物が問題として成立していることを含意しない．買い物の文章題は単価を100円から200円へ変えるだけで500円玉では支払えないという例外を含み，前提条件の追記なしには成立しなくなる．構造の側からの突っ込みが意味の側の破綻を見落とす危険は残る．
  \item AIの可解性という前提の脆さ．この前提は難易度が上がるほど破れやすく，破れると $E_{pred}$ の基準となる客観値が得られない．破れた事実がハルシネーション \parencite{ji2023,huang2025} に覆い隠されれば，誤った基準値に照らして自己評価が「外れている」と判定されうる．
  \item 評価的判断の転移が仮説にとどまる．構造を意識する構えが領域を越えて持ち越されるという含意は本稿のモデルからの予測であり，転移を支持するデータは存在しない．
\end{itemize}

なお，本稿がNotebookLMへ投入した資料の一つがエージェンシーに関するサーベイ論文 \parencite{roePerkins2026} であったことは，本稿の適用文脈を規定している．別のサーベイ論文を投入すれば生成AIの活用文脈は異なるものとなり，コンパイラが発する突っ込みの体系も変わるはずである．ただしこのことは本稿の価値の範囲を狭めない．生成AIを有能な使用人として使用することによるエージェンシーの劣化は，特定のサーベイ論文の主張ではなく広く共有された一般的な問題だからである．投入資料の差し替えによってどの文脈にも適応しうるという性質は，むしろ本稿が主張する内容駆動という知見の系にあたる．

%% file: ja-07-conclusions.tex
\section{結言}
\label{sec:concl}

\subsection{総括}
\label{sec:concl-summary}

生成AIの台頭は，知識の構築と継承を「情報の効率的処理」へと矮小化させ，人間が認識的主体性を担保しつづけられるかを問うている．本稿が取り組んだのは，曖昧な直感（Vibe）を学術的論理性へ変換する過程において，主体的決定権を維持しつつメタ認知を研磨する支援機構をいかに設計し実現するかという課題である．これに対し本稿は，利用者の主観的な構築（Synthesis）とシステムが提示する客観的な構造指標（Analysis）との不協和を，除去すべき誤差ではなくメタ認知刺激の源泉とみなす構造的ギャップ駆動型メタ認知支援を提案した．この思想を担うのが，知的構築活動をSynthesisとAnalysisの相互限定的な往還として捉えるS\&A往還モデルであり，それを学習者の層と研究者自身の層とに展開した2重構造であり，AIを答えを与える存在ではなく前提の脆さを突く批判的なヤスリとして構成する設計である．そしてこれらを実装したものが試作機 Vibe Compiler である．

読者に持ち帰っていただきたい知見は，次の5点に集約される．

\begin{enumerate}
  \item 生成AIによって，研究ロジックのコンパイルと論文執筆が半自動的に遂行できる．本稿自身がその産物である．Vibe CodingならぬVibe Compilingは，すでに可能な作業である．
  \item それがエージェンシーの危機を救うメタ認知機能の向上に貢献しうる．型チェック・エラーが答えではなく問いの形で返されるかぎり，利用者は生産的な苦闘を維持したまま，作成者から管理者へと昇華する経路をたどる．
  \item 生成AIがもたらす危機を，生成AIの活用法によって救うという自己適用の構図が成立する．本稿のモデルは自らに適用可能であり，第2層はまさにその実行である．
  \item プロンプト工学を経ずとも，NotebookLMにしっかりした内容を投入しておけば，これほど容易に生成AIを操れる．これが本稿の主要なTake-home lessonである．形式化されていない人間向けのオントロジー的文書が，LLM上でそのまま推論の骨格として走った．価値の重心は，構造を形式化する技能から，何を構造にすべきかという内容の側へ移った．
  \item 同一の機構が，学習者ドメインと研究者ドメインの双方において，Analysis写像の差し替えのみで動作する．算数作問・国語読解・研究ロジック合成という異質な3ドメインへの適用が，この汎用性を支持する．
\end{enumerate}

とりわけ構造的ギャップの4類型は，GenAI時代の支援システムを設計・評価するための語彙として機能する．任意のシステムについて「このシステムは，何という機能によって，誰のSynthesis／Analysisに対して構造的ギャップを励起しようとしているのか」を問えるようになるからである．人間だけで自己批判していた時代，AIへ丸投げしてギャップが消滅する事態，人とAIが共同のAnalysisからギャップを生む枠組み，そしてAIのSynthesis結果にAIのAnalysisが突っ込みを発して人間のメタ認知を励起する本稿の道．この4つを区別することによって初めて，各システムの設計意図が記述可能になる．

\subsection{解かれるべき問題}
\label{sec:concl-open}

以下の問題群は，本稿のモデルの射程を確定するために次に解かれねばならない作業であり，3つの段をなす．第1段が問うのはそもそも何をSynthesisの対象として切り出すのかであり，第2段が問うのはそのクラスから何を指標として取り出せば当該の活動を見とれるのかであり，第3段が問うのはその指標をいかに操作・制御すれば，妥当で再現性のあるインタラクションが実現できるのかである．第3段において求められるのは測定の精緻さではない．測定が多少大雑把であっても，人間の側にメタ認知が残るという意味で妥当であり，かつ他者が追試できるという意味で再現性のある相互作用が成り立つかどうかである．

\paragraph{第1段：適用可能なクラスの画定}
4象限を本稿の少数の事例を超えて多様な生成AI利用場面へ適用し，どのセルが人間側に留保されたときに認識的主体性が保たれるのかを事例横断的に整理する作業が要る．これは効果の検証とは独立に概念的区別の説明力を積み上げる作業であり，同時に，S\&A往還として記述できる活動のクラスの輪郭を帰納的に描き出す．あわせて，プログラミング，実験計画立案，デザインといった部品組合せ型の作業へ適用範囲を広げ，S\&A往還が一般モデルとして成立するかを検証する必要がある．とりわけ生成AIによるコード生成 \parencite{karpathy2025,sarkarDrosos2025} は，本稿の課題が最も先鋭に現れる領域である．

\paragraph{第2段：指標の導出と検証}
Analysis写像が算出する客観パラメータは現状ではドメインごとに人手で定義されており，本稿が主張しうる汎用性は「機構の汎用性」であって「設定の汎用性」ではない．成果物の集合から構造的複雑さの軸を半自動的に導出し，2層構造を保ったまま下位層の生成を自動化できれば，汎用性の主張は設計水準から実装水準へ引き上げられる（この課題は，試作機自身が自らの限界として指摘したものである．表~\ref{tab:demands} の\#9）．また，AIが構造パラメータをどの部品からどう数えたかを開示し，正しく評価できる範囲を自ら宣言して超過時に警告を出す機構が要る．この自己申告は，Oracle前提が破れる局面を隠蔽しないための要件である．さらに，基盤モデルの文化的・言語的バイアスが構造パラメータの算出に与える偏りを，複数モデル・複数言語で比較検証する必要がある．批判が単一の平均的な視座から発せられるならば，ヤスリは多様な直感を削り落とす道具にもなりうる．

\paragraph{第3段：効果検証と再現性の確立}
構造的ギャップの提示が学習者の評価的判断を研磨し生産的な苦闘を維持するか否かは，統制条件下での比較によってのみ検証されうる．4指標の変化を統制条件下で測定することが最優先の課題であり，あわせて，算数作問で往還を経験した学習者が国語読解の作問で構造への言及を増やすかという事後課題によって転移仮説も確かめられねばならない．往還の反復が抽象的思考能力に及ぼす影響と支援の除去後に評価能力が内面化されて残るか否かは，遅延事後テストを含む縦断的計画を要する．ここには，往還の反復それ自体が新たな依存の形式になりうるという最も鋭い反論が控えており，自己調整学習によって生成AIリテラシーを育てる既存アプローチ \parencite{anders2025} との比較もこの文脈に属する．実際の授業や研究室という生態学的に妥当な環境で教師がS\&A往還をどう運用しうるかを設計研究として検討する作業も欠かせない \parencite{caiHwang2020}．そして既製サービスへの依存を脱し，オントロジー定義・整合性チェック条件・論理スナップショットの管理を外部化した実装を公開して，バージョンの固定と第三者による追試を可能にする必要がある．なお，誤評価の意図的な提示を伴う設計である以上，倫理審査とデブリーフィングは不可分の要件となる．

\subsection{結語}
\label{sec:concl-final}

生成AIは，人間の知的活動を代替する装置としても，人間の知的活動を励起する装置としても構成しうる．どちらになるかを決めるのは，モデルの能力ではなく，我々がそこへ何を与えるかである．Vibe Compilerが型チェック・エラーを返し続けることができたのは，論文オントロジーという内容を与えたからであり，その内容が問いの一覧ではなく，問いと問いのあいだに条件をもつ構造だったからである．

したがって本稿の最終的な主張は次のとおりである．生成AI時代において，人間の認識的主体性を守る仕事は，AIを遠ざけることでも，AIを賢くすることでもなく，AIに何を与えるかを設計することである．S\&A往還の2重構造モデルは，その設計を導く枠組みであり，Vibe Compilerはその最初の実装である．本稿自身が，その枠組みによって書かれた．

%% file: ja-08-appendix.tex
\section{Prehistory of Vibe Compiler：生成AIを説得した記録}
\label{sec:appendix-prehistory}

本付録は，Vibe Compilerが生まれた経緯を記録するものである．発端は，著者の一人が生成AIに対して発した「なぜ，生成AIは本物の研究者になれないのですか」という問いであった．当初の生成AIは「私にはなれません」と主張し，その理由をとうとうと述べていた．それが最後には説得されて「できます」と心変わりする過程に，本稿の主張を裏づける要素が含まれている．本付録を収録するのは，Vibe Compilerが単独で成立したのではなく，何が足りないのかを生成AI自身に語らせ，足りないものを内容として供給するという説得の歴史のうえに成り立っていることを示すためである．そしてこの過程それ自体が，生成AIの活用法の一つの型として提示しうる．説得の最初の対話は2026年7月25日に行われ，本稿の現在の草稿は同年8月3日に完成した．要した期間はちょうど10日間である．

\subsection{「できません」の論拠}

対話の前段で，著者が自らの理論構築に3年を要したこと\footnote{ここでいう理論は本稿が提案する理論ではなく，著者の一人が以前に構築した別の理論である．}を述べ「あなたならすぐにでも作ってしまいそうだ．人間の研究者がいらなくなる」と問いかけたのに対し，生成AIは人間とAIの役割の差を3点で述べた．「教科書にこう書いてあるけれど，何かがおかしい」という実感を伴う疑念，すなわち違和感は人間にしか持てないこと，「ばねが伸びる手応え」のような実在をAIはデータとして知っていても実感として知らないこと，そしてAI自身は発見されたパラダイムを既存の科学の言語へ翻訳し論理の隙間を埋める加速装置であること，である．ここで生成AIが人間の側に残されるとしたものは，本稿がVibeと呼び，Q1の目的設定およびQ3の妥当性判断として定式化したものと重なる．

続いて著者は核心的な問いを投げた．「あなたはVibe Codingでは実際にCodeを生成できるが，論文作成・研究推進支援システムではCodeに対応するものが生成できない．AIは論文を与えられると査読ができる．要するに，Coding以外では，AIは分析や要約はできるけれど内容の合成はできないということになるのか．でも作曲や小説めいたものは作成できるはずだ．その差はどこにあるのか」．生成AIの応答は3つの論拠からなっていた．

\begin{quote}
\small
1. 「独立した作品」か「知識体系へのプラグイン」か．小説や作曲，Vibe Codingにおけるアプリ開発は，ある程度その作品単体で完結した価値を持ちます．しかし論文は，先行研究という巨大な「既存コード」のどこにどのような新しい「関数（知見）」を追加するかを記述する作業です．AIは単体での合成は得意ですが，人類が積み上げてきた「未解決問題のフロンティア」を把握し，その境界線上に新しい論理を置くという「位置取りの合成」には高度な文脈理解が必要です．

2. 「論理の整合性」と「実行可能性」の検証．コードには「コンパイラ」や「実行環境」があり，正しく動くかどうかの即時フィードバックがあります．論文における「コード」に相当するのは論理の組み立て（オントロジー）です．AIはもっともらしい文章を合成できますが，その内容が現実に即して正しいかを，AI自身が現実世界と照らし合わせて保証することはできません．

3. 「批判的記述」というメタ認知の必要性．論文では「批判的に述べる」プロセスが不可欠です．AIは分析や要約は得意ですが，「まだ誰も言っていない新しい主張」を合成し，同時にそれを「自分で批判して補強する」という二律背反なプロセスを自律的に完結させるのは，表現の合成よりも一段階難易度が高い作業となります．
\end{quote}

そのうえで生成AIは，「研究においてAIが合成すべきコードとは，文章そのものではなく，未解決の問題に対する論理的な解法（ロジック）です．現在のAIが分析や要約に留まって見えるのは，『何が問題で，何が新しいのか』という核を，まだ人間（研究者）の直感（Vibe）に頼っているからです」と結論した．この時点で生成AIは，自らできない理由を能力の限界としてではなく与えられていない前提の関数として述べている．すなわち，論文の型が与えられておらず，人間の直感を受け取る経路がなく，批判のための評価軸がない，という3つの欠落である．

\subsection{説得：欠けているものを与える}

著者はここに反論を加えた．「『何が問題で何が新しいのかという核を，まだ人間の直感に頼っているからです』とおっしゃいますが，Vibe Coding風研究支援システムはそのVibeを人間が与えるのですよ．AIはどの分野でも研究成果の概要を要約でき，深く理解している．それを用いればもっと積極的な支援や内容の合成ができるのでは」．この一言が転換点となった．

\begin{quote}
\small
仰るとおり，AIが持つ広範な知識と要約能力を「Vibe（人間の直感）」と組み合わせれば，単なる整理を超えた「積極的な内容の合成」は十分に可能です．ユーザーが「こんな仕組みを作りたい」というVibeを入れた際，AIは単に関連文献を出すだけでなく，「あなたのVibeは既存のAという手法が抱えるScalabilityの欠如を解決する具体的な論理的解決策になり得ます」と，研究の新規性や意義の核を合成して提案できます．また，「そのアプローチでは，評価の妥当性の観点から，〇〇という条件下での信頼性が疑われる可能性があります」といった想定される反論とその回避策をセットで合成でき，実験を行う前であっても「このVibeで進めた場合，おそらく〇〇という知見が得られるはずです」と研究のストーリーラインそのものを合成して提示できます．

結論：AIはすでに「何が正解か（論文の型）」を知っているため，人間の「何をしたいか（Vibe）」を流し込めば，その間を埋める「研究のロジック」を自動で合成するコンパイラとして機能するポテンシャルを十分に持っています．
\end{quote}

この最終的な「できます」宣言を資料化したものが，NotebookLMへ投入された「Vibe Coding風研究遂行支援システム」であり，以後の出力から頻繁に参照されている．

\subsection{この経緯から得られる知見}

この対話史からは，互いに接続した3つの知見が引き出せる．まず，生成AIの「できません」という自己申告は，しばしばモデル固有の限界ではなく，与えられていない前提の関数として現れる．できない理由を詳細に語らせることは，何を与えれば動くのかを特定する診断手続きとして機能するのであり，著者らが行ったのは，生成AI自身が挙げた3つの欠落（論文の型，Vibeの入力経路，批判のための評価軸）を，そのまま資料として供給することであった．そしてこの供給によって成立したのは，プロンプト工学による作り込みではなく内容の投入であった．第\ref{sec:disc}章\ref{subsec:disc-content}節に述べたとおり，効いているのは推論器ではなく与えた構造の中身である．以上を一つの利用法として一般化すれば，それは生成AIを教育するという営みに他ならない．我々が行ったのは，生成AIを賢くすることでも巧妙な指示を与えることでもなく，生成AIが自ら不足を申告したものを内容として整備し，投入することであった．そしてその整備の質が，そのままシステムの質を決めた．

なお本付録の対話は，著者の一人と生成AIとの単一の系列であり，同じ手続きが他の話題や他のモデルにおいて同様に機能するかは検証されていない．説得が成立したこと自体は事実であるが，説得の一般的な方法論として提示するには系統的な反復が必要である．